\documentclass[a4paper,11pt]{article}
\pdfoutput=1 

\usepackage{jheppub} 

\usepackage[T1]{fontenc} 
\usepackage[utf8]{inputenc}
\usepackage{titlesec}

\titleformat{\paragraph}
{\normalfont\normalsize\bfseries}{\theparagraph}{1em}{}
\titlespacing*{\paragraph}
{0pt}{3.25ex plus 1ex minus .2ex}{1.5ex plus .2ex}

\newcommand{\Grad}{\nabla\!\!\!\!\nabla}

\usepackage{bbm}

\usepackage[table,dvipsnames]{xcolor}
\usepackage{array,etoolbox}
\preto\tabular{\setcounter{magicrownumbers}{0}}
\newcounter{magicrownumbers}
\newcommand\rownumber{\stepcounter{magicrownumbers}\arabic{magicrownumbers}}

\title{\boldmath AdS Euclidean wormholes}

\author[a]{Donald~Marolf,}
\author[b]{Jorge~E.~Santos}

\affiliation[a]{Department of Physics, University of California at Santa Barbara, Santa Barbara, CA 93106, U.S.A.}
\affiliation[b]{Department of Applied Mathematics and Theoretical Physics, University of Cambridge, Wilberforce Road, Cambridge, CB3 0WA, UK}

\emailAdd{marolf@ucsb.edu}
\emailAdd{jss55@cam.ac.uk}

\abstract{We explore the construction and stability of asymptotically anti-de Sitter Euclidean wormholes in a variety of models.  In simple ad hoc low-energy models, it is not hard to construct two-boundary Euclidean wormholes that dominate over disconnected solutions and which are stable (lacking negative modes) in the usual sense of Euclidean quantum gravity.  Indeed, the structure of such solutions turns out to strongly resemble that of the Hawking-Page phase transition for AdS-Schwarzschild black holes, in that for  boundary sources above some threshold we find both a `large' and a `small' branch of wormhole solutions with the latter being stable and dominating over the disconnected solution for large enough sources.  We are also able to construct two-boundary Euclidean wormholes in a variety of string compactifications that dominate over the disconnected solutions we find and that are stable with respect to field-theoretic perturbations.  However, as in classic examples investigated by Maldacena and Maoz, the wormholes in these UV-complete settings always suffer from brane-nucleation instabilities (even when sources that one might hope would stabilize such instabilities are tuned to large values).  This indicates the existence of additional disconnected solutions with lower action.  We discuss the significance of such results for the factorization problem of AdS/CFT.
}

\begin{document}
\maketitle
\flushbottom

\section{Introduction}

\begin{figure}[h]
\centering
\includegraphics[width =\textwidth]{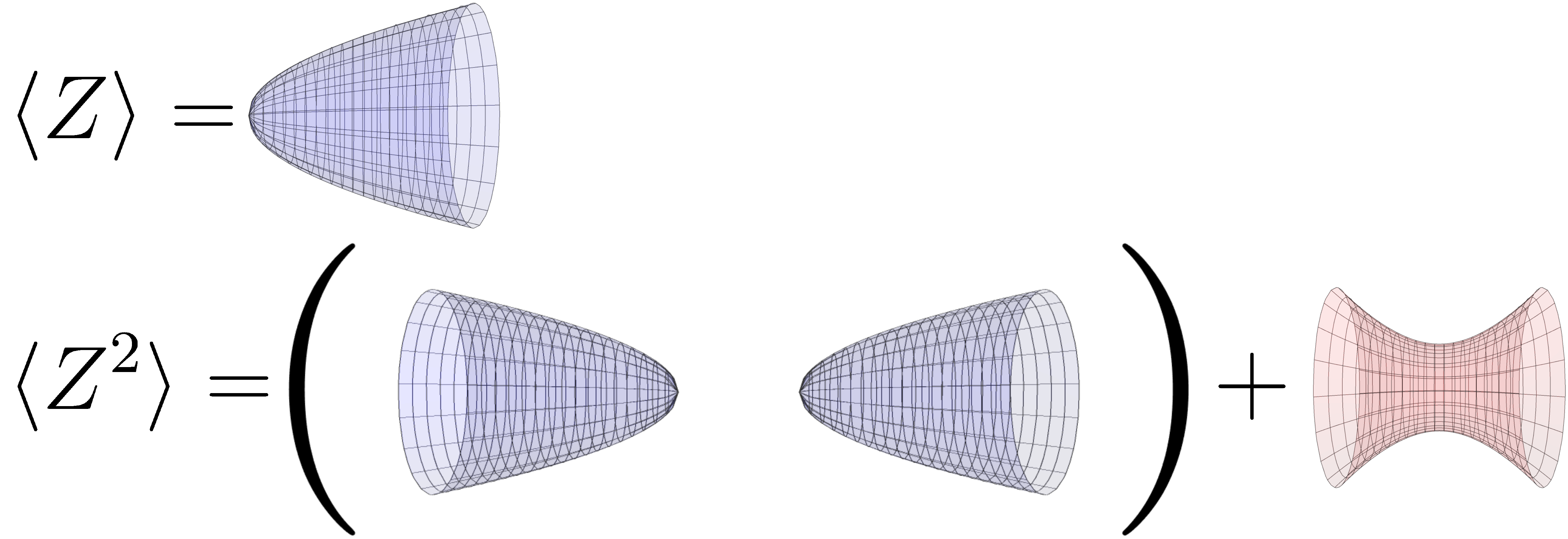}\\
\caption{An example showing failure  of factorization due to spacetime wormholes.  The top line represents a path integral $\langle Z\rangle$. Although we have drawn the configuration as connected, it may include contributions from disconnected spacetimes as well.   In any case, the natural path integral $\langle Z^2 \rangle$ associated with a pair of boundaries yields all terms generated by squaring $\langle Z\rangle$, but also contains additional contributions connecting the two boundaries as indicated by the second term in the bottom line.}
\label{fig:nofac}
\end{figure}

It has long been understood that S-matrices, boundary correlators, or boundary partition functions defined by bulk gravitational path integrals may fail to display familiar factorization properties due to contributions from spacetime wormholes \cite{Lavrelashvili:1987jg,Hawking:1987mz,Hawking:1988ae,Coleman:1988cy,Giddings:1988cx,Giddings:1988wv}.  For our purposes, it is convenient to define spacetime wormholes as connected geometries whose boundaries have more than one compact connected component.  This definition includes real geometries of any signature as well as those that are intrinsically complex.  Using an AdS/CFT language, the point is that a  boundary partition function $Z$ is naively represented by a bulk path integral over configurations with a single compact boundary.  Similarly, a product of boundary partition functions (say,  $Z^2$) is naively represented by a bulk path integral over configurations with two disconnected boundaries.  But contributions from spacetime wormholes suggest that the latter path integral (which we call $\langle Z^2 \rangle$) is not necessarily the square of the former path integral (which we call $\langle Z \rangle$); see figure \ref{fig:nofac}.

Such failures of factorization would clearly require the standard picture (see e.g. \cite{Maldacena:1997re,Gubser:1998bc,Witten:1998qj}) of AdS/CFT duality to be modified; see e.g. \cite{Coleman:1988cy,Giddings:1988cx,Giddings:1988wv,Maldacena:2004rf,Betzios:2019rds} for related discussions.
A possible resolution is that the bulk path integral is in fact dual to an ensemble of boundary theories, and our notation $\langle Z \rangle$, $\langle Z^2 \rangle$ is chosen to reflect this idea.  A non-zero ``connected correlator'' $\langle Z^2 \rangle - \langle Z \rangle^2$ is then interpreted as describing $\delta Z^2$ for fluctuations $\delta Z$ that allow the partition function $Z$ in any particular element of the ensemble to differ from the ensemble-mean $\langle Z \rangle$.    Such an effective description was derived\footnote{While these works pre-date the discovery of AdS/CFT, their arguments apply immediately to that context.}  in \cite{Coleman:1988cy,Giddings:1988cx,Giddings:1988wv} under certain locality assumptions, though this assumption can be dropped by using the argument of \cite{Marolf:2020xie}. In addition, dualities of this kind have been explicitly constructed between (appropriate completions of) various versions of Jackiw-Teitelboim (JT) gravity and corresponding double-scaled random matrix ensembles \cite{Saad:2018bqo,Saad:2019lba,Stanford:2019vob}. See also \cite{Betzios:2020nry} for discussion of related issues for  $c=1$ matrix models, and \cite{Garcia-Garcia:2020ttf} for discussion of ensembles of Sachdev-Ye-Kitaev  models \cite{Sachdev:1992fk,Kitaev} and a proposed relation to wormholes in Jackiw-Teitelboim gravity \cite{Jackiw:1984je}
\cite{Teitelboim:1983ux} coupled to matter fields. Discussions of off-shell wormholes in both JT gravity and pure gravity in AdS$_3$ can be found in \cite{Cotler:2020ugk,Maxfield:2020ale}; see also \cite{Afkhami-Jeddi:2020ezh,Maloney:2020nni} for related discussions of averaging 2d conformal field theories.

However, it is far from clear that this ensemble interpretation will hold for the most familiar examples of AdS/CFT.  In particular, such cases involve bulk theories with large amounts of supersymmetry, and this supersymmetry should be reflected in each member of the boundary ensemble\footnote{This property holds in the matrix model examples of \cite{Stanford:2019vob}. A general argument follows from the fact that the full bulk system admits an algebra of asymptotic SUSY charges, and that these charges must act trivially on the `baby universe sector' of the theory (i.e., on the ${\cal H}_{BU}$ of \cite{Marolf:2020xie,Marolf:2020rpm}). Thus the SUSY algebra acts within each bulk superselection sector.  The boundary dual interpretation is then that each member of the associated ensemble has a well-defined SUSY algebra.}. However, in more than two boundary dimensions $d$ the set of local boundary theories with large amounts of supersymmetry is expected to be very limited; e.g., for $d=4$ and ${\cal N}=4$ SUSY, super Yang-Mills theory is known to be the unique local maximally-supersymmetric theory that admits a weakly-coupled limit\footnote{This follows from classifying the supersymmetric marginal deformations of free field theory.}, and may thus be the unique local theory.  One might thus expect that -- at least when the full physics associated with the UV-completion of bulk gravity is taken into account -- the ensemble of dual boundary theories degenerates in this case so as to contain only a single physically-distinct theory (here $d=4, {\cal N}=4$ SYM); see related discussions in \cite{Maldacena:2004rf,Buchel:2004rr,ArkaniHamed:2007js,Marolf:2020xie,McNamara:2020uza}.  On the other hand,  this would require strong departures from the low-energy semi-classical description of the bulk and would thus render unclear the status of recent apparent successes \cite{Penington:2019npb,Almheiri:2019psf,Almheiri:2019qdq,Penington:2019kki} in using semi-classical bulk physics to resolve issues in black hole information.

We thus return to the question of whether partition functions defined by higher-dimensional bulk path integrals should factorize across disconnected boundaries; i.e., whether in such cases we should in fact find $\langle Z^2 \rangle = \langle Z\rangle^2$.  Such factorization would require either some rule or effect to forbid spacetime wormholes from appearing in the path integral, or alternatively that in all computations one finds precise cancellations when one sums over all connected spacetimes with given disconnected boundaries.  We note that, while it may at first appear somewhat contrived, the latter option is precisely what occurs in one of the baby-universe superselection sectors described in \cite{Coleman:1988cy,Giddings:1988cx,Giddings:1988wv,Marolf:2020xie} and in the limit of the eigenbranes described in \cite{Blommaert:2019wfy} where one fixes all of the eigenvalues of the relevant matrices.  Of course, in both of those cases one has carefully tuned some extra ingredient (the baby universe state or the eigenbrane source) to make such cancellations occur.  Furthermore, this option is difficult to see explicitly unless one can solve the theory in detail.

It will come as no surprise that a complete study of the higher dimensional gravitational bulk path integral is far beyond the scope of this work.  We will therefore resort to the usual crutch of studying bulk saddle points.  Assuming that the contour of integration can be deformed to pass through our saddles, they should dominate over contributions from non-saddle configurations in the limit of small bulk Newton constant $G$.

Before proceeding, it is useful to review the literature concerning asymptotically AdS spacetime-wormhole saddle-points.  The simplest context in which one might imagine such saddles to arise would be Euclidean pure AdS-Einstein-Maxwell theory with two spherical boundaries (each $S^d$ for a $d+1$-dimensional bulk).  However, such solutions are forbidden by the results of \cite{Witten:1999xp}, which showed that spacetime-wormhole saddle-points cannot arise in Euclidean pure AdS-Einstein-Maxwell theory \cite{Witten:1999xp} when the scalar curvature of the boundary metric is everywhere positive.  On the other hand, although none of these solutions have all of the properties that one might desire, a variety of Euclidean spacetime-wormhole saddles have been constructed by allowing the boundary metric to be negative or by adding certain types of matter \cite{Maldacena:2004rf,Buchel:2004rr,ArkaniHamed:2007js} (with the latter based on the zero cosmological constant constructions of \cite{Giddings:1987cg}); see also \cite{Garcia-Garcia:2020ttf} in  the context of JT gravity with matter. We also refer the reader to the very interesting story of constrained instantons (off-shell wormholes) described in \cite{Cotler:2020lxj}, though at least as of now their stability has been analyzed only in theories of pure gravity (and JT gravity for $d=2$).

In particular, the spacetime-wormhole solutions of \cite{Maldacena:2004rf,Buchel:2004rr,ArkaniHamed:2007js}  may be divided into four categories.  The first are the Euclidean solutions of \cite{ArkaniHamed:2007js} which have spherical (and thus positive-curvature) boundaries but follow \cite{Giddings:1987cg} in using axionic matter.  The axion kinetic term can become negative in Euclidean signature, though in this case it is positive definite but has a surprising zero at a radius that depends on angular momentum. This unusual kinetic term and a potential that is negative in the region where the kinetic term is small combine to  allow  saddles to suffer from bulk negative modes associated with non-trivial bulk angular momentum\footnote{The situation is even worse if one makes a further analytic continuation of the axion according to $\phi \rightarrow i\phi$, as the kinetic term then becomes negative definite.  One might expect the same to be true in a reformulation writing the axion in terms of the Hodge-dual 2-form potential, but we have not carried out a detailed analysis.} \cite{Hertog:2018kbz}.  It is thus hard to argue that the dominate over non-saddles even at small $G$.  Other spacetime wormholes with field-theoretic bulk negative modes were described in section 4.2 of \cite{Maldacena:2004rf}.

The second are Euclidean wormholes with negative-curvature boundaries (perhaps compact hyperbolic manifolds).  As discussed in \cite{Witten:1999xp}, in the simplest AdS/CFT examples such solutions have string-theoretic negative modes associated with the nucleation of D-branes.  Indeed, these negative modes render the entire theory unstable in the UV.  While the UV issues can be stabilized by appropriately deforming the CFT \cite{Maldacena:2004rf} (and, in particular, breaking conformal invariance), and while this will forbid any brane nucleation instability close to the AdS boundary, it was found in \cite{Buchel:2004rr} that -- at least in the model studied there -- the wormholes remain unstable to the nucleation of D-branes at finite locations in the interior.

The third class consists of Euclidean wormholes that have no known negative modes, and which can even have lower action than disconnected solutions, but which have no known embedding in string theory.  Examples include the large $\alpha$ solutions in section 4.2 of \cite{Maldacena:2004rf}. Finally, the fourth category (section 5 of \cite{Maldacena:2004rf}) contains Euclidean wormholes with known string-theoretic embeddings and no known negative modes, but where the corresponding disconnected solution is not yet known so that it is unknown which saddle dominates at large $G$.

We should also mention that the so-called `double-cone' solutions of \cite{Saad:2018bqo} define a 5th class of spacetime wormhole solutions.  These solutions are constructed by starting with e.g. the complexification of a two-sided AdS-Schwarzschild black hole and taking the quotient by a discrete Lorentz-signature time translation.  The real Lorentz-signature section of this quotient
is connected and has two compact boundaries, though it is also singular at the bifurcation surface.  However, there are non-singular complex sections (on which the quotient acts freely) that can be used to connect the same two real boundaries.  As a result, we prefer to think of the double-cone as an inherently complex solution.  This is in no way a fundamental problem, but we will instead discuss real Euclidean solutions below.  We hope to return later to a more detailed analysis of stability in the (complex) double cones.

Our goal here is thus to expand the class of known Euclidean spacetime wormholes with simple disconnected boundary metrics, which we will take to consist either of two copies of a (possible squashed) sphere ($S^d$) or two copies of the torus ($\mathbb{T}^d$). In particular, we wish to identify saddles without field-theoretic negative modes, where the saddles can be embedded in simple compactifications of string theory, and where the Euclidean wormholes dominate over disconnected saddles.   We will also explore negative modes associated with brane nucleation, and we will analyze the structure of any the phase transition associated with the exchange of dominance between Euclidean wormholes and disconnected saddles.

We begin in section \ref{sec:FLRW} by describing how a general class of potential Euclidean wormholes may be interpreted as homogeneous isotropic Euclidean cosmologies; i.e., as Euclidean Friedmann–Lemaitre–Robertson–Walker (FLRW) solutions.  This point of view provides useful intuition for why Euclidean wormholes with positive curvature boundaries are forbidden in pure AdS-Einstein-Hilbert gravity, and also for what sort of ingredients are required to overcome this obstacle.

Sections \ref{sec:u13} and \ref{sec:scalars} then study toy models in four bulk dimensions with simple bulk matter content that concretely illustrate the construction suggested by the FLRW analysis of section \ref{sec:FLRW} with $S^3$ boundaries.  As examples of bulk matter we consider both $U(1)$ gauge fields (section \ref{sec:u13}) and non-axionic scalar fields (section \ref{sec:scalars}).  These models do not directly embed in string theory, but turn out to be similar to some that do.  In the toy models we identify connected (wormhole) solutions that are free of bulk field-theoretic negative modes and which dominate over disconnected (non-wormhole) solutions in appropriate regimes.  As a function of the matter sources, we find the associated space of solutions to have structure much like that of the well-known Hawking-Page transition, including in particular a first-order phase transition associated with exchanging dominance between the connected and disconnected saddles. This is precisely the structure that was  found in studies of wormholes in JT gravity coupled to matter \cite{Garcia-Garcia:2020ttf} and in studies of constrained wormholes \cite{Cotler:2020lxj} in theories of pure gravity. Because these are only ad hoc low-energy theories or JT models, they do not contain fundamental branes.  Thus there can be no notion of brane-nucleation instability to explore in such models.

We therefore turn in sections \ref{sec:11dEU12}, \ref{sec:massdeformed}, and \ref{sec:IIB} to studying truncations of string-theory or $M$-theory. In particular, section \ref{sec:11dEU12} considers a truncation of 11-dimensional supergravity that reduces to $U(1)^2$ Maxwell-Einstein theory on AdS$_4$, section \ref{sec:massdeformed} examines a mass-deformed version of the ABJM theory \cite{Aharony:2008ug}, and section \ref{sec:IIB} investigates a truncation of type IIB string theory to an Einstein-scalar system on AdS$_4$.  In each of these cases we require spherical boundaries and construct wormhole solutions.  The structure of the space of solutions is generally much as in the toy models, with wormholes dominating over the most obvious disconnected solution at large values of the relevant boundary source, and with such wormholes being free of field-theoretic negative modes.  However, in all cases we find such would-be dominant wormholes to suffer from brane-nucleation instabilities.  In sections \ref{sec:11dEU12} and \ref{sec:IIB} the brane-nucleation occurs only in a finite region of the bulk and does not occur in the deep UV. Thus the theory itself remains stable with these boundary conditions and only the particular wormhole solution is destabilized.  In contrast, section \ref{sec:massdeformed} studies a context with hyperbolic boundaries of the form reviewed above, but with a deformation parameter $\mu$ that is expected to stabilize the theory at large enough $\mu$.  While it does in fact appear to do so, we find wormholes only in the small-$\mu$ regime where the theory remains unstable. The results of section \ref{sec:massdeformed} are thus similar to those found in \cite{Buchel:2004rr} for mass deformations of $\mathcal{N}=4$ SYM.

We close with some interpretation and discussion of open issues in section \ref{sec:disc}.  The main text is supplemented by various appendices with additional technical details.  This includes appendix \ref{app:torusboundaries}, which describes studies of two UV-complete models with torus boundaries (with results broadly similar to those associated with spherical boundaries).  It also includes appendix \ref{app:table}, which lists results for a larger set of 14 low energy models and 22 string/M-theory compactifications in a variety of dimensions that we have studied at least briefly but whose investigation we may not have chosen to describe in detail.  While the explorations of those models were not always as complete as the ones described in the main text (see appendix \ref{app:table} for details), they suggest that the results of sections \ref{sec:11dEU12}, \ref{sec:massdeformed}, \ref{sec:IIB} are typical.

\section{FLRW approach \label{sec:FLRW}}
Consider any Euclidean wormhole whose boundary consists of two copies of a maximally-symetric Euclidean geometry $\Sigma_d$; i.e.,  where $\Sigma_d$ is a sphere $S^d$, a Euclidean plane $\mathbb{R}^d$, or a hyperbolic plane $\mathbb{H}^d$.  If the metric of the full bulk solution preserves this symmetry, then the wormhole admits a preferred slicing which again preserves the symmetry. We may thus describe the wormhole as a $d+1$-dimensional homogeneous isotropic Euclidean cosmology with $S^d$, $\mathbb{R}^d$, or $\mathbb{H}^d$ slices and with Euclidean time running transverse to each slice.

We may thus write the metric in the (Euclidean) FLRW form
\begin{equation}
\mathrm{d}s^2 = \mathrm{d}\tau^2 + a^2(\tau) \mathrm{d}\Sigma_d^2,
\end{equation}
with $\mathrm{d}\Sigma_d^2$  the standard metric on $S^d$, $\mathbb{R}^d$, or $\mathbb{H}^d$, where in the case of $S^d$ or $\mathbb{H}^d$ we take the spaces to have unit radius for simplicity. Labelling the 3 cases by $k=1,0,-1$ as usual, it is well known that the Einstein equation
\begin{equation}
R_{ab}-\frac{g_{ab}}{2}R=8\pi G T_{ab}
\end{equation}
with $T_{ab}$ the stress energy tensor for bulk matter, reduces to the Friedman equation\footnote{The so-called second Friedman equation follows from the time derivative of \eqref{eq:F1} and conservation of stress-energy.}
\begin{equation}
\label{eq:F1}
\left(\frac{\dot{a}}{a}\right)^2 = -\frac{16 \pi G}{d(d-1)}\rho + \frac{1}{L^2} + \frac{k}{a^2}.
\end{equation}
Here $\dot{a} = \frac{\mathrm{d}a}{\mathrm{d}\tau}$, $L$ is the bulk AdS scale and the term $\frac{1}{L^2} $ encodes the explicit effects of the negative cosmological constant, and $\rho$ is the standard (Lorentz-signature) energy density of any matter fields
\begin{equation}
\rho \equiv -T_{\tau\tau}\,.
\end{equation}
Due to the Euclidean signature this equation differs from its familiar Lorentz-signature counterpart by an overall sign. Of course, we can also take quotients of the above solutions and thus use this formalism when $\Sigma_d$ is a torus $\mathbb{T}^d$ or a compact hyperbolic manifold $\mathbb{H}^d/\Gamma$ for some properly discontinuous isometry group $\Gamma$ of $\mathbb{H}^d$.

Let us now briefly investigate what \eqref{eq:F1} implies for the existence of Euclidean wormholes.  For $\tau \rightarrow \pm \infty$, we ask that $a(\tau) \rightarrow \infty$ to satisfy asymptotically AdS boundary conditions.  As a result, on any wormhole solution $a(\tau)$ must have some minimum $a_0$ where $\dot a=0$.  This clearly requires the right-hand side of \eqref{eq:F1} to vanish.  Without matter we have $\rho=0$, so this condition can be satisfied only when $k=-1$; i.e., when the boundary metric has negative scalar curvature.

However, we immediately notice two encouraging features.  First, for $k=0$ the above failure is only marginal.  Multiplying \eqref{eq:F1} by $a^2$ and setting $k=0$, we see that $\dot{a}$ can vanish at $a=0$.  Indeed, there is a $k=0$ vacuum solution $a(\tau) = L\,e^{\tau/L}$.  One may think of two copies of this solution as describing a degenerate limit of Euclidean wormholes where the neck of the wormhole has been stretched to become both infinitely long and infinitely thin.

Second, for any $k$ it is clear that the $\dot{a}=0$ constraint can be satisfied by adding matter with positive Lorentz-signature energy $\rho$.  For $k=0$ an arbitrarily small amount of such matter will do, but for $k=1$ we require $\rho$ to exceed a critical threshold.  We thus expect $k=0$ wormholes to appear with arbitrarily small matter sources on the boundaries, while for $k=1$ wormholes will arise only when the scalar sources are sufficiently large.

Many familiar kinds of matter yield positive Lorentz-signature energy densities $\rho$.  However, especially since the matter energy density $\rho$ will be large for $k=1$, it is useful to choose matter which is gravitationally attractive in Lorentz signature.  Wick rotation to Euclidean signature then gives gravitational repulsion, which helps to make $\ddot{a}$ positive at $a=a_0$ and also helps to satisfy the asymptotically AdS boundary conditions.  In particular, it would {\it not} be useful to use the potential energy of a scalar field, where the condition $\rho >0$ would effectively require adding a new positive cosmological constant to cancel the old (negative) one.

Furthermore, if the energy at $a=a_0$ comes from time-derivatives, then it is naturally positive if the Lorentz-signature field is real.  But that will make $\tau$-derivatives imaginary at $a_0$, and thus tend to give imaginary (or complex) fields in Euclidean signature.  The kinetic terms of such fields then tend to give {\it negative} contributions to the Euclidean action, and are thus a likely source of negative modes.  This is the essential reason why the axion solutions of \cite{Giddings:1987cg,ArkaniHamed:2007js} have many negative modes\footnote{As discussed above, the kinetic term of \cite{Hertog:2018kbz} does not become negative.  But it does have a surprising zero.} \cite{Hertog:2018kbz}.  We thus wish to take all $\tau$-derivatives to vanish at $a_0$.  Assuming that the scalar sources are identical on the two boundaries, this is equivalent to requiring the entire wormhole to be invariant under a corresponding $\mathbb{Z}_2$ symmetry.

As a result, we study solutions below with a surface $a=a_0$ invariant under such a $\mathbb{Z}_2$ symmetry and where $\rho >0$ at this surface due to {\it spatial} gradients of the matter fields.  Such kinetic energy is indeed gravitationally attractive in Lorentz signature, and thus gravitationally repulsive in Euclidean signature, but is consistent with real Euclidean solutions.  Creating such gradients requires similar gradients in the scalar sources we choose at the two boundaries.  For $k=0$, we expect Euclidean wormhole solutions with arbitrarily small such sources.  For $k=1$, we expect Euclidean wormhole solutions to appear once the boundary sources exceed some critical threshold.

Note that the above analysis and statement of expectations applies only when the surfaces of constant Euclidean time are homogeneous and isotropic.  Since the matter fields have spatial gradients, we will need to choose a finely-tuned matter \emph{Ansatz} to achieve this.  We may expect similar behavior for more general solutions that violate homogeneity\footnote{For $k=0$, spacetime wormholes should require non-zero gradients along each leg of the torus.  Otherwise one could Wick-rotate along a leg with translational symmetry and find a Lorentz-signature solution with two disconnected non-interacting boundaries linked by a traversable wormhole (and thus violating boundary causality).  We thanks Douglas Stanford for discussions on this point.}, but finding solutions would then require the solution of partial differential equations.  We thus save such analyses for future work.

\section{\label{sec:neggen}The general strategy for analysing field theoretical negative modes}
Let us consider a general Euclidean partition function $Z$ associated with a Euclidean action $S_E\left[\vec{\phi};\vec{\phi}_{\partial \mathcal{M}}\right]$ with some collection of fields $\vec{\phi}$ and corresponding boundary conditions $\vec{\phi}_{\partial \mathcal{M}}$. For the case of an Einstein-Scalar theory, $\vec{\phi}$ would contain all $(d+1)(d+2)/2$ independent metric components and the scalar field. Such a partition function can be schematically represented as
\begin{equation}
Z[\vec{\phi}_{\partial \mathcal{M}}] = \int \mathcal{D}\vec{\phi}\;e^{-S_E\left[\vec{\phi};\vec{\phi}_{\partial \mathcal{M}}\right]}\,.
\end{equation}

In the saddle-point approximation, we can expand $Z$ as
\begin{equation}
Z[\vec{\phi}_{\partial \mathcal{M}}] \approx e^{-S_E\left[\vec{\phi}^0;\vec{\phi}_{\partial \mathcal{M}}\right]}\times \int \mathcal{D}\vec{\delta\phi}\;e^{-S^{(2)}_{E}\left[\vec{\delta\phi};0\right]}+\ldots\,,
\end{equation}
where $\vec{\phi}^0$ are classical solutions of the equations of motion derived from $S_E\left[\vec{\phi};\vec{\phi}_{\partial \mathcal{M}}\right]$, and we obtain $S^{(2)}_{E}\left[\vec{\delta\phi};0\right]$ by expanding the fields as $\vec{\phi}=\vec{\phi}^0+\vec{\delta\phi}$ and keeping all terms up to second order in $\vec{\delta\phi}$. A first order term is absent in the expansion above, because $\vec{\phi}^0$ satisfies the classical equations of motion derived from $S_E\left[\vec{\phi};\vec{\phi}_{\partial \mathcal{M}}\right]$.

Let us imagine for a moment that $S^{(2)}_{E}\left[\vec{\delta\phi};0\right]$ is not positive definite. In that case the saddle $\vec{\phi}^0$ is not a local minimum of the Euclidean action and does not dominate over integration over nearby configurations if the integral is performed along the real Euclidean contour. In this case we say that $\vec{\phi}^0$ has field-theoretic negative modes.

For pure gravity $S^{(2)}_{E}\left[\vec{\delta\phi};0\right]$ is infamously not positive definite \cite{Gibbons:1976ue,Gibbons:1978ji}. In fact one can show that the conformal factor of the metric always has the wrong sign for the kinetic term. This is the conformal factor problem of Euclidean quantum gravity. One way around this is to Wick rotate the conformal factor, leading to a convergent Gaussian integral. This procedure, although slightly ad hoc in \cite{Gibbons:1976ue,Gibbons:1978ji}, was justified at the level of linearized gravity in \cite{Hartle:1988xv} and has been recently backed up by detailed dual field theory calculations \cite{Anninos:2012ft,Cotler:2019nbi,Benjamin:2020mfz} in the context of gauge/gravity duality. It was also shown in \cite{Dasgupta:2001ue} that a version of this Wick rotation can be performed at the non-linear level.

The case of gravity coupled to matter is more delicate.  In particular, it is no longer obvious that the conformal factor is the right variable to Wick rotate \cite{Monteiro:2008wr} since perturbations of the conformal factor, i.e. trace-type metric perturbations, will generically couple to other scalar matter perturbations. In addition, if matter is present, the trace free part of the metric can also couple with the trace itself.

Here we follow the procedure outlined in \cite{Kol:2006ga} which was used in \cite{Monteiro:2008wr} to investigate the negative mode of an asymptotically flat Reissner-Nodstr\"om black hole. It turns out that in all the cases we studied, the action can be decomposed as
\begin{equation}
S^{(2)}_{E}\left[\vec{\delta\phi};0\right]=\hat{S}^{(2)}_{E}\left[\vec{\delta\hat{\phi}};0\right]+\tilde{S}^{(2)}_{E}\left[\vec{\delta\tilde{\phi}};0\right]\,
\end{equation}
where  together the perturbations $\vec{\delta\hat{\phi}}$ and $\vec{\delta\tilde{\phi}}$ span the space of the original perturbations $\vec{\delta\phi}$ and both $\hat{S}^{(2)}_{E}\left[\vec{\delta\hat{\phi}};0\right]$ and $\tilde{S}^{(2)}_{E}\left[\vec{\delta\tilde{\phi}};0\right]$ have been written in first order form. The variables $\vec{\delta\hat{\phi}}$ turn out to be non-dynamical, i.e. the action $\hat{S}^{(2)}_{E}\left[\vec{\delta\hat{\phi}};0\right]$ contains no derivatives of $\vec{\delta\hat{\phi}}$. It is in this sector that we find a mode with a non-positive action. Furthermore, the fact that the variables $\vec{\delta\hat{\phi}}$ enter the action algebraically is a consequence of the Bianchi identities, and in an appropriate canonical formalism these variables would become Lagrange multipliers that enforce constraints. Let us denote by $\{\vec{\delta \hat{\phi}}\}^0$ the problematic mode. This is the mode that we Wick rotate as $\{\vec{\delta \hat{\phi}}\}^0\to i\,\{\vec{\delta \hat{\phi}}\}^0$. The Gaussian integral over $\vec{\delta\hat{\phi}}$ can now be performed and we reabsorb it in the measure. We are then left to study  the positivity properties of $\tilde{S}^{(2)}_{E}\left[\vec{\delta\tilde{\phi}};0\right]$. At this stage we introduce gauge invariant variables $\vec{\check{q}}$ which can be used to write $\tilde{S}^{(2)}_{E}\left[\vec{\delta\tilde{\phi}};0\right]$ solely as a function of $\vec{\check{q}}$ and their first derivatives. It is then $\tilde{S}^{(2)}_{E}\left[\vec{\check{q}};0\right]$ whose positivity properties we investigate. Note that the dimensionality of $\vec{\check{q}}$ is necessarily smaller than that of $\vec{\delta \tilde{\phi}}$ because of gauge invariance.

The procedure outlined above is consistent with the studies performed in \cite{Gratton:1999ya,Gratton:2000fj,Gratton:2001gw} which were used in \cite{Hertog:2018kbz} to show the existence of multiple negative modes on wormholes sourced by axions; see also \cite{Khvedelidze:2000cp}.

\section{\label{sec:u13}Einstein-$U(1)^3$ theory with $S^3$ boundary}

We now proceed to study a simple AdS$_4$ Einstein-Maxwell model that illustrates both key elements of the physics and our main techniques.  We assume spherical symmetry, and in particular a spherical boundary metric.   We begin with an overview of our model and then discuss disconnected solutions in section \ref{sec:EMDS}, and  connected wormhole solutions in section \ref{sec:EMWS}. In particular, we will see that connected wormholes can have lower action than the disconnect solution.   Finally, we show in section \ref{sec:EMNM} that these low-action wormholes are stable in the sense that they have no Euclidean negative modes.  Since this is merely an ad hoc low-energy model not derived from string theory (or any other UV-complete theory), the discussion in section \ref{sec:EMNM} concerns only field-theoretic negative modes.  There is no possible notion of a brane-nucleation negative modes as the theory does not contain branes (nor does it contain non-singular magnetic monopoles).

As described below, choosing the model to contain three distinct Maxwell fields will help us to arrange a cohomogeneity-1 solution; i.e., a solution that is homogeneous at each `Euclidean time' in the FLRW sense described above.
We will thus search for wormhole and disconnected solutions to the equations of motion derived from the following action:
\begin{equation}
S_{U(1)^3}= -\int_{\mathcal{M}} \mathrm{d}^4 x\sqrt{g}\left(R+\frac{6}{L^2}-\sum_{I=1}^3 F^{(I)}_{ab}F^{(I)\;ab}\right)-2\int_{\partial \mathcal{M}} \mathrm{d}^3 x\sqrt{h}\;K+S_{\mathcal{B}}\,,
\label{eq:simpleu13}
\end{equation}
where $L$ is the AdS length scale, $K$ is the trace of the extrinsic curvature associated with an outward-pointing normal to $\partial \mathcal{M}$, $h$ the determinant of the induced metric $h_{\mu\nu}$ on $\partial \mathcal{M}$ and $F^{(I)}=\mathrm{d}A^{(I)}$. Here and throughout this paper will take units in which $16\pi G=1$.

The second term in \eqref{eq:simpleu13} is the so-called Gibbons-Hawking-York term. The final term $S_{\mathcal{B}}$ includes a number of counterterms that render the Euclidean on-shell action finite and are functions of the intrinsic geometry on $\partial \mathcal{M}$ only and are dimension dependent. For the above theory in four bulk spacetime dimensions these turn out to be given by
\begin{equation}
S_{\mathcal{B}}=\frac{4}{L}\int_{\partial\mathcal{M}}\mathrm{d}^3 x\sqrt{h}+L\int_{\partial\mathcal{M}}\mathrm{d}^3 x\sqrt{h}\,\mathcal{R}\,,
\end{equation}
where $\mathcal{R}$ is the intrinsic Ricci scalar on $\partial \mathcal{M}$. One might wonder whether we need additional boundary terms associated with $F^{(I)}$ such as the ones reported in \cite{Hawking:1995ap}. However, as noted in \cite{Hawking:1995ap}, no such terms are needed if we are interested in fixing the leading value of $A^{(I)}$ as we approach the conformal boundary, i.e. work in the grand-canonical ensemble. These are precisely the boundary conditions we choose.
The equations of motion derived from (\ref{eq:simpleu13}) read
\begin{subequations}
\begin{align}
& R_{ab}+\frac{3}{L^2}g_{ab}=2\sum_{I=1}^3\left(F^{(I)}_{ac}F^{(I)\phantom{b}c}_{b}-\frac{g_{ab}}{4}F^{(I)}_{cd}F^{(I)\;cd}\right)\,,
\\
& \nabla_a F^{(I)\;ab}=0\,.
\end{align}
\end{subequations}
subject to the boundary conditions that on $\partial M$, the induced metric $h$ is fixed as well as $A^{(I)}$.

We are interested in finding solutions for which the metric has the same isometries as a round three-sphere, i.e. spherical symmetry, and in particular $SO(4)$, but where the Maxwell fields explicitly break such symmetry. An easy way to do so is to write the 3-sphere in terms of left-invariant 1-forms $\{\hat \sigma_1,\hat \sigma_2,\hat \sigma_3\}$ such that the metric on the unit round three sphere reads
\begin{equation}
\mathrm{d}\Omega_3^2 =\frac{1}{4}\left(\hat \sigma_1^2+\hat \sigma_2^2+ \hat \sigma_3^2\right)\,,
\end{equation}
with
\begin{equation}
\mathrm{d}\hat \sigma_I = \frac{1}{2}\;\varepsilon_{IJK}\;\hat \sigma_J\wedge \hat \sigma_K\,.
\end{equation}

In terms of standard Euler angles, we can choose
\begin{subequations}
\begin{align}
&\hat \sigma_1 = -\sin \psi\,\mathrm{d}\theta+\cos \psi\,\sin \theta\,\mathrm{d}\hat \varphi
\\
&\hat \sigma_2 = \cos \psi\,\mathrm{d}\theta+\sin \psi\,\sin \theta\,\mathrm{d}\hat \varphi
\\
&\hat \sigma_3 =\mathrm{d}\psi+\cos \theta\,\mathrm{d}\hat \varphi
\end{align}
\label{eqs:leftinvariant}
\end{subequations}
with $\psi \in(0,4\pi)$, $\theta\in(0,\pi)$ and $\hat \varphi \in(0,2\pi)$.

We then search for solutions of the form
\begin{equation}
\mathrm{d}s^2 = \frac{\mathrm{d}r^2}{f(r)}+g_{S^3}(r)\mathrm{d}\Omega^2_3\,,
\label{eq:gen0}
\end{equation}

For the Maxwell fields, we choose
\begin{align}
&A^{(I)} = L\,\frac{\hat{\sigma}_I}{2} \Phi(r)\,,\quad\text{for} \quad I\in\{1,2,3\}\,.
\label{eq:gen1}
\end{align}
\subsection{Disconnected solutions}
\label{sec:EMDS}

A primary question will be whether connected wormhole solutions dominate over disconnected solutions.  We thus begin here by constructing the simpler disconnected solution for comparison, deferring discussion of wormhole solutions to section \ref{sec:EMWS} below.

As is often the case, it is convenient to fix the gauge in Eq.~(\ref{eq:gen0}) by choosing
\begin{equation}
g_{S^3}(r)=r^2\,.
\end{equation}
We take $r\in(0,+\infty)$, with $r=0$ describing the smooth center where the round $S^3$ shrinks smoothly to zero size and $r=+\infty$ the location of the asymptotic conformal boundary. Regularity of $F^{(I)}$ at the origin demands that
\begin{equation}
\left.\frac{\mathrm{d} \Phi(r)}{\mathrm{d}r}\right|_{r=0}=0\,,
\label{eq:ss}
\end{equation}
with $\Phi(0)$ being a constant, whereas regularity of the metric at the point $r=0$ demands
\begin{equation}
f(0)=1\,.
\end{equation}
Note that regularity of $A^{(I)}$ seen as a 1-form demands that $\Phi(r)=\mathcal{O}(r^2)$, which is stronger than the condition expressed by Eq.~(\ref{eq:ss}).

With our choice of boundary conditions, we find a unique solution given by
\begin{equation}
f(r)=1+\frac{r^2}{L^2}\quad \text{and}\quad \Phi(r) = \Phi_0 \frac{\sqrt{L^2+r^2}-L}{\sqrt{L^2+r^2}+L}\,.
\end{equation}
It turns out that each member of this one-parameter family of solutions is self-dual, in the sense that
\begin{equation}
\star F^{(I)}=F^{(I)}\,,
\end{equation}
where $\star$ is the Hodge dual operation in four spacetime dimensions. Recall that for self-dual solutions the stress energy tensor induced by $F^{(I)}$ is identically zero, which is why the metric is identically Euclidean AdS for any value of $\Phi_0$.  Note also that the above solution has  $\Phi(r)=\mathcal{O}(r^2)$ near $r=0$ so that our boundary condition is satisfied.

It is straightforward to evaluate the Euclidean on-shell action on these solutions. It must of course be a function of $\Phi_0$ only, and we find the particular form
\begin{equation}
S^D_{U(1)^3} = 8 \pi ^2 L^2 \left(1+3 \Phi _0^2\right).
\end{equation}
Here the upper-script $D$ on the left-hand-side denotes the on-shell action of the disconnected solution.
\subsection{Wormhole solutions}
\label{sec:EMWS}
Having found our disconnected solution, we now turn to the study of smooth connected wormholes.  Since such solutions will have a minimal sphere of some non-zero area $4\pi r_0^2$, we now choose to fix the gauge in Eq.~(\ref{eq:gen0}) by writing
\begin{equation}
g_{S^3}(r)=r^2+r_0^2\, .
\end{equation}
Now $r\in\mathbb{R}$, with the two asymptotic boundaries located at $r\to\pm\infty$. We also impose a global $\mathbb{Z}_2$ symmetry that relates the two spheres of given $r\neq r_0$ and which leaves the minimal sphere fixed.  We shall see that the parameter $r_0$ will be a function of $\Phi_0$ only.  Without loss of generality we will take $r_0>0$.

Again, we can integrate our equations of motion to find the full space of such solutions. Note that our $\mathbb{Z}_2$ symmetry requires
\begin{equation}
\label{eq:Z21S}
\left.\frac{\mathrm{d} \Phi(r)}{\mathrm{d}r}\right|_{r=0}=0\,.
\end{equation}

From the equations for the Maxwell fields we find
\begin{equation}
f(r) = \frac{C+4 \Phi(r)^2}{(r^2+r_0^2)\Phi^\prime(r)^2}\,,
\end{equation}
where $C$ is a constant to be determined later and $^\prime$ denotes differentiation with respect to $r$. Consistency of the $rr$ and $S^3S^3$ components of the Einstein equation then demands
\begin{equation}
\label{eq:pprimeS1}
\Phi '(r)^2-\frac{L^2 r^2 \left[C+4 \Phi (r)^2\right]}{\left(r^2+r_0^2\right) \left[C L^4+\left(r^2+r_0^2\right) \left(L^2+r^2+r_0^2\right)\right]}=0\,.
\end{equation}
Since this gives an explicit result for $[C+4 \Phi (r)^2]/\Phi'$ it allows us to write $f(r)$ in the form
\begin{equation}
f(r)=\frac{C L^4+\left(r^2+r_0^2\right) \left(L^2+r^2+r_0^2\right)}{L^2 r^2}\,.
\end{equation}
Thus, we find a singularity at $r=0$, unless we set
\begin{equation}
C=-\frac{L^2 r_0^2 +r_0^4}{L^4}\,,
\end{equation}
With this choice for $C$ one obtains
\begin{equation}
f(r)=\frac{L^2+r^2+2 r_0^2}{L^2}\,,
\end{equation}
which is smooth at $r=0$ as desired. Using our choice of the constant $C$ in \eqref{eq:pprimeS1} yields
\begin{equation}
\Phi^\prime(r)^2-\frac{4 L^4 \Phi (r)^2-r_0^2 \left(L^2+r_0^2\right)}{L^2 \left(r^2+r_0^2\right) \left(L^2+r^2+2 r_0^2\right)}=0\,.
\end{equation}
Since we require $\Phi^\prime(0)=0$ at $r=0$ we want to impose, the above fixes $r_0$ in terms of $\Phi(0)\equiv\Phi_\star$ to be
\begin{equation}
r_0 = b\,L\quad\text{with}\quad a\equiv (1+16 \Phi _\star^2)^{1/4}\quad \text{and}\quad b\equiv  \sqrt{\frac{a^2-1}{2}}\,.
\label{eq:radius}
\end{equation}

With these choices, the equation for $\Phi$ can be readily solved to give
\begin{equation}
\Phi(r)= \Phi_{\star} \cosh \left[\frac{2}{b}F\left(\mathrm{arctan}\left(\frac{r}{L\,a}\right)\Big |1-\frac{a^2}{b^2}\right)\right]\,,
\label{eq:sol}
\end{equation}
where $F(\phi|m)$ is the elliptic integral of the first kind.

\begin{figure}[h]
\centering
\includegraphics[width=0.6\linewidth]{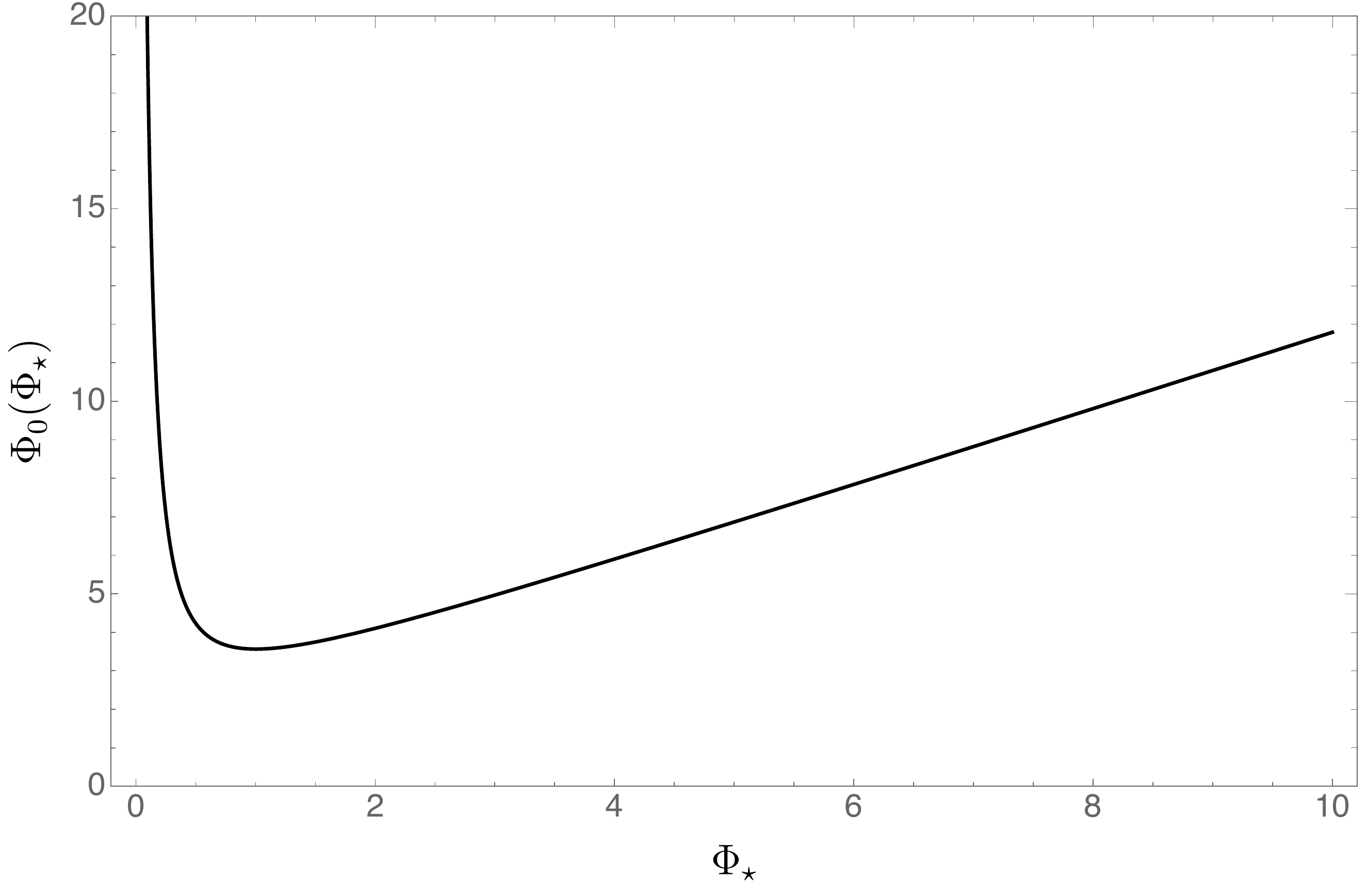}
\caption{The source for the Maxwell fields $A^{(I)}$ $\Phi_0$ as a function of $\Phi_{\star}$. There is a minimum value of $\Phi_0$, $\Phi_0^{\min}\approx 3.563349$, above which two types of wormhole solutions exist.}
\label{fig:source}
\end{figure}

To determine the source $\Phi_0$ in terms of $\Phi_\star$ we simply expand the above $\Phi(r)$ at large $r$, to find
\begin{equation}
\Phi_0(\Phi _\star) = \Phi _\star \cosh \left[\frac{2}{b} K\left(1-\frac{a^2}{b^2}\right)\right]\,,
\end{equation}
where $K(m)$ is the complete elliptic integral of the first kind. We stress that $\Phi_0$ is the actual chemical potential for $A^{(I)}$, but that we find it more convenient to parameterize the solutions in terms of $\Phi_\star$. The reason for this is that there can be more than one solution for  a given value of $\Phi_0$, but that solutions are uniquely determined by their value of $\Phi_*$.  This is best illustrated by looking a plot of $\Phi_0(\Phi _\star)$ (see Fig.~\ref{fig:source}). From this plot it is clear that wormhole solutions can only exist if $\Phi_0\geq \Phi_0^{\min}\approx 3.563349$, for which $\Phi_{\star}=\Phi_{\star}^{\min}\approx 1.002373$.

\begin{figure}[h]
\centering
\includegraphics[width=0.6\linewidth]{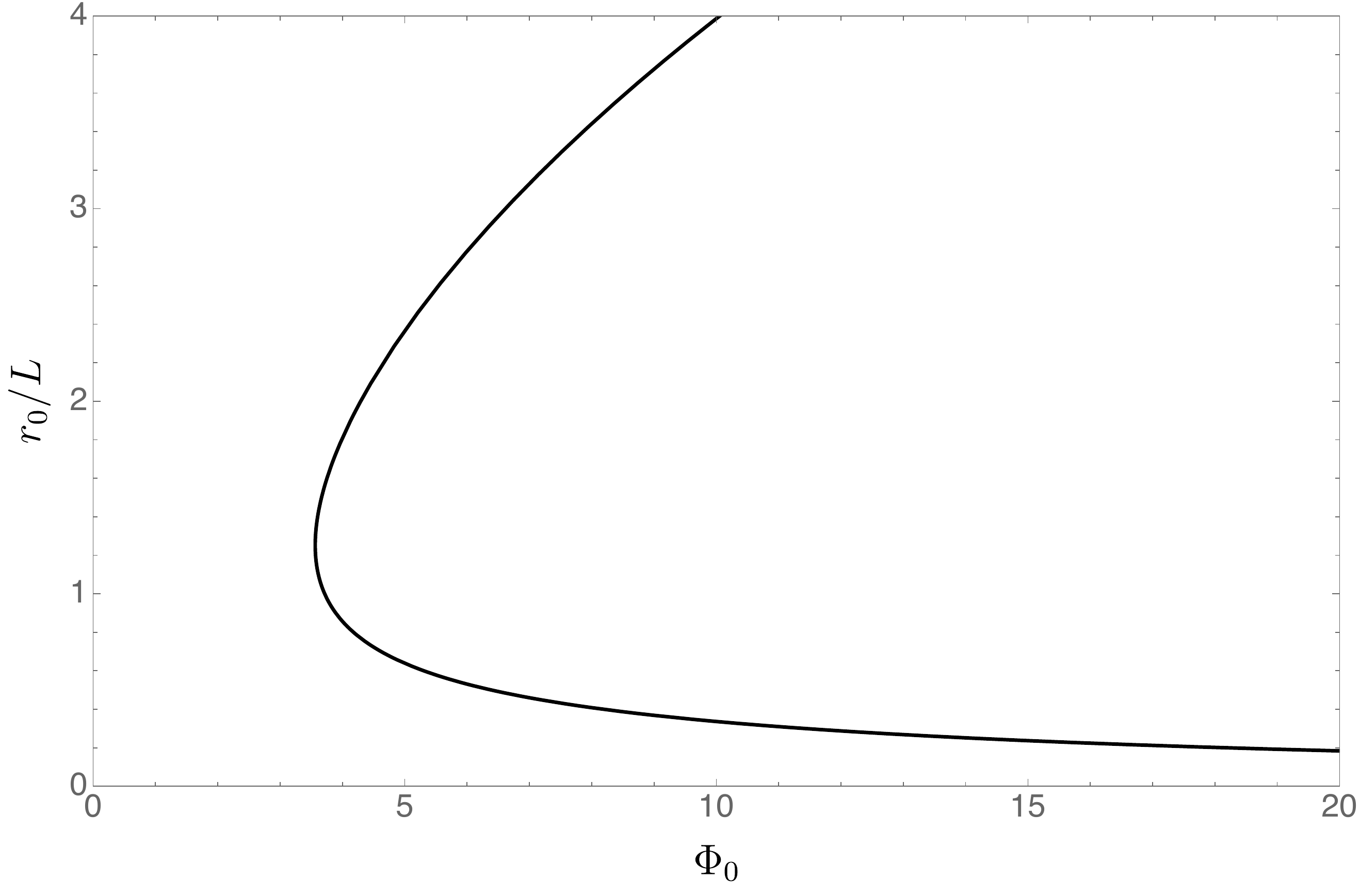}
\caption{Radius of the wormhole solutions as a function of the Maxwell source $\Phi_0$. For fixed value of $\Phi_0>\Phi_0^{\min}$ two wormhole solutions exist.}
\label{fig:radiusmaxwell}
\end{figure}

But what distinguishes the two wormholes with a given value of $\Phi_0$? Perhaps the best way to see the answer is to study the radius $r_0$ of the wormhole throat as a function of $\Phi_0$ shown in Fig.~\ref{fig:radiusmaxwell}. For any fixed value of $\Phi_0>\Phi_0^{\min}\approx 3.563349$, two wormhole solutions exist: a large wormhole and a small wormhole. For $\Phi_0=\Phi_0^{\min}$ we have $r_0=r_0^{\min}\approx 1.251462\,L$, and both the large and small branches merge. We shall see that these solutions behave just like small and large Euclidean Schwarzschild black holes in global AdS. In particular, we will show that the small wormhole branch has a negative mode, and the large wormhole branch does not.

One can also evaluate the Euclidean on-shell associated with our wormhole solutions, which can be written in terms of complete elliptic integrals of the first and second kind in the form
\begin{multline}
S^{C}_{U(1)^3} = \frac{8 L^2 \pi ^2}{(X-1)^{3/2}} \Bigg[2 (X-1) E(-X)-(X-2) K(-X)
\\
+\frac{3 X}{4\sqrt{X-1}}\sinh \left(4 \sqrt{X-1}K(-X)\right)\Bigg]\,.
\end{multline}
Here we defined
\begin{equation}
X\equiv 1+\frac{L^2}{r_0^2},
\end{equation}
and $E(m)$ is the complete elliptic integral fo the second kind.

\begin{figure}[h]
\centering
\includegraphics[width=0.6\linewidth]{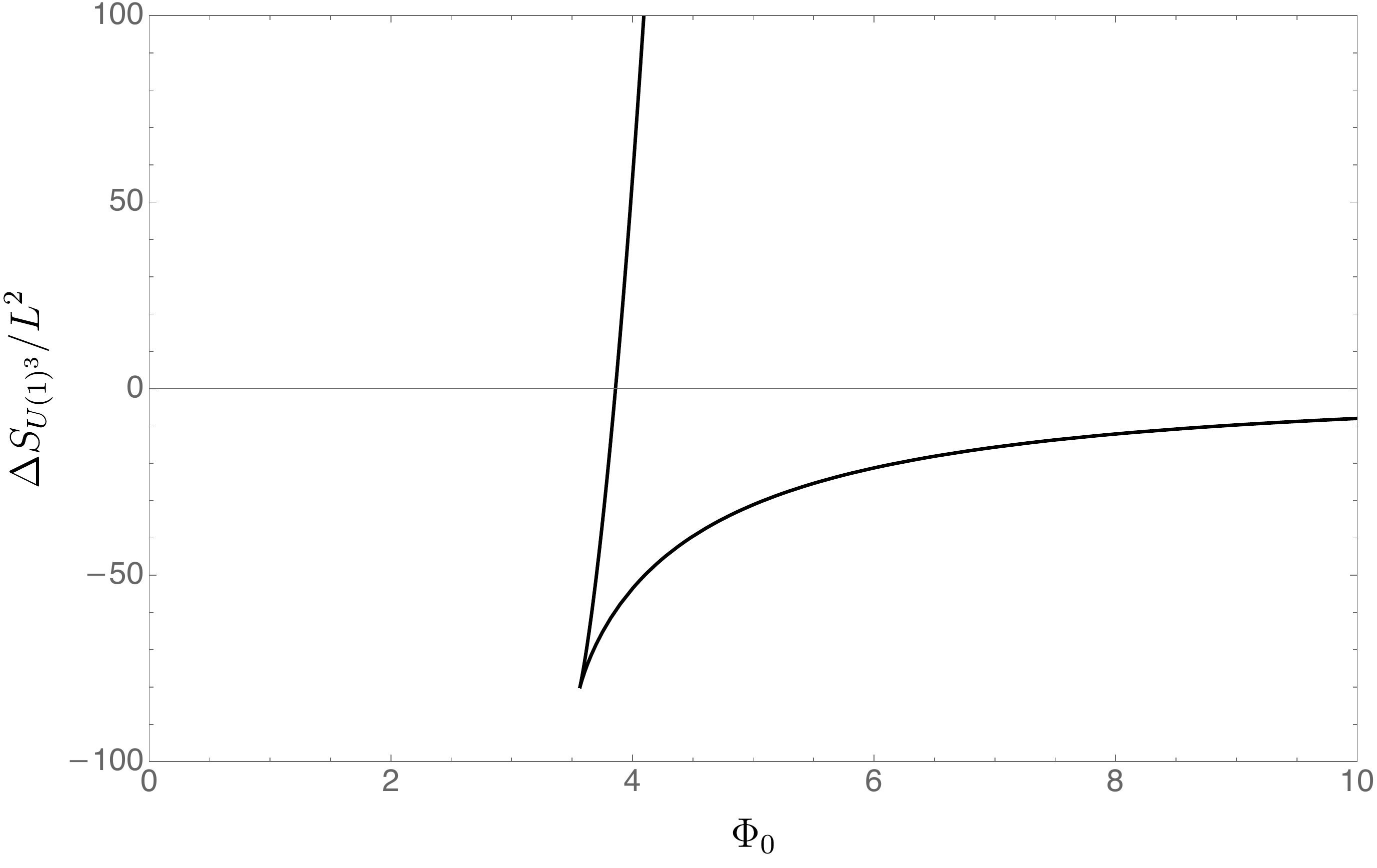}
\caption{The difference in Euclidean action between the disconnected and connected solutions.  At any $\Phi_0$, the larger value of $\Delta S_{U(1)^3}$ corresponds to the large wormhole and the smaller value corresponds to the small wormhole.  Due to the similarity to the familiar Hawking-Page transition, we use $\Phi_{\mathrm{HP}}$ to denote the value of $\Phi_0$ at which $\Delta S_{U(1)^3}=0$ for the large wormhole. For $\Phi_0>\Phi_{\mathrm{HP}}$, the large wormhole solution becomes dominant while the small wormhole solution is always subdominant.}
\label{fig:meritu13}
\end{figure}

We can finally plot one of our figures of merit, namely
\begin{equation}
\Delta S_{U(1)^3}\equiv 2 S^D_ {U (1)^3} - S^C_ {U (1)^3}\,.
\end{equation}
If $\Delta S_{U(1)^3}$ is positive, the wormhole solution has lower Euclidean action than the disconnected solution with the same value of $\Phi_0$. If, on the other hand, $\Delta S_{U(1)^3}<0$, it must be that the wormhole solution is subdominant. We find that the large wormhole solutions are dominant for $\Phi_0>\Phi_{\mathrm{HP}}\approx3.859673$, and subdominant otherwise. We denote the transition value by $\Phi_{\mathrm{HP}}$ due to the similarity to the familiar Hawking-Page transition.  The small wormhole solutions are always subdominant. These two behaviours are displayed in Fig.~\ref{fig:meritu13}. A similar structure was found previously for wormholes in JT gravity coupled to matter \cite{Garcia-Garcia:2020ttf}.
Having found dominant saddles, we now proceed to determine their stability.

We now note that all the solutions we found, either connected or disconnected in the bulk, satisfy a Euclidean version of the first law{involving the Euclidean action $S_E$. According to standard lore in AdS/ CFT, one can find the expectation value of the operators dual to $A^I$ by simply taking a functional derivative of the action with respect to the corresponding boundary value of $A^I$
\begin{equation}
\langle \mathcal{J}_{\mu}^I \rangle = \frac{\delta S_E}{\delta A^{I\,\mu}}\,.
\end{equation}
where Greek indices run over boundary coordinates. These can be easily evaluated on arbitrary on shell solution and it turns out that
\begin{equation}
\langle \mathcal{J}_{\mu}^I \rangle = 8\pi^2\,L^3\,\tilde{J}_{\mu}^I\,.
\end{equation}
with $\tilde{J}_{\mu}^I$ being given by
\begin{equation}
A^I=A^I_{a}\,\mathrm{d}x^a = \Phi^I_{\mu}\,\mathrm{d}x^{\mu}-z\,\tilde{J}^I_{\mu}\,\mathrm{d}x^{\mu}+\mathcal{O}(z^2)
\end{equation}
where $z$ is a Fefferman-Graham coordinate \cite{Fefferman:2007rka}. This in turn implies that
\begin{equation}
\mathrm{d}S_E = \varepsilon\,J^I_{\mu}\,\mathrm{d}\Phi^{I\,\mu}\,,
\end{equation}
where $\varepsilon=1$ for disconnected solutions and $\varepsilon=2$ for wormholes with two boundaries. We have checked that our solutions satisfy this relation. Perhaps more importantly, appropriate generalisation of this type of first law also arise when studying scalar wormhole solutions, which we were only able to study numerically. We have checked that all our numerical solutions satisfy the above relations to better than $10^{-10}\%$ accuracy.
\subsection{Negative modes}
\label{sec:EMNM}
We now discuss perturbations around our wormholes, and in particular, the possible existence of negative modes. We will take advantage of the $SO(4)$ symmetry of the $S^3$ to decompose the perturbations into spherical harmonics. Perturbations then will come into three different classes: scalar-derived perturbations, vector-derived perturbations and tensor-derived perturbations. These are built from scalar, vector and tensor harmonics on the $S^3$. We shall label each of these structure functions by $\mathbb{S}^{\ell_S}$, $\mathbb{S}^{\ell_V}_i$ and $\mathbb{S}^{\ell_T}_{ij}$, with $\ell_S=0,1,2,\ldots$, $\ell_V=1,2,\ldots$ and $\ell_{T}=2,3,\ldots$ and $i,j$ running over the sphere directions.

These structure functions are chosen so that they are orthogonal to each other in the absence of background fields that might break the $SO(4)$ symmetry. Unfortunately, the Maxwell fields do break $SO(4)$, so we will need more structure. Nevertheless, we will be able to use these building blocks to study the negative modes. When there is $SO(4)$ background symmetry of the background, orthogonality only occurs if we take  $\mathbb{S}^{\ell_V}_i$ to be divergence free and $\mathbb{S}^{\ell_T}_{ij}$ to be traceless-transverse. All these operations are, of course, done with respect to the metric on the round three-sphere. The scalars in addition satisfy
\begin{subequations}
\label{eqs:harmonics}
\begin{equation}
\Box_{S^3}\mathbb{S}^{\ell_S}+\lambda_{S}\mathbb{S}^{\ell_S}=0\,,
\end{equation}
with $\lambda_S = \ell_S(\ell_S+2)$, the vectors
\begin{equation}
\Box_{S^3}\mathbb{S}_i^{\ell_V}+\lambda_{V}\mathbb{S}^{\ell_V}_i=0\,,
\end{equation}
with $\lambda_V=\ell_V(\ell_V+2)-1$ and the tensors
\begin{equation}
\Box_{S^3}\mathbb{S}_{ij}^{\ell_T}+\lambda_{T}\mathbb{S}^{\ell_T}_{ij}=0\,,
\label{eq:tensor}
\end{equation}
with $\lambda_T=\ell_T(\ell_T+2)-2$.
\end{subequations}
\subsubsection{Scalar-derived perturbations: the $\ell_S=0$ sector}
This is the only sector where we find a negative mode, and it occurs only for the small wormhole branch. In fact, the threshold for the existence of this mode coincides precisely with $r_0=r_{0}^{\min}$. This is akin of what happens with Schwarschild-AdS, where a negative mode exists for small black holes, but not for large black holes \cite{Prestidge:1999uq}. The threshold can be found analytically by inspecting when the Schwarzschild-AdS black holes become locally thermodynamically stable, i.e. when the specific heat becomes positive. Note that in Schwarzschild-AdS this transition occurs before the Hawking-Page transition, so that when the large black hole branch dominates over pure thermal AdS, the negative mode is no longer presence. We shall see a similar behaviour with the spherical wormholes.

We start with an \emph{Ansatz} for the $\ell=0$ sector. Since there are no vector or tensor perturbations on the $S^3$ with $\ell=0$, our \emph{Ansatz} preserves the same symmetries as the background solution. We thus search for negative modes which take the same form as Eqs.~(\ref{eq:gen0})-(\ref{eq:gen1}) with
\begin{equation}
g(r)=\overline{g}+\delta g(r)\,,\quad f(r)=\overline{f}+\delta f(r)\quad\text{and}\quad \Phi(r)= \overline{\Phi}(r)+\delta\Phi(r)\,,
\end{equation}
where $\overline{\Phi}$ is given in Eq.~(\ref{eq:sol}) and
\begin{equation}
\overline{g}=r^2+r_0^2\,,\quad\text{and} \quad \overline{f}=\frac{L^2+r^2+2 r_0^2}{L^2}\,.
\end{equation}
We expand the action (\ref{eq:simpleu13}) to second order in $\delta g$, $\delta f$ and $\delta \Phi$. The terms linear in $\delta g$, $\delta f$ and $\delta \Phi$ vanish by virtue of the background equations of motion. It is then a simple exercise to recast the action as a function of $\delta g$, $\delta f$ and $\delta \Phi$ and their first derivatives only. Doing so involves integrating by parts, and the resulting boundary term precisely cancels the Gibbons-Hawking-York term. We shall denote this quadratic action by $S^{(2)}$. Furthermore, we find that $\delta f$ enters the action algebraically as it should by virtue of the Bianchi identities. This makes it straightforward to integrate out $\delta f$ (since the action is quadratic in $\delta f$ and thus the path integral is Gaussian). The result of this procedure yields an action that is quadratic in $\delta \Phi$ and $\delta g$ and their first derivatives. We have also checked that the path integral in $\delta f$ has the correct sign for a meaningful integration, i.e. the coefficient of the term proportional to $\delta f^2$ is negative definite. We denote the resulting action by $\tilde{S}^{(2)}$.

We wish to work with gauge invariant perturbations. In order to do this, we must first understand how $\delta g$, $\delta f$ and $\delta \Phi$ transform under an infinitesimal gauge transformation $\xi = \xi_r(r) \partial/\partial r$. This is easily seen by recalling that a metric perturbation $h$ and gauge field perturbation $a$, transforms under infinitesimal gauge transformations $\xi$ as
\begin{equation}
\Delta h=\pounds_{\xi} \overline{g}\,,\quad \Delta a=\pounds_{\xi} \overline{A}
\end{equation}
where $\pounds_{\xi}$ is the Lie derivative along $\xi$, and $(\overline{g},\overline{A})$ are the metric and gauge potential background fields.

We then find
\begin{equation}
\Delta \delta f = \xi_r \overline{f}^\prime-2\overline{f}\xi_r^\prime\,,\quad \Delta \delta g = 2 r \xi_r\quad\text{and}\quad \Delta \delta\Phi = \xi_r \overline{\Phi}^\prime\,.
\end{equation}
As a result, we can then build the gauge invariant quantity
\begin{equation}
q=\delta \Phi-\overline{\Phi}^\prime \frac{\delta g}{2r}\,.
\label{eq:q}
\end{equation}
It is a trivial exercise to show that $\Delta q=0$. We can use this definition to write the quadratic action $\tilde{S}^{(2)}$ for $\delta g$ and $\delta \Phi$ as a function of $q$. This is done by effectively solving Eq.~(\ref{eq:q}) for $\delta \Phi$ and substituting the resulting expression for $\delta \Phi$ in the quadratic action $\tilde{S}^{(2)}$. The dependence in $\delta g$ completely cancels out, as it should, due to gauge invariance. We are are thus left with $\tilde{S}^{(2)}$ written in terms of $q$, $q$ and its first derivative only:
\begin{subequations}
\begin{equation}
\tilde{S}^{(2)}=2 \pi ^2\int_{-\infty}^\infty \mathrm{d}\,r\,\sqrt{\frac{\overline{g}}{\overline{f}}}\Bigg[\overline{f}\,K\,{q^\prime}^2+V\,q^2\Bigg],
\label{eq:finalss}
\end{equation}
where
\begin{equation}
K=\frac{6 L^2 r^2}{r^2-\overline{g} L^2 {\overline{\Phi }^\prime}^2}\quad \text{and}\quad V = \frac{4 K}{\overline{g}}\left[\frac{2 \overline{g}}{L^2 \overline{f} r }\frac{L^2 \left(r-L^2 \overline{\Phi}\,\overline{\Phi }^\prime\right)+\overline{g} \left(r-2 L^2 \overline{\Phi } \,\overline{\Phi }^\prime\right)}{r^2-\overline{g} L^2 {\overline{\Phi }^\prime}^2}-1\right].
\end{equation}
\end{subequations}

Two comments regarding boundary terms are now in order. First, to show that $\delta g$ drops out one needs to integrate by parts twice. This generates two boundary terms. These two terms precisely cancel the counterterms in Eq.~(\ref{eq:simpleu13}). Second, in order to ensure that terms proportional to $q q^\prime$ do not show up in the final form of the action, we again had to integrate by parts. It is easy to show that the resulting boundary terms vanish so long as $q\sim o(r^{4})$ near the conformal boundary. As we shall see below, our boundary conditions for $q$ will require that $q$ vanishes at this rate near the boundary, so these terms can be safely neglected.

To search for negative modes, we integrate the first term in Eq.~(\ref{eq:finalss}) by parts to write
\begin{equation}
\tilde{S}^{(2)}=2 \pi ^2\int_{-\infty}^\infty \mathrm{d}\,r\,\sqrt{\frac{\overline{g}}{\overline{f}}}\,q\,\Bigg\{-\sqrt{\frac{\overline{f}}{\overline{g}}}\left[\sqrt{\overline{f}\,\overline{g}}\,K\,q^\prime\right]^\prime+V\,q\Bigg\}\,.
\end{equation}
The resulting boundary term can be neglected so long as $q\sim o(r^{-1/2})$, and we will verify below that such boundary conditions may be imposed.  The negative mode equation simply becomes
\begin{equation}
-\sqrt{\frac{\overline{f}}{\overline{g}}}\left[\sqrt{\overline{f}\,\overline{g}}\,K\,q^\prime\right]^\prime+V\,q=\lambda\,q\,.
\end{equation}
If we can find values of $\lambda<0$ for which this equation admits non-trivial solutions, then fluctuations about the saddle will make large contributions and the Euclidean solution is locally unstable. Finally, we still need to check whether the possible behaviours that this equation admits near the conformal boundary are consistent with imposing a boundary condition that requires $q\sim o(r^{-1/2})$, so that we can indeed neglect the above boundary terms. A Frobenius analysis close to the conformal boundary reveals that
\begin{equation}
q \sim r^{-\Delta_{\pm}}\quad\text{with}\quad \Delta_{\pm}=\frac{1}{2}\pm \sqrt{\frac{1}{4}-\frac{\lambda }{6}}\,,.
\label{eq:final}
\end{equation}
We see that for $\lambda<0$ the $\Delta_+$ branch satisfies $q\sim o(r^{-1/2})$.  It is thus consistent to require this as a boundary condition and to then neglect the above-mentioned boundary terms.

Since $V(r)$ is symmetric around $r=0$, we can decompose our modes into modes with $q(0)=0$ or $q^\prime(0)=0$.  For numerical convenience we also define
\begin{equation}
q = \frac{L^\Delta_+}{(r^2+r_0^2)^{\Delta_+/2}}\hat{q}\quad \text{and} \quad r=\frac{r_0\,y}{1-y}\,
\end{equation}
so that the conformal boundary is located at $y=1$ and the origin at $y=0$. Solving Eq.~(\ref{eq:final}) off the conformal boundary yields $\hat{q}^\prime(1)=0$ for the choice $q\sim r^{-\Delta_+}$.

Using the numerical methods first outlined in \cite{Monteiro:2009ke} and reviewed in \cite{Dias:2015nua} we search for negative modes with the above boundary conditions. For $\hat{q}(0)=0$ we find no negative modes for any value of $r_0/L$. On the other hand, for $\hat{q}^\prime(0)=0$ we find exactly one negative mode, which becomes positive when $r=r_0^{\min}$ (see Fig.~\ref{fig:meritu13zeromode}). This is precisely when the transition between small and large wormholes occurs, in complete analogy with spherical Schwarzschild-AdS black holes.
\begin{figure}[h]
\centering
\includegraphics[width=0.6\linewidth]{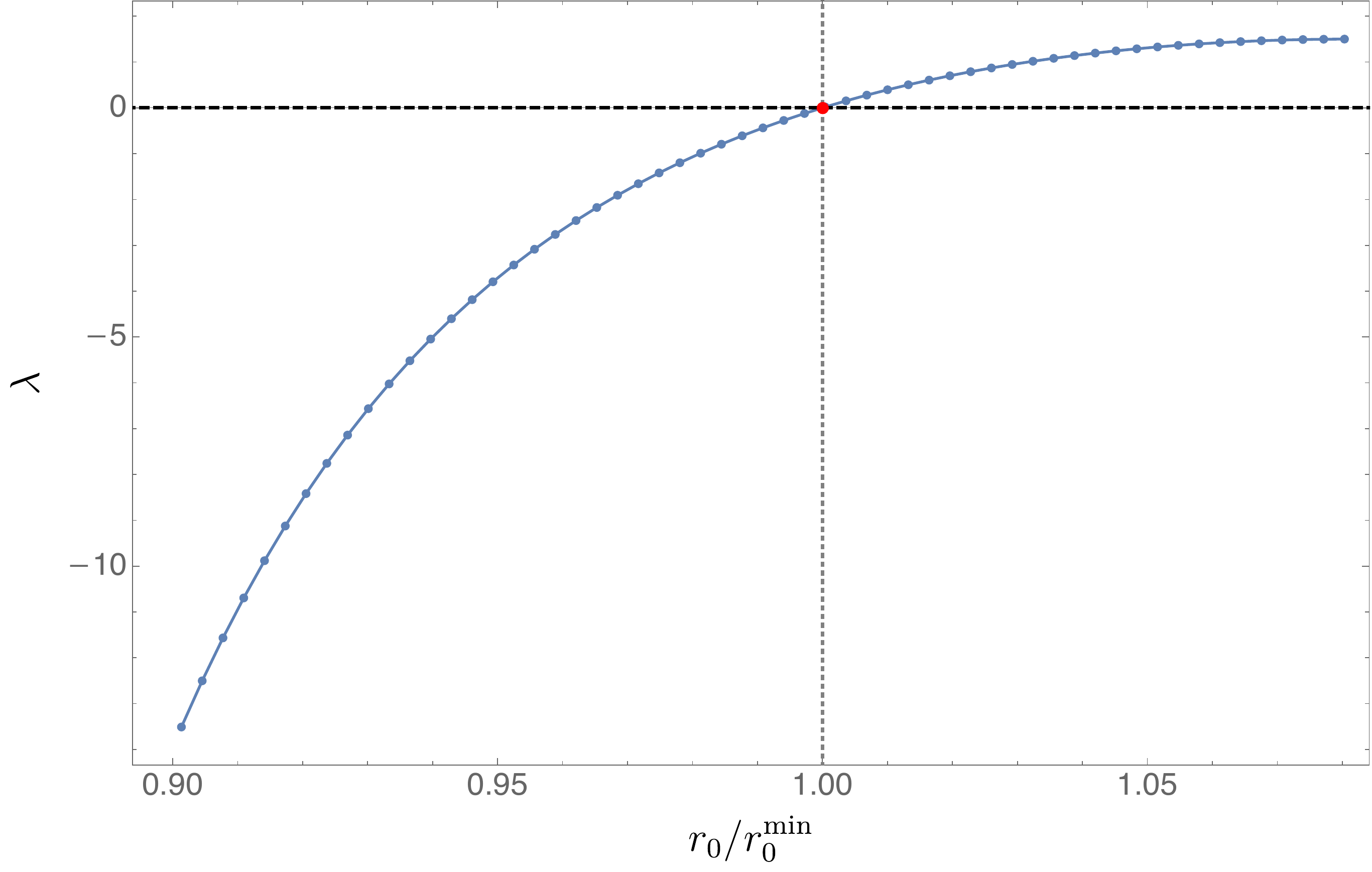}
\caption{The homogeneous negative mode with $\ell=0$ as a function of the black hole radius, measured in units of $r_0^{\min}$. At $r=r_0^{\min}$ the negative mode vanishes, and becomes positive thereafter.}
\label{fig:meritu13zeromode}
\end{figure}
\subsubsection{Scalar-derived perturbations: the $\ell_S\geq2$ sector}
\label{sec:EMscalar}
Here we partially follow the pioneering work of Kodama and Ishibashi in \cite{Kodama:2003kk}. For the metric we take our perturbations to have the form
\begin{equation}
\delta \mathrm{d}s^2_{\ell_S} = h^{\ell_S}_{rr}(r)\,\mathbb{S}^{\ell_S}\,\mathrm{d}r^2+2 h^{\ell_S}_r(r)\,\Grad^i\mathbb{S}^{\ell_S}\,\mathrm{d}r\, \mathrm{d}y^{i}+H_T^{\ell_S}(r)\,\mathbb{S}^{\ell_S}_{ij}\,\mathrm{d}y^i \mathrm{d}y^j+H_L^{\ell_S}(r)\,\mathbb{S}^{\ell_S}\,\mathbbm{g}_{ij}\mathrm{d}y^i\mathrm{d}y^j
\label{eq:grand}
\end{equation}
where $\Grad$ is the covariant derivative on $S^3$, $y$ are coordinates on the $S^3$, lower case Latin indices live on $S^3$, $\mathbbm{g}$ is the round metric on $S^3$ and
\begin{equation}
\mathbb{S}^{\ell_S}_{ij} = \Grad_i \Grad_j \mathbb{S}^{\ell_S}-\frac{\mathbbm{g}_{ij}}{3} \Grad^k \Grad_k \mathbb{S}^{\ell_S}\,.
\end{equation}
Here $\mathbb{S}^{\ell_S}_{ij}$ is by construction trace free. The perturbation of the gauge fields is more intricate. Since the background Maxwell fields break the full $SO(4)$, we expect that their perturbations will strongly depend on the background field. Note that $\mathbb{S}^{\ell_S}_{ij}$ vanishes identically for $\ell_S=1$ so that this mode must be treated separately; see section \ref{sec:ellS1}.

It is our objective to list all independent vector fields on $S^3$ that are linear in $\mathbb{S}^{\ell_S}$ and can be built with the background fields $A_I$ and metric $\mathbbm{g}$. These turn out to be
\begin{equation}
\delta A^{\ell_S}_I = A^{\ell_S}(r)\,S^{\ell_S}\,A_I+B^{\ell_S}(r)\tilde{\pounds}_{\Xi^{\ell_S}} A_I+C^{\ell_S}(r)\,A^i_I \Grad_i S^{\ell_S}\,\mathrm{d}r+D_I^{\ell_S}(r)\mathrm{d}r+E_I^{\ell_S}(r)\Grad_i S^{\ell_S}\mathrm{d}y^i\, ,
\end{equation}
where $\tilde{\pounds}$ is the Lie derivative acting on $S^3$ and
\begin{equation}
\Xi^{\ell_S}=\Grad^i \mathbb{S}^{\ell_S}\frac{\partial}{\partial y^i}\,.
\end{equation}
One might think that one would need to include additional terms of the form
\begin{equation}
\iota_{\Xi^{\ell_S}}\mathrm{d}A_I\,,
\end{equation}
but these can be reabsorbed into the coefficients $A^{\ell_S},B^{\ell_S},C^{\ell_S}$ with a $U(1)$ gauge transformation on $\delta A^{\ell_S}_I$.

In order to check that our \emph{Ansatz} is nontrivial, we have linearized the corresponding Einstein-Maxwell equations and found that there are three dynamical second order gauge invariant equations that govern such perturbations.  Since we want to work with gauge invariant perturbations, we must find how these perturbations transform under infinitesimal gauge transformations (both $U(1)$ and diffeomorphisms). In order to do this, we need to sort out how to write infinitesimal diffeomorphisms in terms of the scalar harmonics $\mathbb{S}^{\ell_S}$. Let $\xi^{\ell_S}$ be the infinitesimal diffeomorphism associated with $\mathbb{S}^{\ell_S}$. Following \cite{Kodama:2003kk} we write $\xi^{\ell_S}$ as
\begin{equation}
\xi^{\ell_S}= \xi^{\ell_S}_r(r)\,\mathbb{S}^{\ell_S}\,\mathrm{d}r+\xi^{\ell_S}_V(r) \Grad_i \mathbb{S}^{\ell_S}\,\mathrm{d}y^i\,.
\end{equation}

Our perturbations then transform as
\begin{align}
& \delta h_{rr}^{\ell_S} = \xi^{\ell_S}_r\frac{f^\prime}{f}+2 \xi^{\ell_S\,\prime}_r\,, \quad \delta h_{r}^{\ell_S} = \xi^{\ell_S}_r-\xi^{\ell_S}_V\frac{g^\prime}{g}+\xi^{\ell_S\,\prime}_V\,,\nonumber
\\
& \delta H_{T}^{\ell_S} =2\xi^{\ell_S}_V\,,\quad \delta H_{L}^{\ell_S} =f\,g^\prime \xi^{\ell_S}_r-\frac{2}{3}\ell_S(\ell_S+2)\xi^{\ell_S}_V\,, \nonumber
\\
& \delta A^{\ell_S}= \frac{f\,\Phi^\prime}{\Phi} \xi^{\ell_S}_r\,,\quad \delta B^{\ell_S}= \frac{\xi^{\ell_S}_V}{g}\nonumber
\\
& \delta C^{\ell_S}= \xi^{\ell_S\,\prime}_V-\frac{g^\prime}{g}\xi^{\ell_S}_V\,,\quad \delta D_I^{\ell_S}=0\,,\quad \text{and}\quad \delta E_I^{\ell_S}=0\,.
\label{eq:gaugegravells2}
\end{align}

Similarly, under $U(1)$ transformations $\delta A^{\ell_S}_I\to \delta A^{\ell_S}_I+\mathrm{d}\chi^{\ell_S}_I$ with gauge parameter $\chi^{\ell_S}_I=\hat{\chi}^{\ell_S}_I(r)\,\mathbb{S}^{\ell_S}$ we find
\begin{align}
& \delta h_{rr}^{\ell_S} = \delta h_{r}^{\ell_S} =\delta H_{T}^{\ell_S} =\delta H_{L}^{\ell_S} =\delta A^{\ell_S}=\delta B^{\ell_S}=\delta C^{\ell_S}= 0\,, \nonumber
\\
&\delta D^{\ell_S}_I=\hat{\chi}_I^{\ell_S\,\prime}\,,\quad \text{and}\quad \delta E^{\ell_S}_I=\hat{\chi}_I^{\ell_S}\,.
\end{align}
Just as we did for the $\ell_S=0$ mode, we now substitute our \emph{Ansatz} into the Einstein-Maxwell action (\ref{eq:simpleu13}) and expand to second order in the perturbations
\begin{equation}
\hat{\mathbf{q}}\equiv \{h^{\ell_S}_{rr},\,h^{\ell_S}_r,\, H^{\ell_S}_L,\, H^{\ell_S}_T,\, A^{\ell_S},\,B^{\ell_S},\,C^{\ell_S},\,D_I^{\ell_S},\,F_I^{\ell_S}\}\,.
\end{equation}
The resulting action, $S^{(2)}[\hat{\mathbf{q}},\hat{\mathbf{q}}^\prime]$ depends on both $\hat{\mathbf{q}}$ and its first derivative $\hat{\mathbf{q}}^\prime$ \footnote{To bring the action to this form, we integrated by parts term proportional to $H_T^{\ell_S\,\prime\prime}$ and $H_L^{\ell_S\,\prime\prime}$, with the boundary terms cancelling with the Gibbons-Hawking-York boundary term in (\ref{eq:simpleu13})\,.}. Crucially, the dependence in the angular coordinates drops out, as it should. Upon further inspection, one notes that by performing further integrations by parts we can in fact write $S^{(2)}$ in a form that depends only algebraically on $h^{\ell_S}_{rr}$ (i.e., it does not depend on derivatives of $h^{\ell_S}_{rr}$). The associated boundary terms again either cancel with the Gibbons-Hawking-York term or with one of the boundary counter-terms. This means we can easily integrate out $h^{\ell_S}_{rr}$ from the action (though this requires the Wick-rotation described in section \ref{sec:neggen}) and find a reduced action $\tilde{S}^{(2)}$ that does not depend on $h^{\ell_S}_{rr}$.  Upon further scrutiny, one finds that, upon a couple of integration by parts, $\tilde{S}^{(2)}$ does not depend on derivatives of $h_r^{\ell_S}$ so again it can be integrated out. At this stage we are left with a quadratic action $\hat{S}^{(2)}$ that depends only on
\begin{equation}
\{H^{\ell_S}_L,\, H^{\ell_S}_T,\, A^{\ell_S},\,B^{\ell_S},\,C^{\ell_S},\,D_I^{\ell_S},\,F_I^{\ell_S}\}\,.
\end{equation}
and their first derivatives. At this stage, we introduce gauge-invariant variables with respect to the $U(1)$. Under such gauge transformations the variables $\{H^{\ell_S}_L,\, H^{\ell_S}_T,\, A^{\ell_S},\,B^{\ell_S},\,C^{\ell_S}\}$ are already invariant.   However $D_I^{\ell_S}$ and $F_I^{\ell_S}$ transform non-trivially, so we define the invariant combination
\begin{equation}
f_I^{\ell_S} \equiv D_I^{\ell_S}-E_I^{\ell_S\,\prime}\Rightarrow D_I^{\ell_S}=f_I^{\ell_S} +E_I^{\ell_S\,\prime}\,.
\end{equation}
Using this definition, we find that $\hat{S}^{(2)}$ depends only on
\begin{equation}
\{H^{\ell_S}_L,\, H^{\ell_S}_T,\, A^{\ell_S},\,B^{\ell_S},\,C^{\ell_S},\,f_I^{\ell_S}\}\,
\end{equation}
and their first derivatives (as one would expect from the $U(1)$ gauge invariance). Furthermore, $f_I^{\ell_S}$ decouples from all other variables and appears in $\hat{S}^{(2)}$ as
\begin{equation}
\hat{S}^{(2)}=\check{S}^{(2)}+4\pi^2\frac{\ell_S(\ell_S+2)}{\ell_S+1}\sum_{I=1}^{3} \int_{-\infty}^{+\infty}\mathrm{d}r\,\sqrt{f\,g}\,(f_I^{\ell_S})^2
\label{eq:crazyone}
\end{equation}
where $\check{S}^{(2)}$ depends only on
\begin{equation}
\{H^{\ell_S}_L,\, H^{\ell_S}_T,\, A^{\ell_S},\,B^{\ell_S},\,C^{\ell_S}\}\,.
\end{equation}
This means that $f_I^{\ell_S}$ are non-dynamical in the sense that it enters the action algebraically and contributes positively to $\hat{S}^{(2)}$. We may thus focus our attention on $\check{S}^{(2)}$.

At this stage we introduce gauge invariant variables with respect to the diffeomorphisms (\ref{eq:gaugegravells2}). Since $\xi^{\ell_S}_r$ and $\xi_V^{\ell_S}$ enter the gauge transformations for $H_L^{\ell_S}$ and $H_T^{\ell_S}$ algebraically, we can use $H_L^{\ell_S}$ and $H_T^{\ell_S}$ to easily construct gauge invariant variables as follows. Define
\begin{subequations}
\begin{align}
& Q_1^{\ell_S} = A^{\ell_S}-\frac{\Phi^\prime}{6\,r\,\Phi}\left[3\,H_L^{\ell_S}+\ell_S(\ell_S+2)H_T^{\ell_S}\right]\,,
\\
&Q_2^{\ell_S}=B^{\ell_S}-\frac{H_T^{\ell_S}}{2\,g}\,,
\\
&Q_3^{\ell_S}=C^{\ell_S}-\frac{H_T^{\ell_S\,^\prime}}{2}+\frac{g^\prime}{g}H_T^{\ell_S}\,.
\end{align}
\end{subequations}
Using the transformations \ref{eq:gaugegravells2} it is relatively simple to see that $\delta Q_1^{\ell_S}=\delta Q_2^{\ell_S}=\delta Q_3^{\ell_S}=0$. Using this relations, we can solve for $A^{\ell_S}$, $B^{\ell_S}$ and $C^{\ell_S}$ and insert those expressions into $\check{S}^{(2)}$. After some integrations by parts, the terms with $H_L^{\ell_S}$ and $H_T^{\ell_S}$ drop out, as they should due to gauge invariance. At this stage, $\check{S}^{(2)}$ is a function of
\begin{equation}
\{Q_1^{\ell_S},Q_2^{\ell_S},Q_3^{\ell_S},Q_1^{\ell_S\,\prime},Q_2^{\ell_S\,\prime}\}\,.
\end{equation}
Remarkably, and for reasons we don't fully understand, $Q_3^{\ell_S}$ only enters the action algebraically (and with positive coefficient for the quadratic term).  This allows us to perform the Gaussian integral over $Q_3^{\ell_S}$ and be left with an effective action for $\{Q_1^{\ell_S},Q_2^{\ell_S}\}$ which we denote by $S^{(2)}_F$.

It turns out to be beneficial to perform one final change of variable and write
\begin{equation}
Q_1^{\ell_S} = \frac{2}{L\,\Phi}\left[q_1^{\ell_S}-\frac{g \lambda_S  \left(r+4 L^2 \Phi  \Phi^\prime\right)}{2 r \left(g \lambda_S +24 L^2 \Phi ^2\right)}q_2^{\ell_S}\right]\,,\quad \text{and}\quad Q_2^{\ell_S} = -\frac{1}{2\,L\,\Phi}q_2^{\ell_S}\,.
\end{equation}
The final action then takes the following form
\begin{equation}
S^{(2)}_F = \int_{-\infty}^{+\infty}\,\mathrm{d}\,r \frac{g^{3/2}}{8\,\sqrt{f}}\left[f (\mathcal{D}q^{\ell_S})_I\mathbb{K}^{IJ}(\mathcal{D}q^{\ell_S})_J+q^{\ell_S}_I\mathbb{V}^{IJ}q^{\ell_S}_J\right]
\end{equation}
where $I,J\in\{1,2\}$, $(\mathcal{D}q^{\ell_S})_I = q^{\ell_S\,^\prime}_I+\varepsilon_{I}^{\phantom{I}J}q^{\ell_S}_J$, $\varepsilon_{I}^{\phantom{I}J}=\varepsilon_{IJ}\mathbb{K}^{IJ}$,
\begin{equation}
\varepsilon_{IJ}=\left[\begin{array}{cc}
0 & \lambda
\\
0 & 0
\end{array}\right]\, ,
\end{equation}
and
\begin{multline}
\lambda = \frac{(\ell_S +1)}{32 \pi ^2 (\lambda_S -3) L^2 r f\left(g \lambda_S +24 L^2 \Phi ^2\right)} \Big[12 L^4 r \Phi  f \Phi^\prime (g-4 L^2 \Phi ^2)
\\
-4 L^2 \Phi ^2 (3 g^2+g \lambda_S  L^2+3 g L^2+3 L^2 r_0^2+3 r_0^4)+g \lambda_S  r_0^2 (L^2+r_0^2)+96 L^6 \Phi^4\Big]\,.
\end{multline}
Finally, the symmetric matrix $\mathbb{V}$ is given in appendix \ref{sec:crazy} and the symmetric matrix $\mathbb{K}$ is more easily expressed in terms of its inverse
\begin{subequations}
\begin{multline}
(\mathbb{K}^{-1})_{11}=\frac{g (\ell _S+1)}{256 \pi ^2 r^2 (\lambda _S-3) (g \lambda _S+24 L^2 \Phi ^2)} \Bigg[2 g^2 L^2 \lambda _S \left(\Phi^\prime-\frac{2 r \Phi}{g}\right)^2+
   \\
+g r^2 (\lambda _S-3) \lambda _S+\frac{16 g \Phi ^2 (g+L^2) (\lambda _S-3)}{f} \left(1-\frac{4 L^4 \Phi ^2}{g \left(g+L^2\right)}\right)\Bigg]\,,
\end{multline}
\begin{align}
& (\mathbb{K}^{-1})_{22}=\frac{(\ell_S +1)}{64 \pi ^2 \lambda_S(\lambda_S-3)} \left(\lambda_S g+24 L^2 \Phi^2\right)\,,
\\
& (\mathbb{K}^{-1})_{12}=0\,.
\end{align}
\end{subequations}
It is not hard to show that $\mathbb{K}$ is positive definite for $r_0/L>3^{1/4}/2^{1/2}\approx 0.930605$. First we note that $(\mathbb{K}^{-1})_{22}$ is positive definite, second we note that is is only the last term in $(\mathbb{K}^{-1})_{11}$ that is not positive definite. However, it is a simple exercise to show that $1-\frac{4 L^4 \Phi^2}{g \left(g+L^2\right)}>0$ for $r_0/L>3^{1/4}/2^{1/2}$. Since $3^{1/4}/2^{1/2}<r_0^{\min}/L$, all large wormholes have positive definite $\mathbb{K}$.

The non-existence of negative modes can then be investigated by looking at the properties of $\mathbb{V}$. As the reader can see in appendix \ref{sec:crazy}, $\mathbb{V}$ is a rather complicated matrix, whose eigenvalues we can only compute numerically as a function of $r$, for particular values of $\ell$ and $r_0/L$. This allows us to exclude portions of the $(r_0/L,\ell_S)$ plane as potential regions with negative modes. In Fig.~(\ref{fig:negell2}) we plot the regions of parameter space where $\mathbb{V}$ is positive definite.  It appears that if
$\mathbb{V}$ is positive definite for given $(r_0/L,\ell_S)$, then it remains positive definite if we increase $\ell_S$ while holding $r_0/L$ fixed.
 Since we are looking at large wormholes, we took $r_0/L>1$.

 We see, perhaps counter-intuitively, that the larger the value of $r_0/L$, the larger value of $\ell_S$ we have to achieve to see $\mathbb{V}$ being positive definite. This might at first look worrying, but in fact it is natural to expect angular momentum to have less effect at larger $r_0$ (since the gradients associated with given $\ell_S$ are smaller at large $r_0$).  Indeed,
as $r_0/L$ increases we find that the most negative eigenvalue of $\mathbb{V}$ moves toward zero. Thus $\mathbb{V}$ becomes less and less negative at large $r_0$.
\begin{figure}[h]
\centering
\includegraphics[width =0.5\textwidth]{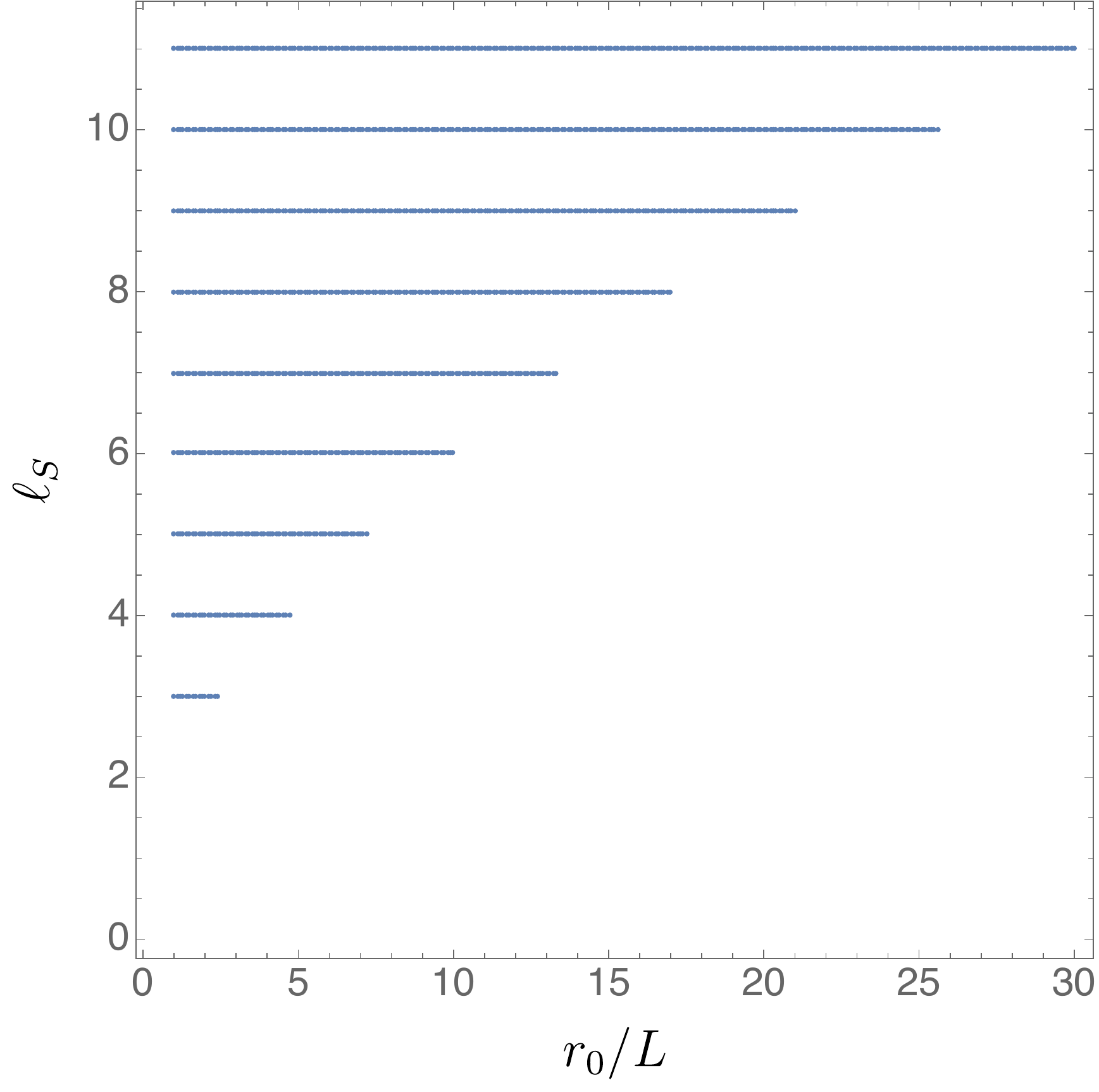}
\caption{Disks represent regions of moduli space in the $(r_0/L,\ell_S)$ plane where $\mathbb{V}$ is positive definite, and thus no negative mode exists.}
\label{fig:negell2}
\end{figure}

Note that if $\mathbb{V}$ is \emph{not} positive definite, it does not necessarily mean that negative modes exist. Indeed, for the complementary region we resort to solving for the spectrum numerically.  After doing so, we find no evidence for the existence of negative modes for $r_0>L$ and for $\ell_S\geq2$. We have performed a thorough search in parameter space by scanning the large wormhole branch to about $r_0\sim 10^4$ and up to $\ell_S \sim 10^3$.
\subsubsection{Scalar-derived perturbations: the $\ell_S=1$ sector}
\label{sec:ellS1}
This mode is special, because $H_T^{\ell}$ no longer appears in the metric perturbation. Apart from that, the construction is similar to what we have done for $\ell_S\geq2$. Perhaps the end result is somehow surprising. Again, we find that the second order action for the perturbations can be brought to a form similar to (\ref{eq:crazyone}) with again $f_I^{1}$ contributing positively to the action. However, we find that the second order action for the remaining variables vanishes identically.  Thus in the linearized theory these additional variables are pure gauge.  This in turn means that they are the linearization of a pure-gauge mode in the full theory, as the method applied by \cite{Alcubierre:2009ij} to showing that Einstein-Maxwell theory admits a symmetric-hyperbolic formulation will also apply to our system.  This in turn means that  no gauge degrees of freedom can remain in the linearized theory once the gauge symmetries of the full theory have been fixed.  We thus find that our wormhole is stable with respect to gauge-invariant perturbations in this sector.
\subsubsection{Vector-derived perturbations: the $\ell_V\geq2$ sector}
Things get more complicated, perhaps unexpectedly, when we move on to study vector-derived or tensor-derived perturbations. Note that when $A_I=0$, these sectors are really easy to study! The issue is that one can contract the fundamental vector harmonics $\mathbb{S}_i$ with $A^I_i$ and build a scalar harmonic. So, in general, the vectors harmonics couple to scalar-derived perturbations.  Thus their treatment will require all of the complications discussed above in the context of scalar-derived perturbations as well as treatment of the vector harmonics.

An explicit discussion is thus extremely tedious but can be performed using precisely the same techniques as in section \ref{sec:EMscalar}.  We suppress the details, but provide the following remarks.   It turns out that the vector derived perturbations only excite a few scalar-derived perturbations and that they do not excite tensors-derived perturbations. For a given vector harmonic $\mathbb{S}_i^{\ell_V}$ with $\ell_V\geq 2$ we find that we need to consider a sum of three scalar harmonics of the form $\mathbb{S}^{\ell_V-1}$ and three harmonics of the form $\mathbb{S}^{\ell_V+1}$. In each of these sectors, one of the harmonics is proportional to $\cos \psi$, $\sin \psi$ or has no $\psi$ dependence. It is also possible to find the exact differential map between these harmonics. It is then an incredibly tedious exercise to find the resulting action, and diagonalise accordingly using appropriate numerics. We have done so and find that the action is again positive definite.

However, there is a sector in which this unpleasant coupling does not occur and where one can work purely with vector harmonics.  We will describe the calculations for this simpler case in detail to illustrate the mechanics of working with the vector harmonics themselves. The simple sector involves vector-harmonics of the form
\begin{equation}
\mathbb{S}^{\ell_V}_i\,\mathrm{d}y^i=\left[- \sin(m \psi)\,\mathrm{d}\theta+\cos(m \psi)\,\sin \theta \, \mathrm{d}\hat \varphi\right]\sin^{\left|m-1\right|}\theta\,.
\end{equation}

Regularity at $\theta = 0$ and $\theta = \pi$ demands that $m\in\mathbb{Z}$. The case with $m=1$ is special, and because the background is invariant under $\psi\to-\psi$, $\phi\to-\phi$ we can take $m\geq2$ without loss of generality. It is a simple exercise to show that $\ell_V = 2m-1$.

For the metric perturbation we take
\begin{equation}
\delta \mathrm{d}s^2_{\ell_V} = 2 h^{\ell_V}_r(r)\,\mathbb{S}^{\ell_V}_i\,\mathrm{d}r\, \mathrm{d}y^{i}+H_T^{\ell_V}(r)\,\mathbb{S}^{\ell_V}_{ij}\,\mathrm{d}y^i \mathrm{d}y^j
\end{equation}
with
\begin{equation}
\mathbb{S}^{\ell_T}_{ij} = \Grad_i \mathbb{S}^{\ell_V}_j+ \Grad_j \mathbb{S}^{\ell_V}_i\,.
\end{equation}
While for the gauge field perturbations we take
\begin{equation}
\delta A^{\ell_V}_I = A^{\ell_V}(r)\tilde{\pounds}_{\Xi^{\ell_V}} A_I+B^{\ell_S}(r)\,A^i_I \mathbb{S}^{\ell_V}_i\,\mathrm{d}r+C_I^{\ell_V}(r)\mathbb{S}^{\ell_V}_i\mathrm{d}y^i\,
\end{equation}
where $\tilde{\pounds}$ is the Lie derivative acting on $S^3$ and
\begin{equation}
\Xi^{\ell_V}=\mathbb{S}^{i\;\ell_V}\frac{\partial}{\partial y^i}\,.
\end{equation}

Just as before we can ask how these perturbations behave under infinitesimal coordinate transformations. Here, the infinitesimal diffeomorphisms are parameterized via
\begin{equation}
\xi^{\ell_V} = L^{\ell_V}(r) \mathbb{S}^{i\;\ell_V}\frac{\partial}{\partial y^i}\,,
\end{equation}
and it turns out this induces the following transformations
\begin{align}
&\delta h_r^{\ell_V} = L^{\ell_V\;\prime}-L^{\ell_V}\frac{g^\prime}{g}\,,\quad \delta H_T^{\ell_V}=L^{\ell_V}\,,\quad \delta A^{\ell_V}=\frac{L^{\ell_V}}{g}\,,\nonumber
\\
&\delta B^{\ell_V}= L^{\ell_V\;\prime}-L^{\ell_V}\frac{g^\prime}{g}\,\quad\text{and}\quad \delta C^{\ell_V}=0\,.
\label{eq:gaugetvector}
\end{align}

From here onwards, the procedures are very similar to what we have seen for the scalars. First, we compute the second order action $S^{(2)}$ which is a function of $h_r^{\ell_V}$, $h_T^{\ell_V}$, $A^{\ell_V}$, $B_r^{\ell_V}$, $C_r^{\ell_V}$ and their first derivatives. Furthermore, it also depends on the second derivatives of $h_T^{\ell_V}$. One can integrate by parts the term proportional to the second derivative of $h_T^{\ell_V}$, reducing $S^{(2)}$ to a function of first derivatives only. The resulting boundary term is cancelled by the Gibbons-Hawking-York boundary action, as it should.

One then notes that $S^{(2)}$ only depends on $h_r^{\ell_V}$, but not on its first derivative. This means we can perform the (here already with the convergent sign) Gaussian integral over $h_r^{\ell_V}$  and find an action $\tilde{S}^{(2)}$ depending on $h_T^{\ell_V}$, $A^{\ell_V}$, $B_r^{\ell_V}$, $C_r^{\ell_V}$  and their first derivatives. At this point we further notice that $C_r^{\ell_V}$ completely decouples from the remaining action, and furthermore that its contribution to $\tilde{S}^{(2)}$ is manifestly positive definite. We are thus left to study an effective action $\hat{S}^{(2)}$ for $h_T^{\ell_V}$, $A^{\ell_V}$, $B_r^{\ell_V}$ and their first derivatives. Finally, we make use of the gauge transformations (\ref{eq:gaugetvector}) and introduce gauge invariant variables of the form
\begin{equation}
q_1^{\ell_V} = A^{\ell_V}-\frac{h_T^{\ell_V}}{g}\quad \text{and} \quad q_2^{\ell_V} = B^{\ell_V}+\frac{g^\prime}{g}h_T^{\ell_V}-h_T^{\ell_V\;\prime}\,.
\end{equation}
It is a simple exercise to check that $\delta q_1^{\ell_V}=\delta q_2^{\ell_V}$. After some integration by parts, the dependence in $h_T^{\ell_V}$ drops out completely (as it should by virtue of diffeomorphism invariance), and $\hat{S}^{(2)}$ is solely written in terms of $q_1^{\ell_V}$, its first derivative and of  $q_2^{\ell_V}$ . Since $q_2^{\ell_V}$ only enters algebraically, we can perform the Gaussian path integral and find an action $S_F^{(2)}$ for $q_1^{\ell_V}$ only. It is convenient to perform one last change of variable and write
\begin{equation}
q_1^{\ell_V}=\frac{Q^{\ell_V}}{\Phi}\,.
\end{equation}
The final action for $Q^{\ell_V}$ reads
\begin{multline}
S_F^{(2)}=\frac{32 \pi ^{5/2} L^2 (m-1)^3 (m+1) \Gamma (m)}{\Gamma \left(m+\frac{1}{2}\right)}\int_{-\infty}^{+\infty}\mathrm{d}r\;\frac{g^{3/2}}{\sqrt{f}}\;\Bigg\{\frac{f}{g (m-1) m+4 L^2 \Phi ^2}{Q^{\ell_V\;\prime}}^2
\\
+\frac{V^{\ell_V}(r)}{[g (m-1) m+4 L^2 \Phi ^2]^2}{Q^{\ell_V}}^2\Bigg\}\,,
\end{multline}
with
\begin{multline}
V^{\ell_V}(r) = \frac{4}{g^2 L^2 (m-1)^2} \Bigg\{2 f g L^4 m\,\Phi \,r \,\Phi^\prime+4 L^2 \Phi ^2 \left[L^2 \left(g m^2-2g m+r_0^2\right)+r_0^4\right]+
\\
g m L^2 \left[m \left(g m^2-g m+r_0^2\right)+r_0^2\right]+m (m+1) g r_0^4\Bigg\}\,.
\end{multline}
For $m\geq2$ all terms appearing in $V^{\ell_V}(r) $ are positive definite (note that $r \,\Phi^\prime$ is positive definite for $r\in \mathbb{R}$), and thus no negative mode exists in this sector as well.
\subsubsection{Vector-derived perturbations: the $\ell_V=1$ sector}
\label{sec:T1}
This mode is again special because $\mathbb{S}^{\ell_V}_{ij}$ vanishes identically, and thus $h_T^{\ell_V}$ does not enter the calculation. Again, we find that $C^{1}$ contributes positively to the action, but the remaining gauge invariant variables have a vanishing action (and thus as in section \ref{sec:ellS1} are in fact pure gauge under some special residual gauge transformations). We thus find that our wormhole is stable with respect to gauge-invariant perturbations in this sector.
\subsubsection{Tensor-derived perturbations: the $\ell\geq2$ sector}
The analysis of general tensor-derived perturbations is even more complicated than in the vector-derived case. The tenor-derived perturbations in general couple to both scalar-derived (with $\ell_S=\ell_T\pm2$) and vector-derived perturbations (with $\ell_V=\ell_T\pm1$). Once again, the general calculation can be performed by combining an analysis of the tensor-derived harmonics with the techniques used above, and doing so yields an action which (with appropriate numerics) can be verified to be positive definite.  As in the vector-derived case we suppress the details of this extremely tedious general study and limit explicit discussion to  a special type of tensor-derived perturbation that does not source either vector- or scalar-derived harmonics.  This allows us to illustrate the treatment of the purely tensor-derived part.  Combining such a treatment with the methods used above for  scalar- and vector-derived perturbations then suffices to treat the general case.

The simple tensor-derived sector is descreibed by metric perturbations of the form
\begin{align}a
\delta \mathrm{d}s^2 & = \frac{g}{2\sqrt{2}L^2}h_T(r)\Big\{(\sigma_2^2-\sigma_1^2)\cos\left[(m-2)\psi\right]+2 \sigma_1 \sigma_2 \sin\left[(m-2)\psi\right]\Big\}\sin^{|m-2|}\theta\nonumber
\\
& \equiv \frac{g}{2\sqrt{2}L^2}h_T(r)\,\mathbb{S}_{ij}\mathrm{d}y^i \mathrm{d}y^j\,,
\end{align}
with $m\in\mathbb{Z}$ and without loss of generality we take $m\geq 2$. For the gauge field perturbation we take
\begin{equation}
\delta A_{I}= \frac{g}{8\Phi L^2}a_T(r) \mathbb{S}_{ij}\mathrm{d}y^i {A_I}^{j}\,.
\end{equation}
Both $h_T(r)$ and $a_T(r)$ are automatically gauge invariant with respect to both infinitesimal diffeomorphisms and gauge perturbations. To show this, recall that $\mathbb{S}_{ij}$ is transverse and trace free. One can readily compute  $\ell_T$ by using Eq.~(\ref{eq:tensor}) and it turns out $\ell_T = 2(m-1)$. Everything is much simpler now because these quantities are gauge invariant. In particular, in order to cast the quadratic action in an adequate form we only need to remove a term proportional to the second derivative of $h_T$. The boundary term readily cancels off the usual Gibbons-Hawking-York term. It turns out we can write the second order action in the form
\begin{equation}
S^{(2)}_F = \frac{\pi ^{5/2} \Gamma (m-1)}{L^4 \Gamma \left(m-\frac{1}{2}\right)}\int_{-\infty}^{+\infty}\mathrm{d}r\,\frac{g^{3/2}}{f^{1/2}}\left[f (\mathcal{D}q^{\ell_S})_I\mathbb{K}^{IJ}(\mathcal{D}q^{\ell_S})_J+q^{\ell_S}_I\mathbb{V}^{IJ}q^{\ell_S}_J\right],
\end{equation}
where $I,J\in\{1,2\}$, $(\mathcal{D}q^{\ell_S})_I = q^{\ell_S\,^\prime}_I+\varepsilon_{I}^{\phantom{I}J}q^{\ell_S}_J$, $\varepsilon_{I}^{\phantom{I}J}=\varepsilon_{IJ}\mathbb{K}^{IJ}$,
\begin{equation}
\varepsilon_{IJ}=\left[\begin{array}{cc}
0 & \beta_1 \Phi^\prime
\\
\beta_2 \Phi^\prime & 0
\end{array}\right]\,.
\end{equation}
where $\beta_1$, $\beta_2$, $\beta_3$ and $\beta_4$ are constants,
\begin{equation}
\mathbb{K}^{IJ}= \left[
\begin{array}{ccc}
1+\frac{L^2}{g}(\beta_4+\beta_3 \Phi)^2 & &\frac{L^2}{g}(\beta_4+\beta_3 \Phi)^2
\\
\\
\frac{L^2}{g}(\beta_4+\beta_3 \Phi)^2 & &\frac{L^2}{g}
\end{array}
\right]
\end{equation}
and
\begin{equation}
h_T = q_1\quad\text{and}\quad a_T = (\beta_4+\beta_3 \Phi)q_1+q_2\,.
\end{equation}
Note that $\mathrm{det}\,\mathbb{K}>0$ and $\mathrm{Tr}\,\mathbb{K}>0$ so that $\mathbb{K}$ is a positive definite symmetric matrix. In addition, we must take $\beta_3 = 2\sqrt{2}-\beta_1+\beta_2$ so that terms of the form $q_I^\prime q_J$ for $I\neq J$ do not feature the action. The symmetric matrix $\mathbb{V}$ is incredibly cumbersome to write down explicitly, and not very illuminating. What is important is that we can choose $\beta_1$, $\beta_2$ and $\beta_4$ so that $\mathbb{V}$ is also positive. A good choice turns out to be
\begin{equation}
\beta_1=-\frac{6}{5} \sqrt{2} (m-1),\quad \beta_2 =-\frac{2}{5} \sqrt{2} (3 m+2)\quad\text{and}\quad \beta_4 = \frac{17 (m-1)^2}{5 \sqrt{2} (6 m-1)}\,.
\end{equation}
For the above choice, we present $\mathbb{V}$ in appendix \ref{sec:b}.

\section{\label{sec:scalars}A Scalar field model with $S^3$ boundary}

Let us now consider a different class of simple asymptotically AdS$_4$ models in which the wormhole is sourced by the stress-energy of scalar fields.  We again assume spherical symmetry, and in particular a spherical boundary metric.  After giving an overview of the model, we briefly describe the ans\"atze we use to study the connected wormhole and the disconnected solution.  Computing the actions require more numerics than in the Einstein-Maxwell model of section \ref{sec:u13}, so we devote a separate subsection to discussing the results of such computations.  A final subsection considers potential negative modes in direct parallel with the discussion of section \ref{sec:EMNM}.  The final results also mirror those of section \ref{sec:u13}, as we again find a Hawking-Page-like structure with two branches of wormholes (large and small).  Moreover, the large wormholes are once again free of negative modes and dominate over the disconnected solution when they are sufficiently large.  However, this model also has no possible notion of a brane-nucleation negative modes as the theory does not contain branes.  In particular, while we will later discuss its relation to certain string-theoretic setups, the current model is not UV-complete.

We will study both conformally coupled scalars and massless scalars, though for the moment we include an arbitrary mass parameter $\mu$. Our action is given by
\begin{equation}
S=-\int_{\mathcal{M}} \mathrm{d}^4 x\sqrt{g}\left[R+\frac{6}{L^2}-2 (\nabla_a \vec{\Pi})\cdot  (\nabla^a \vec{\Pi})^*-2 \mu^2 \vec{\Pi}\cdot \vec{\Pi}^*\right]-2\int_{\partial \mathcal{M}} \mathrm{d}^3 x\sqrt{h}\;K+S^{\mu^2}_{\mathcal{B}}\,,
\label{eq:actionscalargeneral}
\end{equation}
where $L$ is the four-dimensional AdS length scale, $\vec{\Pi}$ describes a doublet of complex scalar fields, and ${}^*$ denotes complex conjugation.  The second term in \eqref{eq:actionscalargeneral} is the usual Gibbons-Hawking term and $S^{\mu^2}_{\mathcal{B}}$ the boundary counter-term to make the action finite and the variational problem well defined.  The precise form of $S^{\mu^2}_{\mathcal{B}}$ will explicitly depend on $\mu^2$.

The Einstein equation and scalar field equation derived from this action read
\begin{subequations}
\label{eqs:eqs}
\begin{align}
&R_{ab}-\frac{R}{2}g_{ab}-\frac{3}{L^2}g_{ab}=2 \,\nabla_{(a} \vec{\Pi}\cdot \nabla_{b)} \vec{\Pi}^*-g_{ab} \nabla_c \vec{\Pi}\cdot \nabla^c\vec{\Pi}^*-\mu^2\,g_{ab}\,\vec{\Pi}\cdot \vec{\Pi}^*\,,
\label{eqs:eqsa}
\\
& \Box \vec{\Pi}=\mu^2 \vec{\Pi}\,.
\end{align}
\end{subequations}
We now note that if we take the trace of the Einstein equation, we find
\begin{equation}
R=-\frac{12}{L^2}+2 (\nabla_a \vec{\Pi})\cdot  (\nabla^a \vec{\Pi})+4 \mu^2 \vec{\Pi}\cdot \vec{\Pi}^*\,.
\end{equation}
As such, the on-shell Euclidean action can be computed by evaluating the following bulk integral
\begin{equation}
S_{\mathrm{on-shell}}=\int_{\mathcal{M}}\mathrm{d}^4 x\sqrt{g}\left[\frac{6}{L^2}-2 \mu^2 \vec{\Pi}\cdot \vec{\Pi}^*\right]-2\int_{\partial \mathcal{M}} \mathrm{d}^3 x\sqrt{h}\;K+S^{\mu^2}_{\mathcal{B}}\,.
\label{eq:actionon}
\end{equation}
The precise form of $S_{\mathcal{B}}$ will be important to evaluate this on-shell action and depends on the boundary conditions that we impose on scalar doublet $\vec{\Pi}$. We will take $\mu^2$ to be zero or to take the conformal value $\mu^2 L^2=-2$.

In standard Fefferman-Graham coordinates \cite{Fefferman:2007rka} the metric takes schematic form
\begin{equation}
\mathrm{d}s^2 = \frac{L^2}{z^2}\left[\mathrm{d}z^2+\hat{g}_{\mu\nu}(x,z)\mathrm{d}x^\mu \mathrm{d}x^\nu\right],
\end{equation}
where $z=0$ marks the location of the conformal boundary and recall that Greek indices run over boundary directions only. One can then show that $\hat{g}_{\mu\nu}(x,z)$ admits a simple expansion in terms of a power series in $z$, possibly with $\log z$ terms depending on the scalar field mass $\mu$. For $\mu^2L^2=-2$ and $\mu=0$ the $\log z$ terms can be shown to be absent and $\hat{g}_{\mu\nu}(x,z)$ can be expanded as
\begin{equation}
\hat{g}_{\mu\nu}(x,z)=g^{0}_{\mu\nu}(x)+z^2 g^{(2)}_{\mu\nu}(x)+z^3 g^{(3)}_{\mu\nu}(x)+ o(z^3)\, ,
\end{equation}
where $g^{0}_{\mu\nu}(x)$ is interpreted as the boundary metric. We can now explain a little bit better with what we mean by $\partial \mathcal{M}$ in Eq.~(\ref{eq:actionscalargeneral}). The surface $\partial \mathcal{M}$ is defined as the $\varepsilon\to0$ limit of hypersurfaces $\partial \mathcal{M}_\varepsilon$ on which $z=\varepsilon$. Furthermore, $h_{\mu\nu}(\varepsilon)$ is the induced metric on $\partial \mathcal{M}_\varepsilon$. In this sense, $\lim_{\varepsilon\to0}\varepsilon^2h_{\mu \nu}/L^2 = g^{0}_{\mu\nu}(x)$.  Note that in the wormhole case $\partial \mathcal{M}$ has two connected components, one at each end of the wormhole.

In FG coordinates the scalar doublet $\vec{\Pi}$ can be expanded as
\begin{equation}
\vec{\Pi} = \vec{\Pi}^{-}(x) z^{\Delta_-}[1+o(1)]+\vec{\Pi}^{+}(x) z^{\Delta_+}[1+o(1)],
\label{eq:general}
\end{equation}
with
\begin{equation}
\Delta_{\pm} = \frac{3}{2}\pm\sqrt{\frac{9}{4}+\mu^2 L^2}\,.
\end{equation}
Throughout this section fix $\vec{\Pi}^{+}$ as a boundary condition (associated with ``standard quantization'' in the language of \cite{Klebanov:1999tb}), so that in AdS/CFT $\vec{\Pi}^{-}(x)$ becomes the expectation value of the operator dual to $\vec{\Pi}$. For the massless and conformal cases we have $\Delta_-=0$ and $\Delta_-=1$ respectively. This choice (partially) dictates  what $S_{\mathcal{B}}$ should be to make the variational problem well defined. That is to say, when deriving the equations of motion we want to ensure that we keep $\vec{\Pi}^{-}(x)$  fixed and that our boundary terms are consistent with such choice. The remaining freedom in choosing $S_{\mathcal{B}}$ is fixed so that the action (\ref{eq:actionscalargeneral}) is finite.

In the massless case we take
\begin{equation}
S^{\mu^2=0}_{\mathcal{B}}= -\frac{4}{L}\int_{\partial \mathcal{M}}\mathrm{d}^3 x\sqrt{h}-L \int_{\partial \mathcal{M}}\mathrm{d}^3 x\sqrt{h}R^h+2 L \int_{\partial \mathcal{M}}\mathrm{d}^3 x\sqrt{h} h^{\mu\nu} \nabla^h_{\mu}\vec{\Phi}\cdot \nabla^h_{\nu}\vec{\Phi}^{\star}\,,
\end{equation}
where $R^h$ is the Ricci scalar associated with $h$ and $\nabla^h$ its metric preserving connection.

For the conformal case we take
\begin{equation}
S^{\mu^2L^2=-2}_{\mathcal{B}}= -\frac{4}{L}\int_{\partial \mathcal{M}}\mathrm{d}^3 x\sqrt{h}-L \int_{\partial \mathcal{M}}\mathrm{d}^3 x\sqrt{h}R^h-\frac{2}{L}\int_{\partial \mathcal{M}}\mathrm{d}^3 x\sqrt{h}\,\vec{\Phi}\cdot \vec{\Phi}^{\star}\,.
\end{equation}

We are interested in finding solutions where the metric enjoys spherical symmetry but the scalars are chosen in such a way that these symmetries are broken. In particular, we might consider
\begin{equation}
\mathrm{d}s^2 = g_{rr}(r)\mathrm{d}r^2+g_{S^3}(r)\mathrm{d}\Omega^2_3\,,
\label{eq:gen}
\end{equation}
where $\mathrm{d}\Omega^2_3$ is the unit round three-sphere, $g_{rr}(r)$, $g_{S^3}(r)$ are to be determined later and $r$ is an arbitrary bulk coordinate.

For the scalar fields, we take
\begin{equation}
\vec{\Pi}=\vec{X}\,\Pi(r)
\end{equation}
with $\Pi(r)\in\mathbb{R}$ and $\vec{X}$ a two dimensional complex unit vector on $S^3$ with $\mathrm{d}\vec{X}\neq0$ and $\vec{X}\cdot \vec{X}^{*}=1$.

In fact, the coordinates of \eqref{eq:gen} turn out to be inconvenient for constructing our solutions.  We thus present two new \emph{Ans\"atze} below corresponding to connected or disconnected solutions and which specify our gauge choice slightly differently in each case.
\subsection{\emph{Ansatz} for the wormhole solutions}
When searching for wormhole solutions, we will take
\begin{equation}
\mathrm{d}s^2 = \frac{L^2}{(1-\tilde{y}^2)^2}\left\{\frac{4f(\tilde{y})\mathrm{d}\tilde{y}^2}{2-\tilde{y}^2}+y_0^2\mathrm{d}\Omega^2_3\right\}\,.
\label{eq:ansatzworm}
\end{equation}
Clearly, Eq.~(\ref{eq:ansatzworm}) falls into the same symmetry class as Eq.~(\ref{eq:gen}), and the factors of $\tilde{y}\in(-1,1)$ where chosen to that asymptotically (as $\tilde{y}\to\pm1$) $f\to1$. From the form of the \emph{Ansatz} above it is clear that $y_0$ is the minimal radius of the $S^3$ in the interior, which is attained at $\tilde{y}=0$. The value of this radius for given boundary sources will be determined numerically. For the scalar field we take
\begin{equation}
\Pi = (1-\tilde{y}^2)^{\Delta_+} q\,.
\end{equation}

Let us describe the boundary conditions at $\tilde{y}=0$ in detail. We wish to impose the reflection symmetry $\tilde y \rightarrow - \tilde y$ which leaves the locus $\tilde y=0$ invariant.  As a result, $f(\tilde{y})=f(-\tilde{y})$ and $q(\tilde{y})=q(-\tilde{y})$. This implies
\begin{equation}
\label{eq:RBCYt}
\left.\frac{\mathrm{d} f}{\mathrm{d}\tilde{y}}\right|_{\tilde{y}=0}=\left.\frac{\mathrm{d} q}{\mathrm{d}\tilde{y}}\right|_{\tilde{y}=0}=0\,.
\end{equation}
The requirement \eqref{eq:RBCYt} in fact motivates us to choose a different coordinate that automatically enforces these conditions. In particular, we use  $y: = \tilde{y}^2$ with $y\in(0,1)$. The line element now reads
\begin{equation}
\mathrm{d}s^2 = \frac{L^2}{(1-y)^2}\left\{\frac{f(y)\mathrm{d}y^2}{(2-y)y}+y_0^2\mathrm{d}\Omega^2_3\right\}\,,
\label{eq:ansatzworm2}
\end{equation}
and the boundary conditions \eqref{eq:RBCYt} become just the statement that $f$ and $q$ are regular at $y=0$ in the sense that  (using further input from the 2nd order equations of motion) both must admit a Taylor series expansion about this locus. The equations in the $y$ coordinates and in this gauge are sufficiently compact to to present here:
\begin{subequations}
\begin{align}
\frac{(1-y)^4 \sqrt{2-y} \sqrt{y}}{\sqrt{f}}\frac{\mathrm{d}}{\mathrm{d}y}\left[\frac{\sqrt{2-y} \sqrt{y}}{\sqrt{f} (1-y)^2}\frac{\mathrm{d}\Pi}{\mathrm{d}y}\right]-\frac{3 (1-y)^2}{y_0^2}\Pi-L^2 \mu ^2 \Pi=0\,,
\\
f = \frac{(2-y)\,y\,y_0^2}{3 \left[(1-y)^2+y_0^2\right]-\Pi^2 \left[3 (1-y)^2+\mu ^2 L^2 y_0^2\right]}\left[3-(1-y)^2 \left(\frac{\mathrm{d}\Pi}{\mathrm{d}y}\right)^2\right]\,.
\end{align}
\end{subequations}
A non-singular wormhole must have $f$ being non-zero and finite everywhere, and in particular at $y=0$. Looking at the expression above for $f$, it follows that at $y=0$ we must have
\begin{subequations}
\begin{equation}
\Pi(0)=\frac{\sqrt{3} \sqrt{1+y_0^2}}{\sqrt{3+\mu ^2 L^2 y_0^2}}\, ,
\end{equation}
where without loss of generality we took $q(0)>0$. Assuming a regular Taylor series for $q$ around $y=0$, it also follows that
\begin{equation}
\left.\frac{\mathrm{d}\Pi}{\mathrm{d}y}\right|_{y=0}=\frac{\sqrt{3} \sqrt{1+y_0^2} \left(3+L^2 y_0^2 \mu ^2\right)^{3/2}}{y_0^2 \left(3-L^2 \mu ^2\right)}\quad\text{and}\quad f(0)=\frac{3+L^2 y_0^2 \mu ^2}{3-L^2 \mu ^2}\,.
\end{equation}
\label{eq:BCSscalars}
\end{subequations}%
Note that the four-dimensional Breitenlohner-Freedman bound ensures that both quantities above are positive definite.

At the conformal boundary we find
\begin{equation}
q = \left\{\begin{array}{l}
V-\frac{3}{2 y_0^2 }V (1-y)^2+\lambda (1-y)^3-\frac{9}{8 y_0^4} V (1-y)^4+\mathcal{O}\left[(1-y)^5\right]\quad \text{for}\quad \mu =0\,,
\\
\\
\frac{V}{y_0}+\kappa (1-y)+\frac{V \left(4+V^2\right)}{2 y_0^3} (1-y)^2+\mathcal{O}\left[(1-y)^3\right] \quad \text{for}\quad \mu^2 L^2 =-2
\end{array}
\right.
\label{eq:expansiony}
\end{equation}
where $\lambda$ and $\kappa$ are constants. The metric (\ref{eq:ansatzworm2}) is not yet in FG coordinates, so we are not yet ready to read off the values of boundary sources. In order to achieve this we need to perform a change of variables from $y$ to $z$. We only require this change asymptotically,
\begin{equation}
y = \left\{\begin{array}{l}
1-y_0 z-\frac{1}{4} \left(1-V^2\right) y_0 z^3+\mathcal{O}(z^5)\quad \text{for}\quad \mu =0\,,
\\
\\
1-y_0 z-\frac{1}{4} \left(1+V^2\right) y_0 z^3-\frac{4}{9} \kappa  V y_0^3 z^4+\mathcal{O}(z^5)\quad \text{for}\quad \mu L^2 =-2\,.
\end{array}\right.
\end{equation}
We can now compare the expansion for $\Pi$ in powers of $z$ with the general expansion (\ref{eq:general}) which gives
\begin{equation}
\vec{\Pi}^{+} = \vec{X}\,V
\end{equation}
for both $\mu=0$ and $\mu^2 L^2=-2$. We thus see that $V$ is to be interpreted as the source of the operator dual to $\vec{\Pi}$.

The numerical procedure to find these solutions is clear. We take a value for $y_0$, and use the boundary conditions (\ref{eq:BCSscalars}) to numerically integrate the equations outwards to $y=1$. Once this is done, we read off $V$, thus finding what source was needed to source a wormhole with minimal size $y_0$. To do this, we write the equations in first order form, and use a Chebyshev collocation grid on Gauss-Lobatto points to perform the numerical integration. Because in our gauge the expansion of $q$ can be shown to be analytic at the singular points $y=0,1$ we obtain exponential convergence in the number of grid points as we approach the continuum limit.

Our integration also determines the values of $\lambda$ and $\kappa$ in (\ref{eq:expansiony}). These are related to the expectation value $\langle \mathcal{O}_{\vec{\Pi}}\rangle$ of the operator dual to $\vec{\Pi}$ via
\begin{equation}
\langle \vec{\mathcal{O}}_{\vec{\Pi}}\rangle=\frac{\delta S}{\delta \vec{\Phi}^{+}}=\vec{X}\,\left\{
\begin{array}{l}
-24 \pi ^2 \lambda  L^2 y_0^3\quad \text{for}\quad \mu =0\,,
\\
\\
8 \pi ^2 \kappa  L^2 y_0^2\quad \text{for}\quad \mu L^2 =-2\,.
\end{array}
\right.
\end{equation}

Finally, we give a few words on how to evaluate (\ref{eq:actionon}) in a numerically stable manner. We first look at how the first term in (\ref{eq:actionon}) diverges in inverse powers of $(1-y)$ and $\sqrt{y}$. We then add and subtract a regulator with precisely the same singularity structure, but one that we can integrate analytically. More precisely we use
\begin{align}
\frac{S}{8 \pi^2\,L^2} & = \int_{0}^{+\infty}\mathrm{d} y\frac{ \sqrt{f}\,y_0^3}{(1-y)^4 \sqrt{2-y} \sqrt{y}} \left(3-L^2 \mu ^2 \Pi^2\right)-2\int_{\partial \mathcal{M}} \mathrm{d}^3 x\sqrt{h}\;K+S^{\mu^2}_{\mathcal{B}}\nonumber
\\
& = \int_{0}^{+\infty}\mathrm{d} y\left[\frac{ \sqrt{f}\,y_0^3}{(1-y)^4 \sqrt{2-y} \sqrt{y}} \left(3-L^2 \mu ^2 \Pi^2\right)-G\right]\nonumber
\\
& \hspace{4cm} +\int_{0}^{+\infty}\mathrm{d} y\,G-2\int_{\partial \mathcal{M}} \mathrm{d}^3 x\sqrt{h}\;K+S^{\mu^2}_{\mathcal{B}},
\label{eq:regu}
\end{align}
where $G$ takes the form
\begin{equation}
G= \frac{G^{(-4)}}{(1-y)^4}+\frac{G^{(-2)}}{(1-y)^2}+G^{(0)}+\frac{G^{(1)}(1-y)}{\sqrt{y}}\,.
\end{equation}
Here all $G^{(i)}$ are constant and are chosen in such a way that the first integrand on the second line of (\ref{eq:regu}) vanishes as $y\to1$ and as $y\to0$. These constants can be found analytically, because we know the expansion for all functions via (\ref{eq:expansiony}). Once this is done, we can analytically perform  the integrations on the last line of (\ref{eq:regu}) and check that the result is finite as desired. The numerical integral that remains  is then manifestly finite, with all the complicated cancelations having been implemented analytically.
\subsection{\emph{Ansatz} for the disconnected solution}
The procedure for the disconnected solutions is similar to what we have just described for the wormhole, so we will be more brief here. The \emph{Ansatz} reads
\begin{equation}
\mathrm{d}s^2 = \frac{L^2}{(1-\tilde{y}^2)^2}\left\{\frac{4\,f(\tilde{y})\mathrm{d}\tilde{y}^2}{2-\tilde{y}^2}+\tilde{y}^2(2-\tilde{y}^2)\mathrm{d}\Omega_3^2\right\}\,,
\label{eq:disco}
\end{equation}
with the regular centre located at $\tilde{y}=0$ and the conformal boundary at $\tilde{y}=1$. Regularity at $\tilde{y}=0$ now demands that
\begin{equation}
f(y)=1+\mathcal{O}(\tilde{y}^2)\quad \text{and}\quad \Pi(y)=Q^{(0)}\tilde{y}[1+\mathcal{O}(\tilde{y}^2)]\,,
\end{equation}
with $Q^{(0)}$ constant. These regularity conditions suggest introducing a coordinate $\tilde{y}^2=y$ just as before, and they also suggest redefining
\begin{equation}
\Pi = \sqrt{y}\sqrt{2-y}(1-y)^{\Delta_+} q\,.
\end{equation}
At the conformal boundary $y=1$ we demand $\Pi(1)=q(1)=V$. For both $\mu=0$ and $\mu^2 L^2=-2$ this coincides with fixing the source for the expectation value dual to $\vec{\Pi}$.

Just as before, the Einstein equation and scalar field equation yield a single equation for $q$ which we can solve numerically given the boundary conditions above. The procedures to regulate the action and to identify the source are very similar to the ones used for the wormhole geometry, so we will suppress this discussion here and pass directly to results below.
\subsection{Results}
Let us define
\begin{equation}
\Delta S = 2 S^{D}-S^{W}\,,
\end{equation}
where $S^D$ is the Euclidean on-shell action for the disconnected solution and $S^W$ the on-shell action for the wormhole solution with the same values of the boundary sources $\vec{\Pi}^{+}$. Note that both of these actions are finite due to the counter-terms.

We first focus on the massless case $\mu=0$. As shown in  Fig.~\ref{fig:tran}, our numerical results indicate that $\Delta S$ crosses zero and becomes positive for $V>V^{\mathrm{HP}}\approx 3.70655(6)$. In addition, for any $V\geq V_{\min}\approx 3.38021(4)$, there are two wormholes for a given value of $V$. At $V_{\min}$ we find $y_0=y^{\min}_0\approx 2.13813(6)$. The solution with larger $\Delta S$ we will call the large wormhole, and the one with lower $\Delta S$ we call the small wormhole. These phases are depicted as blue disks (large wormhole) and orange squares (small wormhole), respectively, in Fig.~\ref{fig:tran}. For the small wormhole phase we find $\Delta S<0$ in all cases.
\begin{figure}[t!]
    \centering
    \includegraphics[scale=0.4]{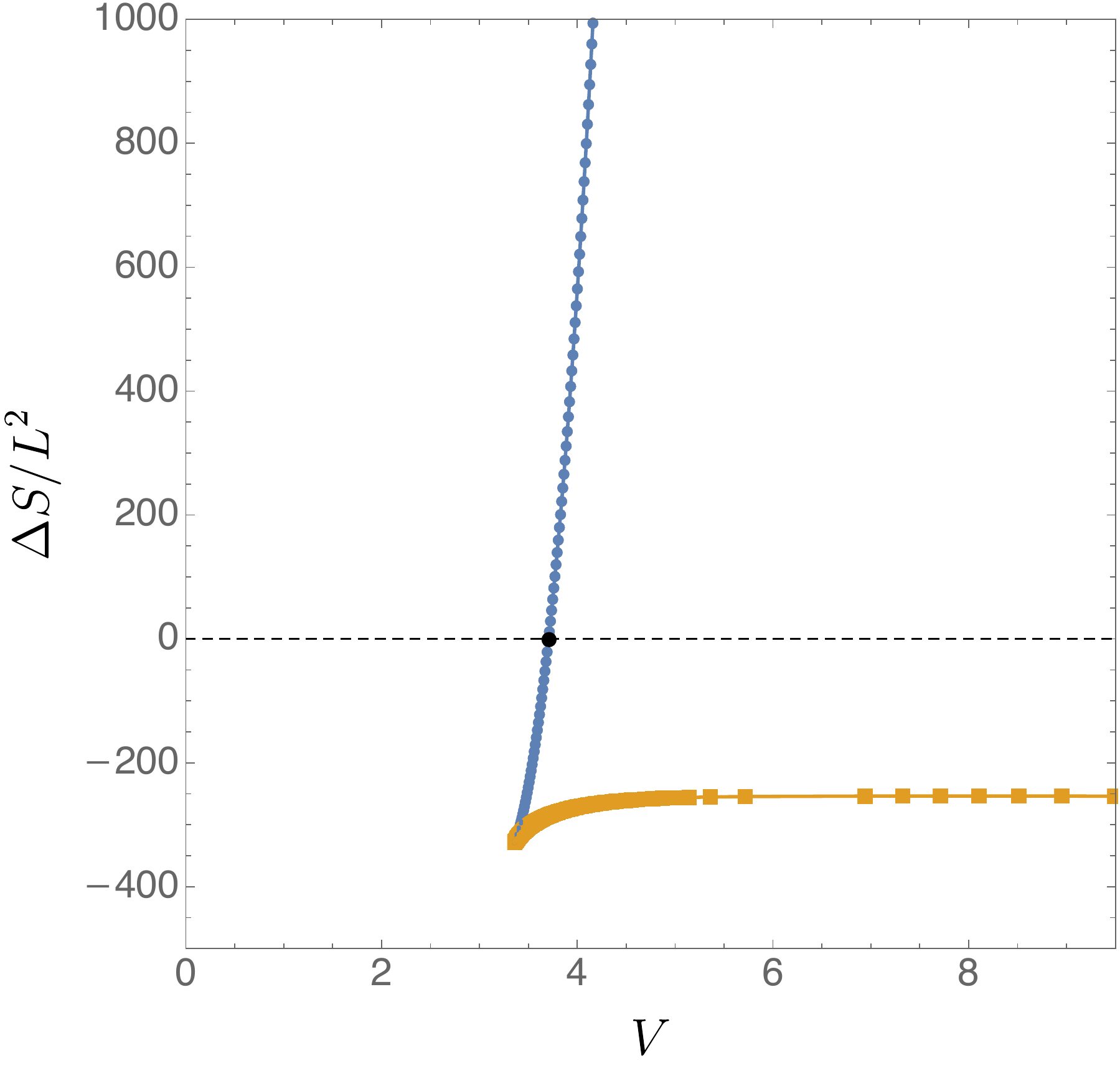}
    \caption{Action difference $\Delta S$ as a function of the source $V$ for $\mu=0$. There is a Hawking-Page transition around $V>V^{\mathrm{HP}}\approx 3.70655(6)$. The large wormholes are depicted as blue disks and the small wormholes as orange squares.}
    \label{fig:tran}
\end{figure}
To backup our nomenclature, we plot the minimum radius $y_0$ of the $S^3$  as a function of $V$ in Fig.~\ref{fig:radius} using the same color coding as in Fig.~\ref{fig:tran}.
\begin{figure}[t!]
    \centering
    \includegraphics[scale=0.4]{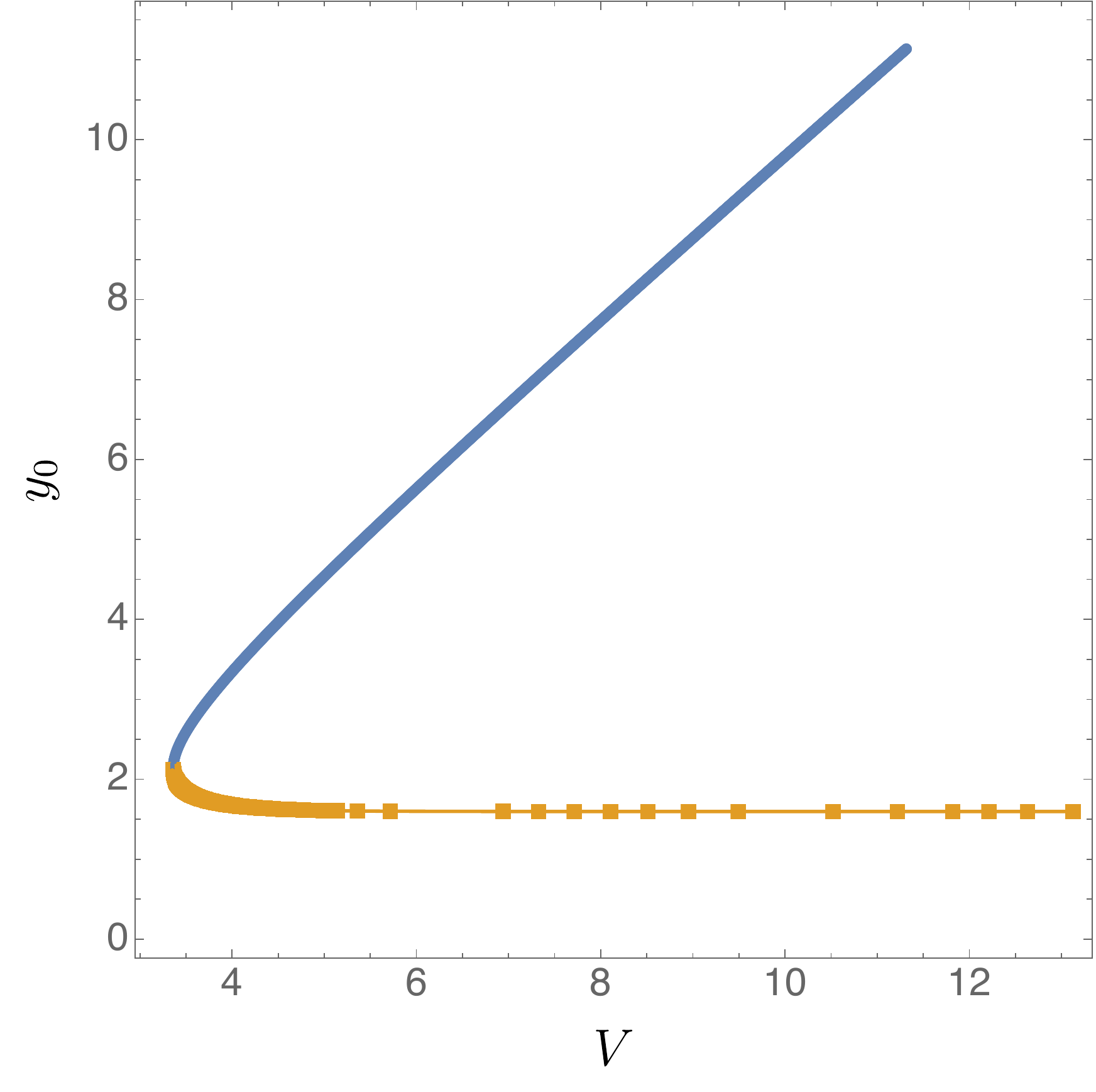}
    \caption{Minimum value of the $S^3$ radius as a function of $V$ for $\mu=0$. The blue disks represent the large wormholes and the orange disks the small wormholes. Wormholes only exist for $V>V_{\min}\approx 3.38021(4)$.}
    \label{fig:radius}
\end{figure}

Although the conformally coupled case is qualitatively similar, it turns out to be more challenging  numerically.  This is because in that case $V$ needs to be very large for the wormholes to exist, and even larger for the large wormholes to dominate (i.e., to see the Hawking-Page-like transition). We find the transition to occur at $S^D\sim 5\times 10^{9}\,L^2$ and $S^W\sim 10^{10}\,L^2$, while the difference is of order $\Delta S\sim L^2$. This means we have to accurately extract the first $15$ digits when evaluating the regulated integrals to determine the transition with good accuracy. At this point the use of high-precision arithmetics was essential. We used octuple precision throughout, keeping track of at least the first $256$ digits.

The resulting phase diagram is similar in structure to what we found in massless case. For $V\geq V_{\min}\approx 179.55054(5)$ we find two wormhole solutions for a given value of $V$. At this point, $y_0=y_0^{\min}\approx 0.8416(7)$. The Hawking-Page transition occurs on the large wormhole branch for $V=V^{\mathrm{HP}}\approx 639.20819(5)$; see Fig.~\ref{fig:tran_conformal}. The two  wormhole solutions for each $V$ can again be distinguished by the different values of $y_0$; see Fig.~\ref{fig:radius_conformal} where we used the same color coding as in Fig.~\ref{fig:tran}.
\begin{figure}
    \centering
    \includegraphics[scale=0.4]{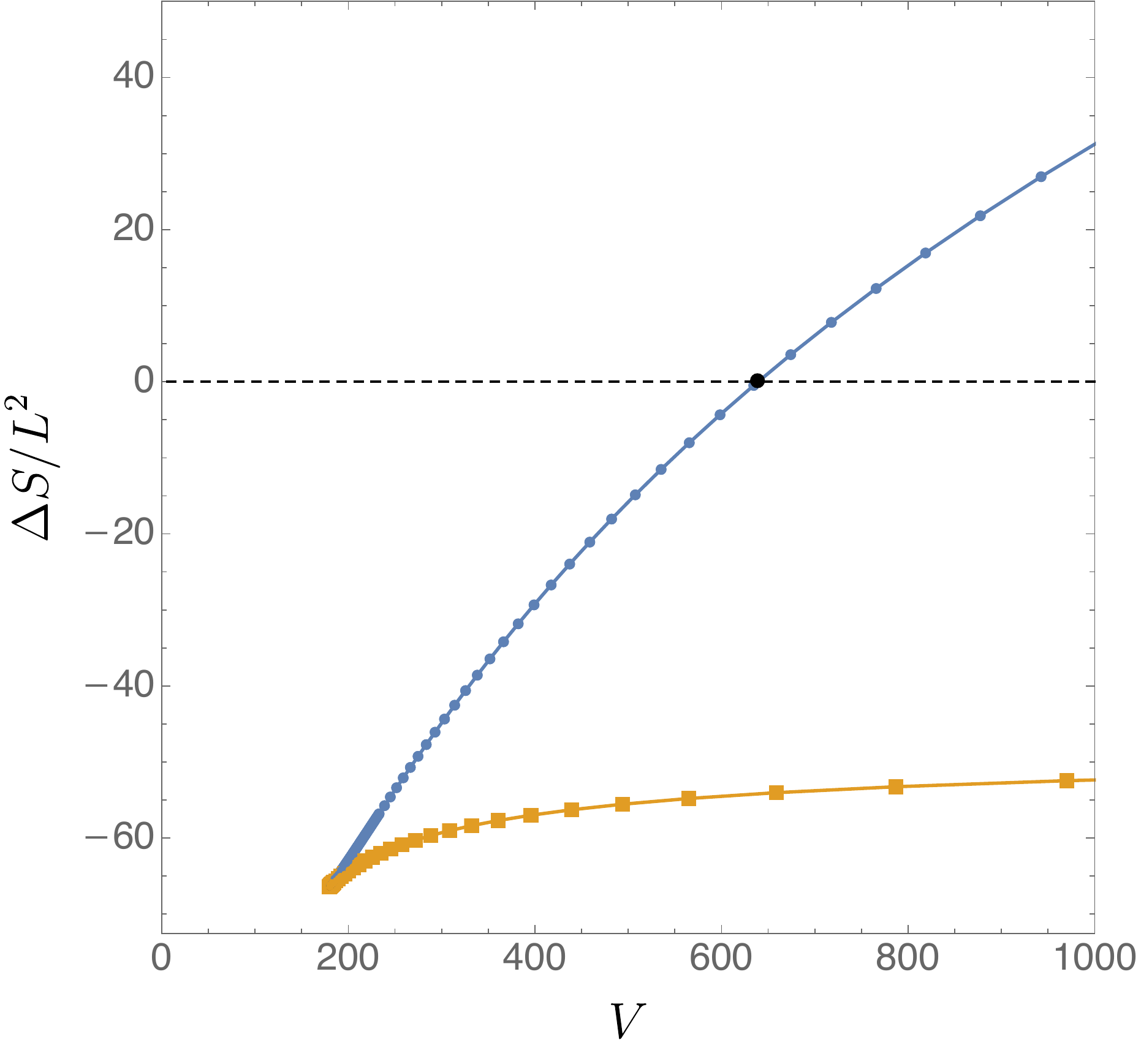}
    \caption{Action difference $\Delta S$ as a function of the source $V$ for $\mu^2 L^2=-2$. There is a Hawking-Page transition around $V>V^{\mathrm{HP}}\approx 639.20819(5)$. The large wormholes are depicted as blue disks and the small wormholes as orange squares.}
    \label{fig:tran_conformal}
\end{figure}
\begin{figure}
    \centering
    \includegraphics[scale=0.4]{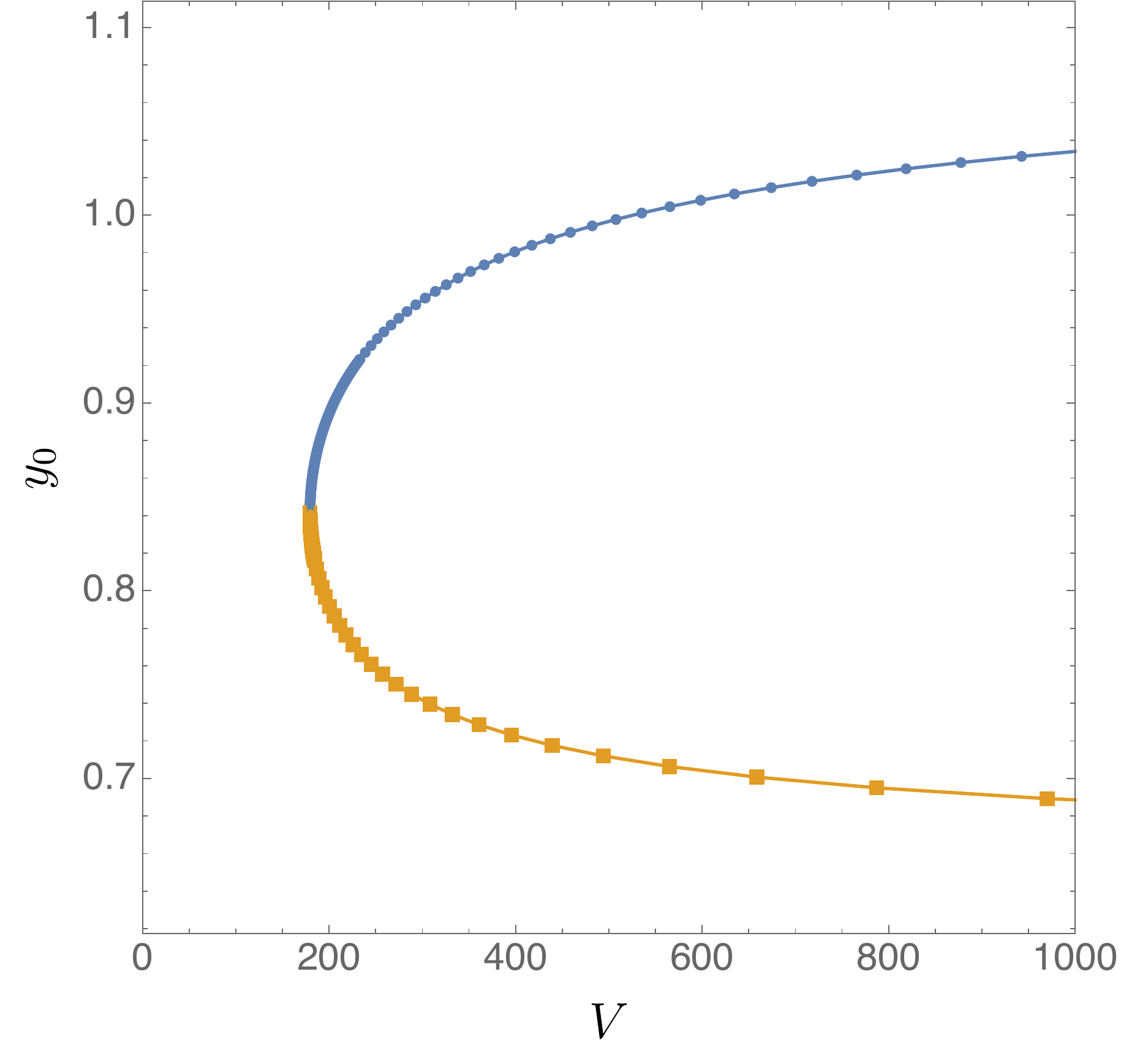}
    \caption{Minimum value of the $S^3$ radius as a function of $V$ for $\mu^2 L^2=-2$. The blue disks represent the large wormholes and the orange disks the small wormholes. Wormholes only exist for $V>V_{\min}\approx 179.55054(5)$.}
    \label{fig:radius_conformal}
\end{figure}
\subsection{\label{sec:negascalarheading}Negative Modes}
We now discuss perturbations around our scalar wormholes, and in particular, the potential existence of negative modes. Just as we did for the $U(1)^3-$Maxwell wormholes, we will take advantage of the $SO(4)$ symmetry of the $S^3$ to decompose the perturbations into the spherical harmonics (\ref{eqs:harmonics}).
\subsubsection{The scalar homogeneous mode : $\ell_S=0$}
The metric perturbations are given by
\begin{subequations}
\begin{equation}
\delta \mathrm{d}s^2 = \frac{L^2}{(1-y)^2}\left[\frac{\delta f(y)\mathrm{d}y^2}{(2-y)y}+\delta g(y)\mathrm{d}\Omega^2_3\right]\,,
\end{equation}
and
\begin{equation}
\delta \vec{\Pi} = \vec{X}\,\delta \Pi(y)\,,
\end{equation}
\end{subequations}
Throughout, we shall work with gauge invariant variables. The most general infinitesimal diffeomorphism compatible with $SO(4)$ takes the form $\xi= L^2\,\xi_y(y) \mathrm{d}y$. This in turn induces the following gauge transformations on the different metric and scalar perturbations
\begin{subequations}
\begin{align}
& \delta f=\left[2 (1-y) \left(2 y^2-4 y+1\right)-\frac{(1-y)^2 (2-y) y f^\prime}{f}\right]\,\xi_y+2 (2-y) y (1-y)^2 \xi _y^\prime\,,
\\
& \delta g = \frac{2 (1-y) (2-y) y}{f} \xi _y\,,
\\
& \delta \Pi = \frac{(1-y)^2 (2-y) y \Pi^\prime}{f} \xi _y\,.
\end{align}
\label{eq:transgaugescalar}
\end{subequations}
where ${}^\prime$ denotes differentiation with respect to $y$.

Our procedure will be similar to the one we used for the $U(1)^3-$Maxwell wormhole in section \ref{sec:u13}. We first evaluate the action (\ref{eq:actionscalargeneral}) to second order in $\{\delta f, \delta g,\delta \Pi\}$ and their derivatives. Let us denote this quadratic action by $S^{(2)}$. Second derivatives of $\delta g$ with respect to $y$ appear in the action, which we integrate by parts to write the action in first order form. This procedure generates a boundary term which cancels the perturbed Gibbons-Hawking-York boundary term. The resulting action $S^{(2)}$ is written in terms of $\{\delta f, \delta g,\delta \Pi\}$ and their first derivatives only. We then note that, after some integration by parts whose surface term vanishes or cancels with the boundary counterterms, $\delta f$ only enters the action algebraically. That is to say, no derivatives act on $\delta f$. This means we can formally perform the Gaussian integral over $\delta f$, and find a new action $\tilde{S}^{(2)}$ that depends only on $\{\delta g,\delta \Pi\}$ and their first derivatives.

At this stage we introduce gauge invariant quantities. Looking at the transformations (\ref{eq:transgaugescalar}), we introduce a gauge invariant variable $Q$ through the relation
\begin{equation}
\delta \Pi = Q+\frac{1}{2} (1-y) q^\prime\,\delta g\,.
\end{equation}
Substituting $\delta \Pi$ into $\tilde{S}^{(2)}$ yields an action for $Q$ and its first derivative $Q^\prime$. To reduce the action to this form, more integrations by parts have to be performed and, remarkably, every non-vanishing term at the wormhole boundaries cancel with contributions from the perturbed boundary counter-terms. It turns out that the action takes a simpler form if we further redefine
\begin{equation}
Q = \frac{\sqrt{3-(1-y)^2 \Pi '^2}}{\sqrt{6}}\tilde{Q}\,.
\label{eq:change}
\end{equation}
The second order action $\tilde{S}^{(2)}$ then takes the following rather explicit form
\begin{equation}
\tilde{S}^{(2)}=4 \pi^2 L^2\int_{0}^{+\infty } \mathrm{d}y \frac{\sqrt{f} y_0^3}{(1-y)^4 \sqrt{2-y} \sqrt{y}}\left[\frac{(1-y)^2 (2-y) y}{f}\tilde{Q}'^2 +V\,\tilde{Q}^2\right]\,,
\end{equation}
where
\begin{multline}
V(y) = \frac{1}{\Pi^2}\Bigg\{\frac{9 f \left[\Pi \left(1-\Pi^2\right) m_y^3 q^\prime+3 \left(m_y^2+y_0^2\right)\right]^2}{(2-y) y y_0^4 z_y^2}+\frac{3 (2-y) y}{f} \left[\left(1+m_y \,\Pi\,\Pi^\prime\right)^2+1\right]
\\
-\frac{8 \Pi^2 \left(1-q^2\right) m_y^4}{y_0^2 z_y} \left[\frac{3 \left(6-\Pi^2\right)}{8 \Pi \left(1-\Pi^2\right) m_y}+\Pi^\prime\right]^2-\frac{18 m_y\,\Pi \,\Pi^\prime}{z_y}+\frac{9 \left(6-\Pi^2\right)^2 m_y^2}{8 \left(1-\Pi^2\right) y_0^2 z_y}-\frac{45 \left(m_y^2+y_0^2\right)}{y_0^2 z_y}\Bigg\}\,,
\end{multline}
with
\begin{equation}
m_y = 1-y\quad\text{and}\quad z_y = 3-(1-y)^2 {\Pi^\prime}^2\,.
\end{equation}
The change of variable (\ref{eq:change}) only makes sense so long as the argument inside the square root is positive. One can show analytically that, so long as $y_0>y_0^{c}\approx 1.94712(2)$ for the massless case and $y_0>y_0^{c}\approx 0.77845(0)$ for the conformally coupled case, the argument inside the square root is indeed positive definite. In both cases $y^c_0<y_0^{\min}$, meaning that this transformation makes sense for all large wormholes, and for a range $y_0\in(y_0^c,y_0^{\min})$ of small wormholes. Recall that we want to establish that the large wormhole branch has no negative modes so, for our purposes, it suffices to study wormholes in the range $y_0>y_0^c$.

To search for negative modes we simply study the eigenvalue equation
\begin{equation}
-\frac{(1-y)^4 \sqrt{2-y} \sqrt{y}}{\sqrt{f}}\left[\frac{\sqrt{2-y} \sqrt{y}}{(1-y)^2 \sqrt{f}}\tilde{Q}^\prime\right]^\prime+V \tilde{Q}=\lambda \tilde{Q}.
\end{equation}
Note that in the $\tilde{y}$ coordinates the potential is an even function of $\tilde{y}$ as expected from the $Z_2$ symmetry. This means that we can study separately those eigenfunctions which are even and odd in $\tilde{y}$. In terms of the $y$ coordinates, this maps to functions that scale as $\tilde{Q}\sim \sqrt{y}(a_0+b_0 y)$ near $y=0$ for the odd case while for the even case we have $\tilde{Q}\sim \tilde{a}_0+\tilde{b}_0 y$ (where $a_0$, $b_0$, $\tilde{a}_0$ and $\tilde{b}_0$ are constants).

Before proceeding, we must specify the boundary conditions at infinity. A Frobenius analysis near the conformal boundary reveals two possible near boundary behaviors,
\begin{equation}
\tilde{Q}\sim (1-y)^{\frac{3}{2}\pm \sqrt{\frac{9}{4}+\mu^2 L^2-\lambda}}\,.
\end{equation}
The integrations by parts we have performed throughout our analysis yield boundary terms that vanish only consistent if we choose the $+$ root. This motivates performing one last change of variable,
\begin{equation}
\tilde{Q}= (1-y)^{\frac{3}{2}+ \sqrt{\frac{9}{4}+\mu^2 L^2-\lambda}}y^{\varepsilon/2}(2-y)^{\varepsilon/2}\tilde{q}\,,
\end{equation}
where we set $\varepsilon=0$ for the even sector and $\varepsilon=1$ for the odd sector of perturbations.

We find no odd negative modes, and a single even negative mode exists in the regime $y_0\in(y_0^c,y_0^{\min})$ but not for other values of $y_0$.  In particular we find no negative modes for the large wormhole branch. Perhaps even more interesting, the negative mode of the small wormhole branch disappears precisely at $y_0=y_0^{\min}$. This can be seen in Fig.~\ref{fig:negative} where we plot the negative eigenvalue $\lambda$ as a function of $y_0/y_{\min}$ for the massless (left panel) and conformally coupled scalars (right panel).
\begin{figure}
    \centering
    \includegraphics[scale=0.4]{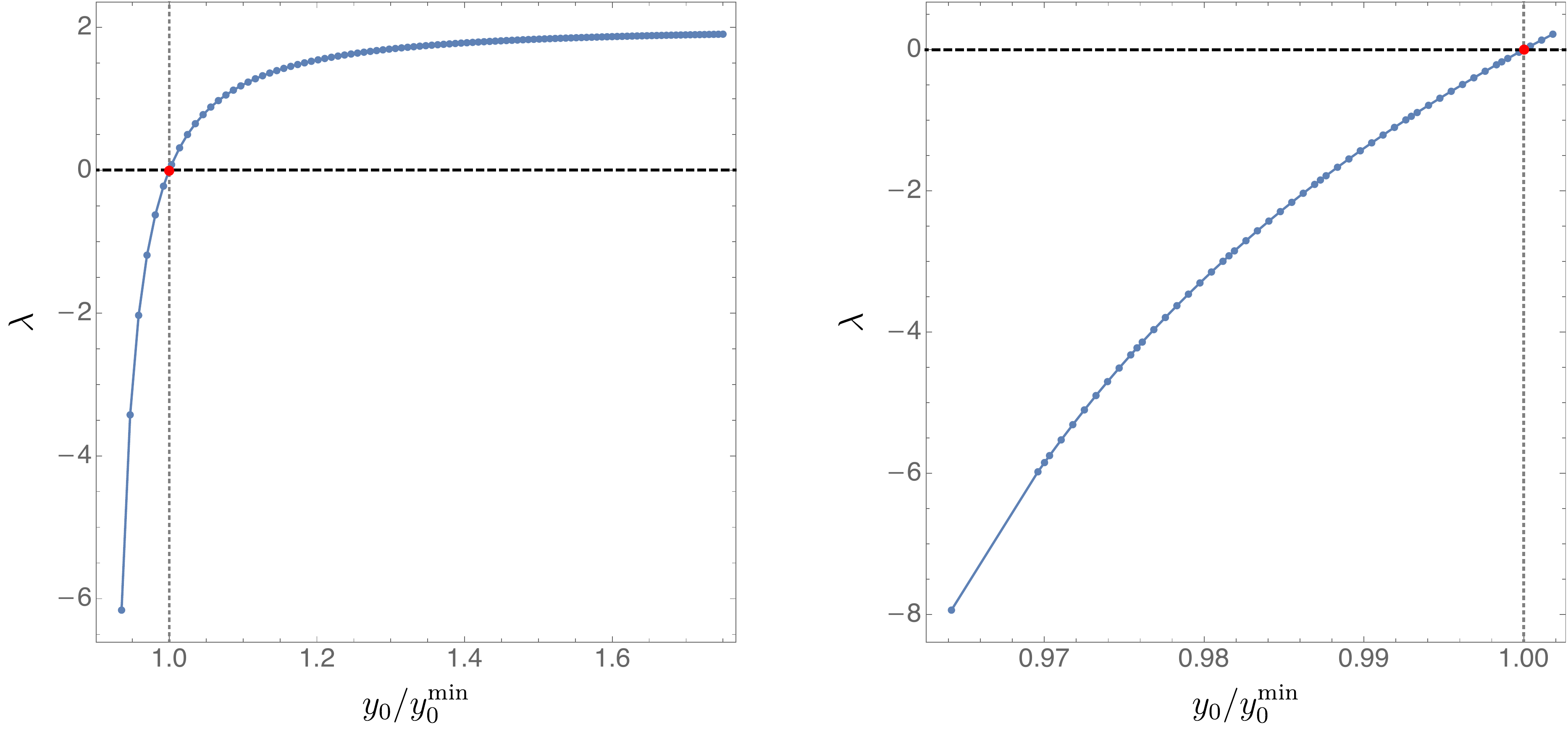}
    \caption{We find a negative mode on the small wormhole branch, which vanishes at $y_0=y_0^{\min}$ On the left panel we plot the negative mode for the massless case and on the right the negative mode of the conformally coupled case. In both cases, the horizontal axis is given by $y_0/y_0^{\min}$.}
    \label{fig:negative}
\end{figure}
\subsubsection{Scalar derived modes with $\ell_S\geq2$}
We now study scalar-derived perturbations in detail for $\ell_S\geq2$. This turns out to be a much easier task than studying perturbations of the $U(1)^3$ Maxwell theory with a spherical boundary metric. For the metric perturbations we take the same \emph{Ansatz} as in Eq.~(\ref{eq:grand})
\begin{subequations}
\begin{equation}
\delta \mathrm{d}s^2_{\ell_S} = h^{\ell_S}_{yy}(y)\,\mathbb{S}^{\ell_S}\,\mathrm{d}y^2+2 h^{\ell_S}_y(y)\,\Grad^i\mathbb{S}^{\ell_S}\,\mathrm{d}y\, \mathrm{d}y^{i}+H_T^{\ell_S}(y)\,\mathbb{S}^{\ell_S}_{ij}\,\mathrm{d}y^i \mathrm{d}y^j+H_L^{\ell_S}(y)\,\mathbb{S}^{\ell_S}\,\mathbbm{g}_{ij}\mathrm{d}y^i\mathrm{d}y^j,
\label{eq:grand1}
\end{equation}
while for the scalar perturbation we choose
\begin{equation}
\delta\vec{\Pi}_{\ell_S} = \vec{X}\,B_{\ell_S}(y)\,\mathbb{S}^{\ell_S}+(\Grad^i\mathbb{S}^{\ell_S}\Grad_i \vec{X})\,A_{\ell_S}(y)\,.
\end{equation}
\end{subequations}

Under an infinitesimal diffeomorphism of the form
\begin{equation}
\xi^{\ell_S} = \xi^{\ell_S}_y(y)\,\mathrm{d}y+L^{\ell_S}_y(y)\,\Grad_i \mathbb{S}^{\ell_S}\,\mathrm{d}y^i,
\end{equation}
the metric and scalar perturbations transform as
\begin{subequations}
\begin{align}
\delta h^{\ell_S}_{yy} & = \left(\frac{1}{y}-\frac{1}{2-y}-\frac{2}{1-y}-\frac{f^\prime}{f}\right)\xi^{\ell_S}_y+2 {\xi^{\ell_S}_y}^\prime,
\\
\delta h^{\ell_S}_{y} & = -\frac{2 L^{\ell_S}_y}{1-y}+\xi^{\ell_S}_y+{L^{\ell_S}_y}^\prime,
\\
\delta H_T^{\ell_S} &= 2 L^{\ell_S}_y,
\\
\delta H_L^{\ell_S} &= -\frac{2}{3} \lambda_S L^{\ell_S}_y+\frac{2\,(2-y)\,y\,y_0^2}{(1-y)f}\xi^{\ell_S}_y,
\\
\delta A^{\ell_S} & = \frac{(1-y)^2\,\Pi}{L^2 y_0^2}L^{\ell_S}_y,
\\
\delta B^{\ell_S}& = \frac{(2-y)(1-y)^2\,y\,\Pi^\prime}{L^2 f}\xi^{\ell_S}_y.
\end{align}
\end{subequations}
Our procedure will again be very similar to the one we used for the $U(1)^3$ Maxwell theory. We first expand the action (\ref{eq:actionscalargeneral}) to quadratic order in the perturbations, using  $S^{(2)}$ to denote the result. This quadratic action involves second derivatives of $H_T^{\ell_S}$ and $H_L^{\ell_S}$, which we readily remove via an integration by parts. The resulting boundary term cancels the perturbed Gibbons-Hawking-York boundary term. At this stage, $S^{(2)}$ is a function of $h_{yy}^{\ell_S}$, $h_{y}^{\ell_S}$, $H_T^{\ell_S}$, $H_L^{\ell_S}$, $A^{\ell_S}$, $B^{\ell_S}$ and their first derivatives with respect to $y$. However, after a few integration by parts, whose boundary terms partially cancel some of the counter-terms, we can write $S^{(2)}$ in a way where $h_{yy}^{\ell_S}$ only enters the action algebraically. As such, it is easy to perform the (correctly-signed) Gaussian integral over this variable and obtain a new action $\tilde{S}^{(2)}$ which is a function of  $h_{y}^{\ell_S}$, $H_T^{\ell_S}$, $H_L^{\ell_S}$, $A^{\ell_S}$, $B^{\ell_S}$ and their first derivatives. Upon a few more integration by parts, we can rewrite $\tilde{S}^{(2)}$  in a manner where $h_{y}^{\ell_S}$ again only enters algebraically. Since it also has the correct sign, we can again perform the Gaussian integral to finally obtain an action $\check{S}^{(2)}$ which is a function of $H_T^{\ell_S}$, $H_L^{\ell_S}$, $A^{\ell_S}$, $B^{\ell_S}$ and their first derivatives.

At this stage we introduce two gauge invariant quantities built using $H_T^{\ell_S}$, $H_L^{\ell_S}$, $A^{\ell_S}$, $B^{\ell_S}$. These are
\begin{subequations}
\begin{align}
Q_1^{\ell_S}&=A^{\ell_S}-\frac{(1-y)^2\Pi}{2L^2 y_0^2}H^{\ell_S}_T\,, \ \ \ \rm{and}
\\
Q_2^{\ell_S}&=B^{\ell_S}-\frac{(1-y)^3\Pi^\prime}{2L^2 y_0^2}\left(H^{\ell_S}_L+\frac{1}{3}H^{\ell_S}_T\right)\,.
\end{align}
\end{subequations}
We then solve for $A^{\ell_S}$ and $B^{\ell_S}$ in terms of $Q_1^{\ell_S}$, $Q_2^{\ell_S}$, $H^{\ell_S}_L$, $H^{\ell_S}_T$, and substitute the resulting expression into $\check{S}^{(2)}$. After some integration by parts (which generate boundary terms that again cancel some of the perturbed counter-terms) we find that all dependence on  $H^{\ell_S}_L$ and $H^{\ell_S}_T$ disappears (as it should due to gauge invariance). At this stage $\check{S}^{(2)}$ is a function of $Q_1^{\ell_S}$, $Q_2^{\ell_S}$ and their first derivatives only.

To proceed, we now treat the massless and conformal cases separately. Strictly speaking, the procedure we will apply to the conformally coupled case also works for the massless case, but as we shall see it is much more cumbersome so we will be more schematic there. For the massless case we take
\begin{align}
Q_1^{\ell_S}&\equiv \frac{(1+\lambda _S)^{1/4}}{8 \sqrt{2} \pi  \sqrt{\lambda _S}}\,\Pi\,q_1^{\ell_S}\,,\nonumber
\\
Q_2^{\ell_S}&\equiv \frac{(1+\lambda _S)^{1/4}}{8 \sqrt{2} \pi  \sqrt{\lambda _S}}(1-y)\Pi^\prime\,q_2^{\ell_S}\,.\label{eq:tra}
\end{align}
The reason why this change of variable is only adequate for the massless case is the presence of the multiplying factor $\Pi^\prime$ in the definition of $q_2^{\ell_S}$. It is a relatively simple exercise to show that $\Pi$ must have an extremum for $y\in(0,1)$ for any $\mu^2 L^2<0$. This means that this redefinition will necessarily include a singularity in those cases and is thus not appropriate to use. For the massless case, however, $\Pi$ is monotonic and as such $\Pi^\prime$ does not vanish in $y\in(0,1)$.

For the massless case, the resulting quadratic action takes the form
\begin{equation}
\check{S}^{(2)}=L^2\,\int_0^{+\infty}\mathrm{d}y\,\frac{\sqrt{f} y_0^3}{(1-y)^4 \sqrt{2-y} \sqrt{y}}\,\left[\frac{(1-y)^2 (2-y) y}{f} {q_I^{\ell_S}}^\prime\mathbb{K}^{IJ} {q_J^{\ell_S}}^\prime+q_I^{\ell_S}\mathbb{V}^{IJ}q_J^{\ell_S}\right]\,,
\end{equation}
where
\begin{equation}
\mathbb{K}^{-1}=\frac{1}{\lambda_S-3}\left[
\begin{array}{cc}
 3 \Pi ^2+\lambda _S-3 & \lambda _S \Pi ^2 \\
 \lambda _S \Pi ^2 & \frac{\lambda_S\Pi ^2 \left[\lambda _S-3+(1-y)^2 \Pi '^2\right]}{(1-y)^2 \Pi '^2} \\
\end{array}
\right]\,.
\end{equation}
Note that $\lambda_S = \ell_S(\ell_S+2)$, and we are focusing on $\ell_S\geq2$, so that $\lambda_S\geq 8$. From the above expression it is clear that $\mathrm{Tr}(\mathbb{K}^{-1})>0$. Furthermore, we have
\begin{equation}
\mathrm{det}(\mathbb{K}^{-1})=\frac{\lambda _S \Pi ^2}{(1-y)^2 \left(\lambda _S-3\right) \Pi '^2} \left[\lambda _S-3+3 \Pi ^2+(1-y)^2 \left(1-\Pi^2\right) \Pi '^2\right]\,.
\end{equation}
It turns out that $\mathrm{det}(\mathbb{K}^{-1})$ is not positive definite for all values of $y_0$, but one can check numerically that for $\ell_S=2$ it is positive so long as $y_0\gtrsim 1.6859(2)<y_0^{\min}$. We thus conclude that $\mathrm{det}(\mathbb{K}^{-1})$ is positive for all large wormholes, and thus that $\mathbb{K}$ is positive definite for all large wormholes. Note that for larger values of $\ell_S$, these values will become even smaller.

We now turn out attention to $\mathbb{V}$, which has a rather complicated expression. However, it turns out that the combination
\begin{subequations}
\begin{equation}
\mathbb{V}_{IJ}=(\mathbb{K}^{-1})_{IK}(\mathbb{K}^{-1})_{JL}\mathbb{V}^{KL}
\end{equation}
has a more manageable form, namely
\begin{equation}
\mathbb{V}_{11}=\frac{m_y^2}{y_0^2 \left(\lambda _S-3\right) \Pi^2} \left(\lambda _S-3+3 \Pi^2\right) \left(\lambda _S-4+4 \Pi^2\right),
\end{equation}
\begin{equation}
\mathbb{V}_{12}=-\frac{m_y}{y_0^2 (\lambda_S-3) \Pi\,\Pi '} \lambda _S \left[2 \left(\lambda _S-3\right)-m_y \Pi  \left(\lambda _S-4+4 \Pi ^2\right) \Pi '\right],
\end{equation}
\begin{multline}
\mathbb{V}_{22}=\frac{\lambda _S}{(2-y) m_y y y_0^4 \left(\lambda _S-3\right) \Pi '^3} \Big\{(2-y) m_y y y_0^2 \lambda _S \left(\lambda _S-4+4 \Pi ^2\right) \Pi '+6 y_0^2 \left(\lambda _S-3\right) \Pi  f
\\
+m_y f \left[3 y_0^2 \left(\lambda _S-2\right)+m_y^2 \lambda _S-\left(6 y_0^2\lambda _S-6 y_0^2+5 m_y^2 \lambda _S\right) \Pi ^2+4 m_y^2 \lambda _S \Pi ^4\right] \Pi '\Big\},
\end{multline}
\end{subequations}%
where one should recall that $m_y=1-y$.

It is now a simple exercise to compute the two real eigenvalues $\lambda_-\leq \lambda_+$ of $\mathbb{V}$ as a function of $y$. We then take the minimum value of $\lambda_-$ in the interval $y\in(0,1)$, and plot it as a function of $y_0$. It turns out a critical value of $y_0$ exists above which $\lambda_-$ is positive definite for all $\lambda_S\geq2$. For the massless case this occurs for $y_0\gtrsim1.7435(6)<y_0^{\min}$ .This establishes that no negative modes exist in the scalar sector with $\ell_S\geq2$ for large wormholes generated by massless scalar sources.

The conformally coupled case is more complicated because we cannot apply (\ref{eq:tra}). Instead, we have to use a procedure more similar to the one we used for  the $U(1)^3$ theory. Here we consider a generic value of $\mu^2 L^2<0$ parametrized by the conformal dimension $\Delta$ satisfying
\begin{equation}
\mu^2 L^2 = \Delta(\Delta-3)\,.
\end{equation}
Instead of (\ref{eq:tra}) we consider
\begin{align}
Q_1^{\ell_S}&\equiv \frac{(1+\lambda _S)^{1/4}}{8 \sqrt{2} \pi  \sqrt{\lambda _S}}\,(1-y)^\Delta\,\left[q_1^{\ell_S}+\frac{(1-y) \Pi \,\Pi '}{\lambda _S-3+(1-y)^2 \Pi'^2}q_2^{\ell_S}\right]\,,\nonumber
\\
Q_2^{\ell_S}&\equiv \frac{(1+\lambda _S)^{1/4}}{8 \sqrt{2} \pi  \sqrt{\lambda _S}}(1-y)^\Delta\,q_2^{\ell_S}\,.\label{eq:tra2}
\end{align}
The second order action now reads
\begin{equation}
\check{S}^{(2)}= L^2\,\int_0^{+\infty}\mathrm{d}y\,\frac{\sqrt{f} y_0^3}{(1-y)^4 \sqrt{2-y} \sqrt{y}}\,\left[\frac{(1-y)^2 (2-y) y}{f} (\mathcal{D}q^{\ell_S})_I\mathbb{K}^{IJ} (\mathcal{D}q^{\ell_S})_J+q_I^{\ell_S}\mathbb{V}^{IJ}q_J^{\ell_S}\right]\,,
\end{equation}
with
\begin{equation}
\mathbb{K}^{-1}=\frac{1}{(1-y)^{2\Delta}}\left[
\begin{array}{cc}
1+\frac{\Pi ^2 \left[3-(1-y)^2 \Pi '^2\right]}{\lambda _S-3+(1-y)^2 \Pi'^2}& 0
\\
0 & \lambda _S \left[1+\frac{(1-y)^2 \Pi '^2}{\lambda _S-3}\right]\,
\end{array}
\right].
\end{equation}
This result is positive so long as $y_0\gtrsim0.694251<y_0^{\min}$, thus again implying that it is positive definite on the large wormhole branch. Furthermore,
\begin{equation}
(\mathcal{D}q^{\ell_S})_I  = {q_I^{\ell_S}}^\prime+\varepsilon_I^{\phantom{I}J}q_J^{\ell_S},
\end{equation}
where $\varepsilon_I^{\phantom{I}J}=\varepsilon_{IK}\mathbb{K}^{KJ}$ and
\begin{multline}
\varepsilon_{IK}=\left\{\left(3 m_y^2+L^2 y_0^2 \mu ^2\right) \Pi -m_y \left[2 m_y^2+3 y_0^2-\left(2 m_y^2+L^2 y_0^2 \mu ^2\right) \Pi ^2\right]\Pi '\right\}
   \\
   \times \frac{m_y^{-1-2 \Delta } \lambda _S \Pi  f }{y_0^2 y (2-y)\left(\lambda _S-3+m_y^2 \Pi '^2\right)}\left[\begin{array}{cc}0 & 1 \\ -1 & 0\end{array}\right]\,.
\end{multline}
The matrix $\mathbb{V}$ turns out to be positive definite so long as $y_0\gtrsim0.75186(7)<y_0^{\min}$, thus rendering the large wormholes branch stable. The expression for $\mathbb{V}$ can be found in appendix \ref{eq:labc}. The second formalism described above can also be used to study the massless case, but the expressions are considerably more complicated than in the approach described for $m=0$ above.
\subsubsection{Scalar-derived perturbations with $\ell_S=1$}
As we have seen, scalar modes with $\ell_S=1$ are excluded from the previous analysis due to the fact that $\mathbb{S}^{1}_{ij}=0$ for this special mode. Our metric perturbation becomes simpler in this case where it takes the form
\begin{subequations}
\begin{equation}
\delta \mathrm{d}s^2_{\ell_S=1} = h^{1}_{yy}(y)\,\mathbb{S}^{1}\,\mathrm{d}y^2+2 h^{1}_y(y)\,\Grad^i\mathbb{S}^{1}\,\mathrm{d}y\, \mathrm{d}y^{i}+H_L^{1}(y)\,\mathbb{S}^{1}\,\mathbbm{g}_{ij}\mathrm{d}y^i\mathrm{d}y^j,
\end{equation}
while the scalar perturbation is essentially unchanged and yields
\begin{equation}
\delta\vec{\Pi} = \vec{X}\,B_{1}(y)\,\mathbb{S}^{1}+(\Grad^i\mathbb{S}^{1}\Grad_i \vec{X})\,A_{1}(y)\,.
\end{equation}
\end{subequations}%
Since scalar derived infinitesimal diffeomorphisms still have two degrees of freedom,
\begin{equation}
\xi^{\ell_S=1}= \xi^{1}_y(y)\,\mathrm{d}y+L^{1}_y(y)\,\Grad_i \mathbb{S}^{1}\,\mathrm{d}y^i\,,
\end{equation}
we only expect a single gauge-invariant master function. We shall see that this is indeed the case. Under such diffeomorphisms the metric and scalar perturbation functions transform as
\begin{subequations}
\begin{align}
\delta h^{1}_{yy} & = \left(\frac{1}{y}-\frac{1}{2-y}-\frac{2}{1-y}-\frac{f^\prime}{f}\right)\xi^{1}_y+2 {\xi^{1}_y}^\prime,
\\
\delta h^{1}_{y} & = -\frac{2 L^{\ell_S}_y}{1-y}+\xi^1_y+{L^{1}_y}^\prime,
\\
\delta H_L^{1} &= -2 L^{1}_y+\frac{2\,(2-y)\,y\,y_0^2}{(1-y)f}\xi^{1}_y,
\\
\delta A^{1} & = \frac{(1-y)^2\,\Pi}{L^2 y_0^2}L^{1}_y,
\\
\delta B^{1}& = \frac{(2-y)(1-y)^2\,y\,\Pi^\prime}{L^2 f}\xi^{1}_y.
\end{align}
\end{subequations}
Just as for the case with $\ell_S\geq2$, both $h_{yy}^{\ell_S}$ and $h_y^{\ell_S}$ can be integrated out, leaving an action which depends only on $ H_L^{1}$, $A_L^{1}$ and $B_L^{1}$. At this stage we introduce a gauge invariant variable
\begin{equation}
Q^1=\Pi\,B^1-(1-y)\Pi^\prime A^1-\frac{(1-y)^3 \Pi\,\Pi^\prime}{2 L^2 y_0^2}H_L^1\,.
\end{equation}
The resulting quadratic action $\check{S}^{(2)}$ can be written entirely in terms of $Q^1$ and its first derivatives and takes the form
\begin{equation}
\frac{\check{S}^{(2)}}{12 \pi^2}=\int_0^{+\infty}\mathrm{d}y \frac{\sqrt{f} y_0^3}{(1-y)^4 \sqrt{2-y} \sqrt{y}}\,\left[\frac{(2-y) y}{f} \frac{m_y^2\Pi^4}{3 \Pi ^2+m_y^2 \left(1-\Pi ^2\right) \Pi '^2}{{Q^1}'}^2+V^1\,{Q^1}^2\right]\,,
\end{equation}
with
\begin{multline}
V^1=\frac{\Pi ^2}{\left[3 \Pi ^2+m_y^2 \left(1-\Pi ^2\right)\Pi'^2\right]^2 y_0^2} \Bigg\{36 \left(m_y^2+y_0^2\right)-\Pi  \Big[36 \left(m_y^2+y_0^2\right) \Pi -9 m_y^2 \Pi ^3
\\
-6 m_y \left(7 m_y^2+6 y_0^2-m_y^2 \Pi ^2\right) \Pi '-3 m_y^2 \Pi  \left(5 m_y^2+2 y_0^2-3 m_y^2 \Pi^2\right)\Pi'^2-2 m_y^5 \left(1-\Pi ^2\right)^2\Pi'^3\Big]
\\
-\frac{2 (2-y) y y_0^2}{f} \Big[9-\Pi  (9 \Pi -18 m_y \Pi '-12 m_y^2 \Pi \Pi'^2+3 m_y^3 \Pi ^2\Pi'^3+3 m_y^4 \Pi \Pi'^4+2 m_y^5 \Pi'^5-m_y^5 \Pi ^2\Pi'^5)\Big]
\\
-\frac{6 \left(m_y^2+y_0^2\right) f}{(2-y) y y_0^2} \left(1-\Pi ^2\right) \left[3 \left(m_y^2+y_0^2\right)+m_y^3 \Pi \left(1-\Pi ^2\right) \Pi '\right]\Bigg\}
\end{multline}
Once more, one can easily verify that $V^1$ and $3 \Pi ^2+m_y^2 \left(1-\Pi ^2\right) \Pi '^2$ are positive definite along large wormhole branch. This thus establishes that $\ell_S=1$ scalar derived perturbations yield no negative modes on the large wormhole branch.
\subsubsection{Vector-derived perturbations with $\ell_V\geq2$}
This sector of perturbations turns out to be much easier to study than in the $U(1)^3-$Maxwell theory. This is because in the current case the vector-derived perturbations do not source scalar-derived perturbations. Our \emph{Ansatz} for the metric perturbations reads
\begin{subequations}
\begin{equation}
\delta \mathrm{d}s^2_{\ell_V} = 2 h^{\ell_V}_y(y)\,\mathbb{S}_i^{\ell_V}\,\mathrm{d}y\, \mathrm{d}y^{i}+H_T^{\ell_V}(y)\,\mathbb{S}^{\ell_V}_{ij}\,\mathrm{d}y^i \mathrm{d}y^j,
\end{equation}
while for the scalar perturbation we choose
\begin{equation}
\delta\vec{\Pi}_{\ell_V} = (\mathbb{S}_i^{\ell_V}\Grad^i \vec{X})\,A_{\ell_V}(y)\,.
\end{equation}
\end{subequations}

The most general vector-derived infinitesimal diffeomorphism can be written
\begin{equation}
\xi^{\ell_V} = L_y^{\ell_V} \mathbb{S}^{\ell_V}_i \mathrm{d}y^i\,.
\end{equation}
This infinitesimal diffeomorphism induces the gauge transformation
\begin{subequations}
\begin{align}
\delta h_y^{\ell_V} &=-\frac{2 L_y^{\ell_V}}{1-y}+{L_y^{\ell_V}}'\,,
\\
\delta H_T^{\ell_V} &=L_y^{\ell_V}\,,
\\
\delta A^{\ell_V}& =\frac{(1-y)^2\Pi}{L^2 y_0^2}L_y^{\ell_V}\,.
\end{align}
\end{subequations}
The procedure now is very similar to what we have have seen for the scalar-derived perturbations. We first derive the second order action $S^{(2)}$ and note that $H_T^{\ell_V}$ appears in the action with terms that involve second derivatives with respect to $y$. We integrate these by parts, with the non-vanishing boundary terms cancelling the perturbed Gibbons-Hawking-York term. Furthermore, after some further integration by parts, $h_y^{\ell_V}$ only enters the action algebraically. This means we can perform the Gaussian path integral and find a new second order action $\check{S}^{(2)}$ which depends on $H_T^{\ell_V} $, $A^{\ell_V}$ and their first derivatives with respect to $y$. At this point, we introduce the gauge invariant variable
\begin{equation}
Q^{\ell_V} = A^{\ell_V}-\frac{(1-y)^2 \Pi}{L^2 y_0^2}H_T^{\ell_V}\,.
\end{equation}
Solving this expression with respect to $A^{\ell_V}$ and substituting it back into  $\check{S}^{(2)}$ gives an action for $Q^{\ell_V}$ and its first derivative with respect to $y$. The dependence in $H_T^{\ell_V}$ completely drops out from the calculations, as it should due to diffeomorphism invariance. For convenience we further define
\begin{equation}
Q^{\ell_V}=\frac{(\lambda _V+2)^{1/4}}{2 \sqrt{2}  \sqrt{\lambda _V-2}}\,\Pi\,\tilde{Q}_{\ell_V}\,.
\end{equation}
The second order action for $\tilde{Q}^{\ell_V}$ reads
\begin{equation}
\check{S}^{(2)}=4\pi^2 L^2 \int_0^{\infty} \mathrm{d}y\,\frac{y_0^3 \sqrt{f} \Pi ^2}{\sqrt{2-y} m_y^2 \sqrt{y}}\left[\frac{(2-y) y}{\left(\lambda _V-2+4 \Pi ^2\right) f}{\tilde{Q}_{\ell_V}^\prime}{}^2+\frac{ \tilde{Q}_{\ell_V}^2}{y_0^2}\right]\,.
\end{equation}
Since the above quadratic action is manifestly positive for any value of $y_0$, there are no negative modes in the vector-derived sector with $\ell_V\geq2$.
\subsubsection{The vector $\ell_V=1$ mode}
The situation here is completely analogous to that of  the $U(1)^3$ Maxwell theory case. This mode turns out to give a zero-mode when written in gauge-invariant variables. So since the theory admits a symmetric hyperbolic formulation, it must be the linearization of a pure-gauge mode and can be ignored.
\subsubsection{The tensor modes with $\ell_T\geq2$}
Finally, we come to the easiest sector, which is the one defined by tensor-derived perturbations. The reason why this sector is easiest is twofold: the scalar perturbations are zero in this sector, and the metric perturbation reduces to a single gauge-invariant variable. Unlike in the $U(1)^3$ case, the tensors to not excite vectors or scalar derived perturbations. The metric perturbation takes the rather simple form
\begin{equation}
\delta \mathrm{d} s^2_{\ell_T}= \frac{2\sqrt{2} y_0^2}{\sqrt{1+\ell_T}}\frac{L^2}{(1-y)^2}\,H_{\ell_T}(y)\mathbb{S}^{\ell_T}_{ij}\,,
\end{equation}
where the factors multiplying $H^{\ell_V}_T$ are only there for later convenience in the presentation of the second order action. The second order action reads
\begin{equation}
\check{S}^{(2)} = 4\pi^2 L^2 \int_0^{\infty} \mathrm{d}y\,\frac{y_0^2 \sqrt{f}}{(1-y)^2 \sqrt{2-y} \sqrt{y}}\left[\frac{(2-y)  y}{f} {H_{\ell_T}^\prime}{}^2+\frac{1}{y_0^2}\left(\lambda _T+2+4 \Pi ^2\right)H_{\ell_T}^2\right]\,,
\end{equation}
which is manifestly positive for all wormhole solutions.
\section{Einstein-$U(1)^2$ wormholes in 11-dimensional supergravity}
\label{sec:11dEU12}

Having explored wormholes in simple but ad hoc low-energy theories in sections \ref{sec:u13} and \ref{sec:scalars}, we now wish to understand the behavior of at-first-sight similar wormholes in various UV-complete theories.  This will be explored in the next few sections, and will in particular raise the important issue of possible brane instabilities.

We begin in the current section by describing a truncation of 11-dimensional supergravity that leads to a four-dimensional action similar to that studied in section \ref{sec:u13}, but with only two Maxwell fields instead of three.  In later sections we will also consider an example in a mass-deformed ABJM setup and a type IIB compactification that leads to a theory of scalar fields in AdS$_3$.

To begin our discussion, recall that bosonic fields of 11-dimensional supergravity are just a metric ${}^{(11)}g$ and a four-form $F_{(4)} = \mathrm{d}A_{(3)}$. The Euclidean action reads
\begin{equation}
S= -\int\left({}^{(11)}R \star_{11} 1-\frac{1}{2}F_{(4)}\wedge \star_{11} F_{(4)}-\frac{i}{6}F_{(4)}\wedge F_{(4)} \wedge A_{(3)}\right)\,,
\label{eq:action11Dasa}
\end{equation}
with ${}^{(11)}R\equiv {}^{(11)}R_{AB}G^{AB}$ being the $11-$dimensional Ricci scalar associated with ${}^{(11)}g$ and ${}^{(11)}R_{AB}$ its Ricci tensor. The equations derived from (\ref{eq:action11Dasa}) are
\begin{subequations}
\begin{align}
& {}^{(11)}R_{AB}-\frac{1}{2}{}^{(11)}g_{AB} {}^{(11)}R= \frac{1}{12}\left[F_{(4)ACDE}F_{(4)\;b}^{\phantom{(4)B}CDE}-\frac{1}{8}\;{}^{(11)}g_{AB}\;F_{(4)CDEF}F_{(4)}^{\phantom{(4)}CDEF}\right]\,,
\\
& \mathrm{d}\star F_{(4)} = \frac{i}{2}F_{(4)}\wedge F_{(4)}\,.
\end{align}%
\label{eq:gen11D}%
\end{subequations}%
Here upper case Latin indices are eleven-dimensional and $\star_{11}$ is the eleven-dimensional Hodge dual operation.
\subsection{Einstein-$U(1)^2-$theory}
We consider an \emph{Ansatz} where the eleven-dimensional fields take the form
\begin{subequations}
\begin{multline}
\mathrm{d}s^2_{11d} \equiv {}^{(11)}g_{AB}\mathrm{d}X^A \mathrm{d}X^B= g_{ab}\;\mathrm{d}x^a\mathrm{d}x^b+
\\
\frac{1}{g^2}\Bigg\{\mathrm{d}\xi^2+\frac{\cos^2\xi}{4}\;\left[\mathrm{d}\theta_1^2+\sin^2\theta_1\,\mathrm{d}\phi_1^2+\left(\mathrm{d}\psi_1+\cos \theta_1\,\mathrm{d}\phi_1-2 g A^{(1)}\right)^2\right]
\\
+\frac{\sin^2\xi}{4}\;\left[\mathrm{d}\theta_2^2+\sin^2\theta_2\,\mathrm{d}\phi_2^2+\left(\mathrm{d}\psi_2+\cos \theta_2\,\mathrm{d}\phi_2-2 g A^{(2)}\right)^2\right]\Bigg\}\,,
\end{multline}
with $\xi\in(0,\pi/2)$, $\theta_i\in(0,\pi)$, $\phi_i\in(0,2\pi)$ and $\psi_i\in(0,4\pi)$, and
\begin{equation}
F_{(4)}=-6\,g\,i\,\mathrm{Vol}_4+i\,\tilde{F}_{(4)}
\end{equation}
with
\begin{multline}
\tilde{F}_{(4)}=\frac{\cos \xi}{2g^2}\left[\sin \xi\;\mathrm{d}\xi \wedge (\mathrm{d}\psi_1+\cos \theta_1\,\mathrm{d}\phi_1-2 g A^{(1)})+\frac{1}{2}\cos \xi\,\sin \theta_1\,\mathrm{d}\theta_1\wedge \mathrm{d}\phi_1\right]\wedge \star F^{(1)}
\\
-\frac{\sin \xi}{2g^2}\left[\cos \xi\;\mathrm{d}\xi \wedge (\mathrm{d}\psi_2+\cos \theta_2\, \mathrm{d}\phi_1-2 g A^{(2)})-\frac{1}{2}\sin \xi\,\sin \theta_2\,\mathrm{d}\theta_2\wedge \mathrm{d}\phi_2\right]\wedge \star F^{(2)}\,,
\end{multline}%
\label{eqs:11D4D}%
\end{subequations}%
where $\star$ is the Hodge dual with respect to the four-dimensional metric $g$, $F^{(I)}=\mathrm{d}A^{(I)}$ with $I=1,2$ and $\mathrm{Vol}_4$ the volume form of $g$. Here, lower case Latin indices are four-dimensional. We also restrict to configurations where
\begin{equation}
F^{(1)}\wedge \star F^{(1)}=F^{(2)}\wedge \star F^{(2)}\,.
\end{equation}

Inserting the \emph{Ans\"atze} for the eleven-dimensional metric ${}^{(11)}g$ and four-form field $F_{(4)}$ into the equations of motion (\ref{eq:gen11D}) induces a set of equations for the four-dimensional metric $g$ and gauge fields $F^{(I)}$ which can be derived from the following somehow familiar four-dimensional action
\begin{equation}
S = -\int_{\mathrm{\mathcal{M}}} \mathrm{d}^4 x\sqrt{g}\left(R+\frac{6}{L^2}-\sum_{I=1}^2 F^{(I)}_{ab}F^{(I)\;ab}\right)-2\int_{\partial \mathcal{M}} \mathrm{d}^3 x\sqrt{h}\;K+S_{\mathcal{B}}\,,
\label{eq:simpleu12}
\end{equation}
where the boundary terms are exactly as in Eq.~(\ref{eq:simpleu13}) and $2\,g=L$. This looks remarkably similar to the Einstein-$U(1)^3$ theory, but with only two Maxwell fields. Any solution of the equations of motion induced by (\ref{eq:simpleu12}) can be uplifted to a solution of eleven-dimensional supergravity via Eqs.~(\ref{eqs:11D4D}). This truncation is a sub-truncation of a more general truncation that (to our knowledge) first appeared in \cite{Cvetic:1999au}. We will later consider in section \ref{sec:massdeformed} yet another sub-truncation of \cite{Cvetic:1999au}.

We are interested in finding solutions whose boundary metric is a round $S^3$. However, we are going to relax the assumption that the $SO(4)$ symmetry is preserved in the bulk. In fact, in the bulk, we will only require our geometry to enjoy $U(2)$ symmetry. The best way to visualise how we are going to do this in a simple manner is to again introduce the left-invariant one forms $\hat{\sigma}_i$ described in section \ref{sec:u13}  and to consider a line element of the form
\begin{equation}
\mathrm{d}s^2 = \frac{\mathrm{d}r^2}{f(r)}+\frac{g(r)}{4}\,\left[h(r)\,\hat{\sigma}_3^2+\hat{\sigma}_1^2+\hat{\sigma}_2^2\right]\,,
\end{equation}
where the $U(2)=U(1)\times SU(2)$ symmetry is manifest (with the $U(1)$ parametrising the angle that rotates $\hat{\sigma}_1$ into $\hat{\sigma}_2$). The functions $f$, $g$ and $h$ are functions of $r$ only. For the gauge fields we take
\begin{equation}
A^{(1)} = \frac{L}{2}\,\Phi(r)\,\hat{\sigma}_1\quad\text{and}\quad A^{(2)} = \frac{L}{2}\,\Phi(r)\,\hat{\sigma}_2\,.
\end{equation}

We want to construct solutions where the dual operator to $A^{(I)}$ has a non-vanishing source. This is obtained by searching for solutions for which
\begin{equation}
\lim_{r\to+\infty}\Phi = \Phi_0\,.
\end{equation}
The objective of the sections below is to construct the phase diagram of the wormhole and disconnected solutions as a function of $\Phi_0$.
\subsection{The disconnected phase}
The disconnected phase is easy to find analytically. Just as for the Einstein-$U(1)^3$ theory, it satisfies
\begin{equation}
F^{(I)}=\pm \star F^{(I)}\,,
\end{equation}
with the lower sign yielding a singular solution. We thus take the upper sign .

The stress energy tensor is then identically zero, and $g$ is just the usual metric on Euclidean AdS with a boundary $S^3$ for which
\begin{equation}
g(r)=r^2\,,\quad f(r)=\frac{r^2}{L^2}+1\,,\quad h(r) = 1\,\quad\text{and}\quad \Phi(r) = \Phi_0\frac{\sqrt{r^2+L^2}-L}{\sqrt{r^2+L^2}+L}\,.
\end{equation}
We were also able to find solutions where $h\neq1$, but it turns out they all lead to boundary metrics that are not round spheres; i.e., the squashing does not disappear on the boundary. We will comment further on this more general case later. It is a simple exercise to evaluate the on-shell action for which we find
\begin{equation}
\Delta S_{U(1)^2}=8 \pi ^2 L^2 \left(1+2 \Phi _0^2\right)\,.
\end{equation}
\subsection{The wormhole phase}
Despite our best efforts we were not able to find an analytic solution for the wormhole phase.  We thus proceed numerically. Our \emph{Ansatz} takes the form
\begin{subequations}
\begin{equation}
\mathrm{d}s^2 = \frac{L^2}{(1-y)^2}\left[\frac{f \,\mathrm{d}y^2}{y(2-y)}+\frac{y_0^2}{4}\left(g\,\hat{\sigma}_3^2+\hat{\sigma}_1^2+\hat{\sigma}_2^2\right)\right],
\end{equation}
with $f$ and $g$ to be determined numerically and depending only on $y$. Just as for the Einstein-$U(1)^3$ theory, $y_0$ measures the minimal size of the wormhole at the neck.

For the Maxwell fields we again take
\begin{equation}
A^{(1)} = \frac{L}{2}\,\Phi(y)\,\hat{\sigma}_1\quad\text{and}\quad A^{(2)} = \frac{L}{2}\,\Phi(y)\,\hat{\sigma}_2\,.
\end{equation}
\end{subequations}

The equations of motion read
\begin{subequations}
\begin{align}
&\frac{\sqrt{g} \sqrt{2-y} \sqrt{y}}{\sqrt{f}}\left[\frac{\sqrt{g} \sqrt{2-y} \sqrt{y}}{\sqrt{f}}\Phi^\prime\right]^\prime-\frac{4}{y_0^2}\Phi=0\,,
\\
&\frac{f g}{n_y y y_0^2}+\frac{8 f m_y^2 \Phi ^2}{n_y y y_0^4 g}+\frac{g'}{g m_y}-\frac{1}{n_y y y_0^2 m_y^2}\left[\left(4 m_y^2+3 y_0^2\right) f-n_y y \left(3 y_0^2-2 m_y^4 \Phi'^2\right)\right]=0\,,
\\
& \frac{3 f g m_y^2}{n_y}-\frac{8 f \Phi ^2 m_y^4}{g y_0^2 n_y}-\frac{f \left(4 m_y^2+3 y_0^2\right)}{n_y}-\frac{y m_y y_0^2 f'}{f}-\frac{m_y y y_0^2}{n_y}+(1+2 y) y_0^2+2 m_y^4 y \Phi '^2=0\,.
\end{align}
\end{subequations}
with $n_y=2-y$ and $m_y=1-y$.

We now discuss the boundary conditions at $y=0$, the wormhole $\mathbb{Z}_2$ plane of symmetry. Demanding that $f$ and $g$ have a regular Taylor expansion around $y=0$ gives the following set of Dirichlet conditions at $y=0$
\begin{align*}
&\Phi (0)=\frac{\sqrt{g(0)} y_0 \sqrt{3 y_0^2+4-g(0)}}{2 \sqrt{2}}\,,
\\
& \Phi^\prime(0)=\frac{y_0 \sqrt{6 y_0^2+8-2 g(0)}}{\sqrt{g(0)} \left[3 y_0^2+4-2 g(0)\right]}\,,
\\
& f(0)=\frac{y_0^2}{3 y_0^2+4-2 g(0)}\,,
\\
&g'(0)+\frac{2 g(0) \left[3 y_0^2+8-5 g(0)\right]}{3 y_0^2+4-2 g(0)}=0\,.
\end{align*}
One can show that $\Phi$, $f$ and $g$ admit a regular Taylor series around $y=0$, with all higher order terms in the series being uniquely fixed by $y_0$ and $g(0)$.

We now turn out attention to the boundary conditions imposed at the conformal boundary located at $y=1$. There, we demand $f(1)=g(1)=1$ since we are primarily interested in solutions with a round $S^3$ at the conformal boundary. $\Phi(1)\equiv \Phi_0$, on the other hand, determines the source. Expansion the fields with a regular Taylor expansion around $y=1$ determines $f$, $g$ and $\Phi$ as a function of four unknowns which we take to be $y_0$, $\Phi(1)$, $\Phi^\prime(1)$ and $g^{\prime\prime\prime}(1)$.

The procedure is now clear: we take a given value of $y_0$ and $g(0)$ and integrate outwards to the conformal boundary and, in general, we find that $g(1)\neq1$. This means for a given value of $y_0$, we need to adjust $g(0)$ so that $g(1)=1$. Once this is the case, we read off the corresponding values of $\Phi_0$, $\Phi^\prime(1)$ and $g^{\prime\prime\prime}(1)$. We thus have a one-parameter family of solutions whose boundary metric is a round three-sphere. This one-parameter family of solutions is, in turn, labelled by $y_0$.

Once we have the desired solution, we can determine its on-shell action numerically just as we did when we studied the Einstein-$U(1)^3$ theory. We call the Euclidean action of the wormhole solution $S^W_{U(1)^2}$.
\subsection{M$2$ branes on wormhole backgrounds}
Before discussing the phase diagram, we will pause for a moment and study probe branes on wormhole backgrounds. The point of this study is that in the limit of weak coupling the action of a probe brane describes the change in the action of the wormhole when a brane is inserted into the solution\footnote{This property, together with the idea that sources at the AdS boundaries should remain fixed, determines the detailed form of the probe brane action and forbids the addition of arbitrary constants.  With this understanding, the sign of the probe-brane action becomes physically meaningful}.  As a result, negative probe-brane actions mean that than our wormhole is not in fact the lowest-action solution.   In this case one would in principle then like to study solutions with such branes to find the true minimum, and to determine whether it remains a connected wormhole or whether it becomes disconnected.  However, this is beyond the scope of the present work. At weak coupling such a true minimum can be achieved only by including a large number of branes, which one might describe as a condensate.  If on the other hand all probe branes have positive action in our wormhole, this would support the idea that our wormhole does indeed dominate the computation of the desired partition functions.

The appropriate M$2$ brane probe action is
\begin{equation}
S_{\mathrm{M}2}=T_{\mathrm{M}2}\int_{\mathcal{M}_{\mathrm{M}2}} \mathrm{d}^3 \sigma \left[\sqrt{\det \tilde{G}}-i\,\varepsilon_{\mathrm{M}2}\mathcal{C}_{\mathrm{M}2}\right]\,,
\end{equation}
with the metric on the world-volume of the M$2$ branes being given by
\begin{equation}
\tilde{G}_{\dot{\mu}\dot{\nu}}= {}^{(11)}g_{AB}\frac{\mathrm{d}x^A}{\mathrm{d}\sigma^{\dot{\mu}}}\frac{\mathrm{d}x^B}{\mathrm{d}\sigma^{\dot{\nu}}}\,,
\end{equation}
and $\varepsilon_{\mathrm{M}2}=\pm1$ for brane anti-brane configurations, respectively. Furthermore, the potential term is given by
\begin{equation}
\mathcal{C}_{\mathrm{M}2}=\frac{1}{3!}\varepsilon^{\dot{\mu}\dot{\nu}\dot{\rho}}A_{(3)\,ABC}\frac{\mathrm{d}x^A}{\mathrm{d}\sigma^{\dot{\mu}}}\frac{\mathrm{d}x^B}{\mathrm{d}\sigma^{\dot{\nu}}}\frac{\mathrm{d}x^C}{\mathrm{d}\sigma^{\dot{\rho}}}\,,
\end{equation}
with $\varepsilon^{\dot{\mu}\dot{\nu}\dot{\rho}}$ being the totally anti-symmetric alternating symbol with $\varepsilon^{\dot{1}\dot{2}\dot{3}}=1$.

We are interested in branes that wrap the $S^3$ so that we take $\sigma^{\dot{\mu}}=\{\psi,\theta,\hat{\varphi}\}$ with the standard Euler angles given in (\ref{eqs:leftinvariant}).
Applying this procedure to the our wormhole \emph{Ansatz} gives
\begin{equation}
S_{\mathrm{M}2}(y;\varepsilon_{\mathrm{M}2})=2 \pi ^2 L^3 y_0^3 T_{\mathrm{M}2}\left[\frac{\sqrt{g(y)}}{(1-y)^3}-3 \varepsilon_{\mathrm{M}2}\,\int_0^y \frac{\sqrt{f\left(\tilde{y}\right)} \sqrt{g\left(\tilde{y}\right)}}{\left(1-\tilde{y}\right)^4 \sqrt{\tilde{y}} \sqrt{2-\tilde{y}}} \,
   d\tilde{y}\right]\,.
   \label{eq:membrane}
\end{equation}
In deriving the above expression we made a choice in determining $A_{(3)}$ from $F_{(4)}$. This choice was such that upon the change of variable
\begin{equation}
r = L\,y_0 \frac{\sqrt{y}\sqrt{2-y}}{1-y}\,
\end{equation}
the action $S_{\mathrm{M}2}(r;\varepsilon_{\mathrm{M}2})$ satisfies
\begin{equation}
S_{\mathrm{M}2}(r;-1)=S_{\mathrm{M}2}(-r;1)\,,
\end{equation}
so that studying $S_{\mathrm{M}2}(r;1)$ covers both the case of brane and anti-brane probes. The question is then whether $S_{\mathrm{M}2}(r;1)$ is positive definite for all values of $r$. If $S_{\mathrm{M}2}(r;1)<0$ for any range of $r$, we would expect brane nucleation to take place and render our wormhole solution unstable.
\subsection{Negative modes}
Studying negative modes of this novel class of geometries turns out to be more complicated than in the Einstein$-U(1)^3$ case as even the background metric fails to enjoy spherical symmetry. However, the current isometry group is now $SU(2)\times U(1)$ remains large enough to reduce the study of the perturbations to ordinary differential equations. We have used this observation to analyze such perturbations by generalizing the techniques used for $SO(4)$ symmetry in the above sections.  However, due to the extremely cumbersome nature of this prcedure, we present only a sketch of the analysis below.  This sketch should suffice to allow dedicated readers who wish to check and reproduce our results to do so.

We first introduce charged scalar harmonics on $\mathbb{CP}^1\equiv S^2$. Following \cite{Dias:2010eu} we define these to be solutions of the following eigenvalue problem
\begin{equation}
\mathcal{D}_i \mathcal{D}^i \mathbb{Y}_{m\,\kappa}+\lambda_{m\,\kappa} \mathbb{Y}_{m\,\kappa}=0\,,
\end{equation}
with $\mathcal{D}=\Grad-i\,m\,\mathbb{A}_{\mathbb{CP}^1}$, $m\in\mathbb{Z}$, $\Grad$ being the standard connection on $\mathbb{CP}^1$ and $\mathbb{A}_{\mathbb{CP}^1}$ is the K\"ahler one-form that relates to the standard K\"ahler 2-form on $\mathbb{CP}^1$, $\mathbb{J}_{\mathbb{CP}^1}$, as $\mathbb{J}=\mathrm{d}\mathbb{A}_{\mathbb{CP}^1}/2$. Regularity then demands
\begin{equation}
\lambda_{m\,\kappa}= \ell(\ell+2)-m^2\quad \text{with}\quad \ell= 2\,\kappa+|m|\quad \text{where}\quad \kappa = 0,1,2,\ldots
\end{equation}

The construction of the metric and gauge field perturbations then conforms with those studied in \cite{Dias:2010eu}. Using numerics, we find that there are no negative modes for any $m\neq0$ and $\kappa>0$. For $m=\kappa=0$, however, for small wormholes we do find a negative mode on which we report further below.
\subsection{Results}
\label{sec:U12results}
Our numerical results are as follows. For each value of the boundary source $\Phi_0>\Phi_0^{\min}\approx 4.0162(5)$ we find two wormhole solutions which we may again call small and large; see left hand side panel of Fig.~\ref{fig:mtheory}. We find no field-theoretic negative modes anywhere on the large wormhole branch, though the small wormhole branch has at least one negative mode in the $\kappa=m=0$ sector. Furthermore, the Euclidean action of the large wormhole branch of solutions eventually becomes smaller than twice the corresponding action of the disconnected solution; see right hand side panel of Fig.~\ref{fig:mtheory} where $\Delta S_{U(1)^2} = 2 S^D_{U(1)^2}-S^W_{U(1)^2}$ - for $\Phi_0>\Phi_0^{\mathrm{HP}}\approx 4.352(8)$.

Finally, we also studied the positivity of the probe M$2$ brane action (\ref{eq:membrane}). On the large wormhole branch, and for large enough boundary sources $\Phi_0>\Phi_0^{\mathrm{M}2}\approx 4.097(2)$, we find a finite range of $r$ (or, equivalently, values of $y$) for which the action from M$2$ branes wrapped on the $S^3$ becomes negative. As a result, all large wormholes which dominate over our disconnected solution turn out to be unstable to brane nucleation. We mark the nucleation threshold at $\Phi_0^{\mathrm{M}_2}$ with a red dot in the right panel of Fig.~\ref{fig:mtheory}. We note, however, that there exists a range $\Phi_0\in(\Phi_0^{\min},\Phi_0^{\mathrm{M}2})$ in which the large wormholes seem to have no pathology, though in this range they are subdominant with respect to our disconnected solution (these non pathological wormholes are represented by the blue disks in Fig.~\ref{fig:mtheory}).
\begin{figure}[h]
\centering
\includegraphics[width =\textwidth]{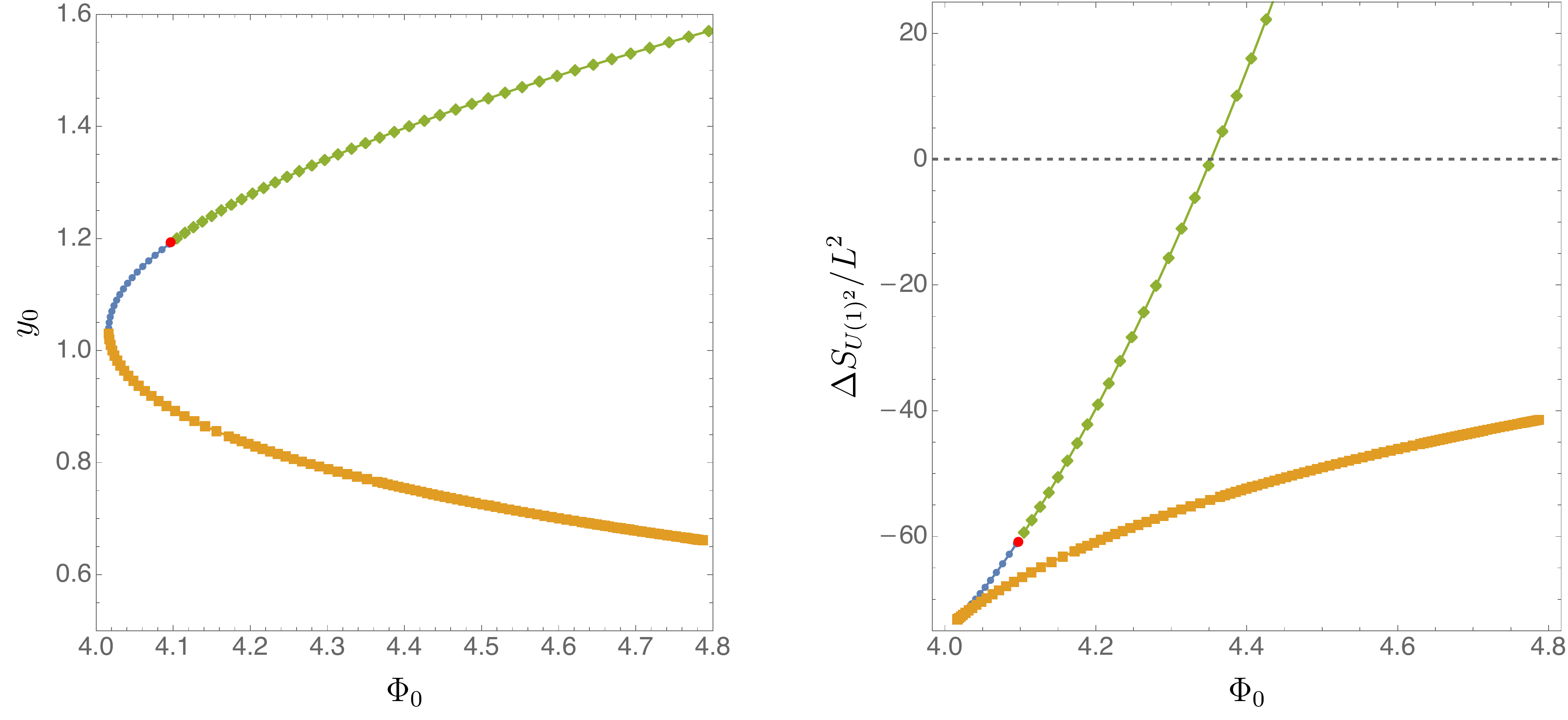}\\
\caption{{\bf Left panel:} radius of the wormhole solutions as a function of the source $\Phi_0$. Wormholes only exists for $\Phi_0>\Phi_0^{\min}\approx 4.0162(5)$. {\bf Right panel:} difference in the Euclidean action $\Delta S_{U(1)^2} = 2 S^D_{U(1)^2}-S^W_{U(1)^2}$ as a function of $\Phi_0$. The wormhole solutions have a lower action for $\Phi_0>\Phi_0^{\mathrm{HP}}\approx 4.352(8)$. The orange squares represent small wormholes, the blue disks correspond to large wormholes where the Euclidean action for M$2$ is positive and the green diamonds indicate large wormholes where the Euclidean action for M$2$ is not positive definite. The red disk indicates the value of $\Phi_0$ in the large wormhole branch above which the Euclidean action (\ref{eq:membrane}) for M$2$ branes is not positive definite.}
\label{fig:mtheory}
\end{figure}

Finally, we comment on a small extension of our result. We have also considered cases where the metric at the boundary is not round, i.e. $g(1)\equiv g_1\neq1$. These constitute a two-parameter family of wormhole solutions which we can parametrise with $(\Phi_0,g_1)$. It turns out that near the conformal boundary, located at $y=1$, one has
\begin{equation}
S_{\mathrm{M}2}(y;1)= \frac{\pi ^2 (4-g_1) \sqrt{g_1} L^3 y_0}{1-y}+\mathcal{O}(1)\,.
\end{equation}
It is thus clear that we need $0<g_1<4$ in order for $S_{\mathrm{M}2}(y;\varepsilon_{\mathrm{M}2})$ to be positive definite near the conformal boundary. So we restricted our attention to $0<g_1<4$. In this range we were not able to find any value of $g_1$ for which wormholes dominate over the disconnected solution, have positive definite $S_{\mathrm{M}2}(y;\varepsilon_{\mathrm{M}2})$, and have no negative modes. In fact, we find that for small enough values of $g_1$ even the small wormhole branch has non positive $S_{\mathrm{M}2}(y;\varepsilon_{\mathrm{M}2})$. The value of $g_1$ corresponding to the largest ratio $\Phi_0^{\mathrm{M}2}/\Phi_0^{\mathrm{HP}}$ is $g_1=4/3$, corresponding to the maximisation of the numerator in the divergent term of $S_{\mathrm{M}2}(y;1)$ near the conformal boundary.

\section{\label{sec:massdeformed}Mass deformation of ABJM}

Perhaps the simplest asymptotically-AdS Euclidean wormholes are the quotients of Euclidean AdS$_d$, which is of course also the hyperbolic plane $\mathbb{H}^d$.    This space admits a foliation by hyperbolic planes $\mathbb{H}^{d-1}$ of dimension $(d-1)$, so that the metric can be written
\begin{equation}
\label{eq:hypslice}
\mathrm{d}s^2 = \frac{\mathrm{d}r^2}{\frac{r^2}{L^2}+1}+(r^2+L^2)\mathrm{d}s^2_{\mathbb{H}^{d-1}}\,,
\end{equation}
with $\mathrm{d}s^2_{\mathbb{H}^{d-1}}$ the metric on the unit ${\mathbb{H}^{d-1}}$.  Here $r$ takes values in $(-\infty, \infty)$.    Now, any compact hyperbolic space of dimension $d-1$ can be written as the quotient of ${\mathbb{H}^{d-1}}$ with respect to an appropriate discrete group $\Gamma$ of ${\mathbb{H}^{d-1}}$ isometries.  From \eqref{eq:hypslice}, we see that taking the corresponding quotient of AdS$_d$ yields a wormhole with compact hyperbolic slices at each value of $r \in (-\infty,\infty)$ and with two separate boundaries at $r = \pm \infty$.  In the obvious conformal frame both boundaries are again compact hyperbolic manifolds.

This construction embeds easily in many UV-complete models.  However, as described in \cite{Maldacena:2004rf}, in simple models it is associated witb a dramatic brane-nucleation instability that occurs even near the AdS boundary.  This makes the entire theory unstable with such boundaries.  The dual field theory interpretation is that conformal field theories require conformal couplings to curvature, and that such couplings naturally generate negative mass terms when the theory is placed on a compact hyperbolic space.

As noted in \cite{Maldacena:2004rf}, this also suggests that such instabilities can be cured by breaking conformal invariance and adding explicit new couplings to the would-be-CFT that give masses to various scalars.  The question is then what happens to the above wormhole solutions under such deformations.  We investigate this issue below using a particular mass-deformation of an AdS$_4$  compactification of 11-dimensional supergravity; i.e., from the dual gauge theory point of view we study a deformation of the ABJM model \cite{Aharony:2008ug}.  Interestingly, at least in this case, we find wormhole solutions only for small mass deformations $\mu$, and in particular only at deformations $\mu<\mu_{max}$ where $\mu_{max}$ is still too small to stabilize the theory.  At such small deformations we find two branches of wormhole solutions, but  they coalesce $\mu=\mu_{max}$.  It thus appears that wormholes do not exist in the stable members of this family of theories.

\subsection{The mass-deformation model}
The deformation of interest can be described using a sub-truncation of the truncation of 11-dimensional supergravity detailed in \cite{Cvetic:1999au}. Our $11-$dimensional metric $G$ takes the form
\begin{multline}
\mathrm{d}s^2_{11d} \equiv {}^{(11)}g_{AB}\mathrm{d}x^A\mathrm{d}x^B=\Xi^{1/3}\,g_{ab}\mathrm{d}x^a\mathrm{d}x^b
\\
+\frac{\Xi^{1/3}}{g^2}\Bigg\{\mathrm{d}\xi^2+\frac{\cos^2\xi}{4\,Z_1}\;\left[\mathrm{d}\theta_1^2+\sin^2\theta_1\,\mathrm{d}\phi_1^2+(\mathrm{d}\psi_1+\cos \theta_1\,\mathrm{d}\phi_1)^2\right]
\\
+\frac{\sin^2\xi}{4\,Z_2}\;\left[\mathrm{d}\theta_2^2+\sin^2\theta_2\,\mathrm{d}\phi_2^2+(\mathrm{d}\psi_2+\cos \theta_2\,\mathrm{d}\phi_2)^2\right]\Bigg\}\,,
\label{eq:11a}
\end{multline}
where
\begin{subequations}
\begin{align}
& Z_1=e^{\Phi}\cos^2\xi+\sin^2\xi\,,
\\
& Z_2 =(e^{-\Phi}+\chi^2 e^{\Phi})\sin^2\xi+\cos^2\xi\,,
\\
&\Xi = Z_1\,Z_2\,.
\end{align}
\end{subequations}
For $F_{(4)}$ we have
\begin{equation}
F_{(4)}=-2\,i\,g\,U\,\mathrm{Vol}_{4}+i\frac{\sin \xi\cos \xi}{g}(\star \mathrm{d}\Phi-\chi\,e^{2\Phi}\star\mathrm{d}\chi)\wedge \mathrm{d}\xi-\mathrm{d}\tilde{A}_{(3)}\,,
\label{eq:11b}
\end{equation}
where $\mathrm{Vol}_4$ is the volume form on $g$ and $\star$ the Hodge operation with respect to $g$. Furthermore,
\begin{multline}
\tilde{A}_{(3)}=\frac{1}{8\,g^2}\chi e^{\Phi}\Big[\frac{\cos^4\xi}{Z_1}\sin \theta_1 (\mathrm{d}\psi_1+\cos \theta_1\mathrm{d}\phi_1)\wedge \mathrm{d}\theta_1\wedge \mathrm{d}\phi_1
\\
-\frac{\sin^4\xi}{Z_2}\sin \theta_2 (\mathrm{d}\psi_2+\cos \theta_2\mathrm{d}\phi_2)\wedge \mathrm{d}\theta_2\wedge \mathrm{d}\phi_2\Big]
\end{multline}
and
\begin{equation}
U = e^{\Phi}\cos^2\xi +(e^{-\Phi}+\chi^2 e^{\Phi})\sin^2\xi +2\,.
\end{equation}
Inserting the above expressions for $G$ and $F_{(4)}$ into the eleven-dimensional equations of motion (\ref{eq:gen11D}) yields four-dimensional equations of motion for $g$, $\chi$ and $\Phi$ that can be derived from the following four-dimensional action
\begin{equation}
S_{4d} = -\int_{\mathcal{M}} \mathrm{d}^4 x\sqrt{g}\left(R+\frac{1}{2}\nabla_a \Phi \nabla^a \Phi-\frac{e^{2\Phi}}{2}\nabla_a \chi\nabla^a\chi-V\right)-2\int_{\partial \mathcal{M}} \mathrm{d}^3 x\sqrt{h}\;K+S_{\mathcal{B}}\,,
\end{equation}
with
\begin{equation}
V= -\frac{1}{L^2}\left(e^{\Phi}+e^{-\Phi}+\chi^2 e^{\Phi}+4\right)\,,
\end{equation}
where $2g = L$. In the above we will not need to specify $S_{\mathcal{B}}$. From the four-dimensional perspective, $\Phi$ and $\chi$ have masses $\mu^2L^2=-2$. To proceed we need to understand what boundary conditions we should choose for these fields. By comparing our \emph{Ansatz} for $F_{(4)}$ with the one presented in \cite{Bena:2000zb}, we conclude that the sources associated with boundary values of the supergravity fields $\Phi$ and $\chi$ parametrise (possibly supersymmetric) mass deformations of ABJM. The supersymmetric mass deformation corresponds to deformations for which $\chi \sim \mu\,z/L$, $\Phi =\mathcal{O}(z^2)$, with $\mu$ being proportional to the mass deformation and $z$ a Fefferman-Graham coordinate. These are the boundary conditions that we will employ.

Rather remarkably, the action above admits yet another sub-truncation in which
\begin{equation}
\chi = \sqrt{1-e^{-2\Phi}}\,.
\label{eq:subtrun}
\end{equation}
Perhaps more interestingly, when we performed our numerical studies, we did not impose the relation above, and yet all our numerically determined solutions were consistent with the above relation.
\subsection{Wormholes}
Our starting point is family of solutions \eqref{eq:hypslice} with $d=4.$  Below, we implicitly assume a quotient by some $\Gamma$ that makes the $r=constant$ slices compact hyperbolic spaces.   As noted above, such solutions are unstable to the nucleation of branes, which the in present context are M$2$ branes.  To understand the effect of the mass deformation on this instability  we will study M$2$ branes on our deformed wormholes. Before doing this, we provide some detail regarding our construction.

As a metric \emph{Ansatz} we take
\begin{equation}
\mathrm{d}s^2 = \frac{L^2}{(1-y)^2}\left[\frac{f\mathrm{d}y^2}{y(2-y)}+y_0^2\mathrm{d}s^2_{\mathbb{H}^3}\right]\,,
\end{equation}
with $f$ a function only of $y\in(0,1)$ and with $y_0$ to be interpreted as the minimal size of the wormhole neck. The neck is located at $y=0$ and the conformal boundary is at $y=1$. For the scalars we take
\begin{equation}
\chi = (1-y)\,q_1\quad\text{and}\quad \Phi = (1-y)^2 q_2\,.
\end{equation}

The procedure is now very similar to what we have seen before: finiteness of $f$ at $y=0$ locks $y_0$ into a relationship with $\Phi(0)$ and $\chi(0)$. We then take these values at the origin $y=0$ and integrate outwards. In general $\Phi=\mathcal{O}[(1-y)]$ so we adjust $\Phi(0)$ so that near the conformal boundary we find $\Phi=\mathcal{O}[(1-y)^2]$.   What remains is a one-parameter family of solutions parameterized by either $y_0$ or $\mu$. 
\subsection{M$2$ probe branes}

We now need to understand what  value of $\mu$ is required to remove the UV brane-nucleation instability by making the action for probe M$2$ branes positive near the AdS boundary.  We use the action (\ref{eq:membrane}) together with the $11-$dimensional \emph{Ans\"atze} (\ref{eq:11a}) and (\ref{eq:11b}). We begin by  wrapping our M$2$ branes on (the relevant quotient of) $\mathbb{H}^3$, and obtain an action as a function of $y$. Near the conformal boundary we can use the near boundary behaviour of our fields to determine $S_{\mathrm{M}2}$ near $y=1$. This turns out to be
\begin{equation}
S_{\mathrm{M}2}=\frac{y_0 L^3\mathrm{Vol}_{\mathbb{H}^3}}{8(1-y)} \left[\mu ^2-12+2 \cos (2 \xi ) \left(\mu ^2-y_0^2 \Phi ''(1)\right)\right]+\mathcal{O}(1)\,,
\end{equation}
where $\mathrm{Vol}_{\mathbb{H}^3}$ is the volume of the relevant quotient of $\mathbb{H}^3$. The condition (\ref{eq:subtrun}) automatically ensures that the term proportional to $\cos(2\xi)$ vanishes (we once more note that we find this condition to be true numerically). We thus see that we must have $\mu^2>12$ to stabilize the theory in the UV, and in particular to have a hope of $S_{\mathrm{M}2}$ being positive definite.
\subsection{Results}
The main result of this section is presented in Fig.~\ref{fig:massdeform}, where we plot the radius of the wormhole $y_0$ as a function of $\mu$. Rather interestingly, we find no wormhole solution for $\mu>\mu_{\max}\approx1.66(2)$, which is smaller than $\sqrt{12}\approx 3.464$, so that this new class of mass deformed wormholes is still unstable to nucleating M$2$ branes. The results of this section are thus similar to those found in \cite{Buchel:2004rr} for mass deformations of $\mathcal{N}=4$ SYM.
\begin{figure}[h]
\centering
\includegraphics[scale=0.4]{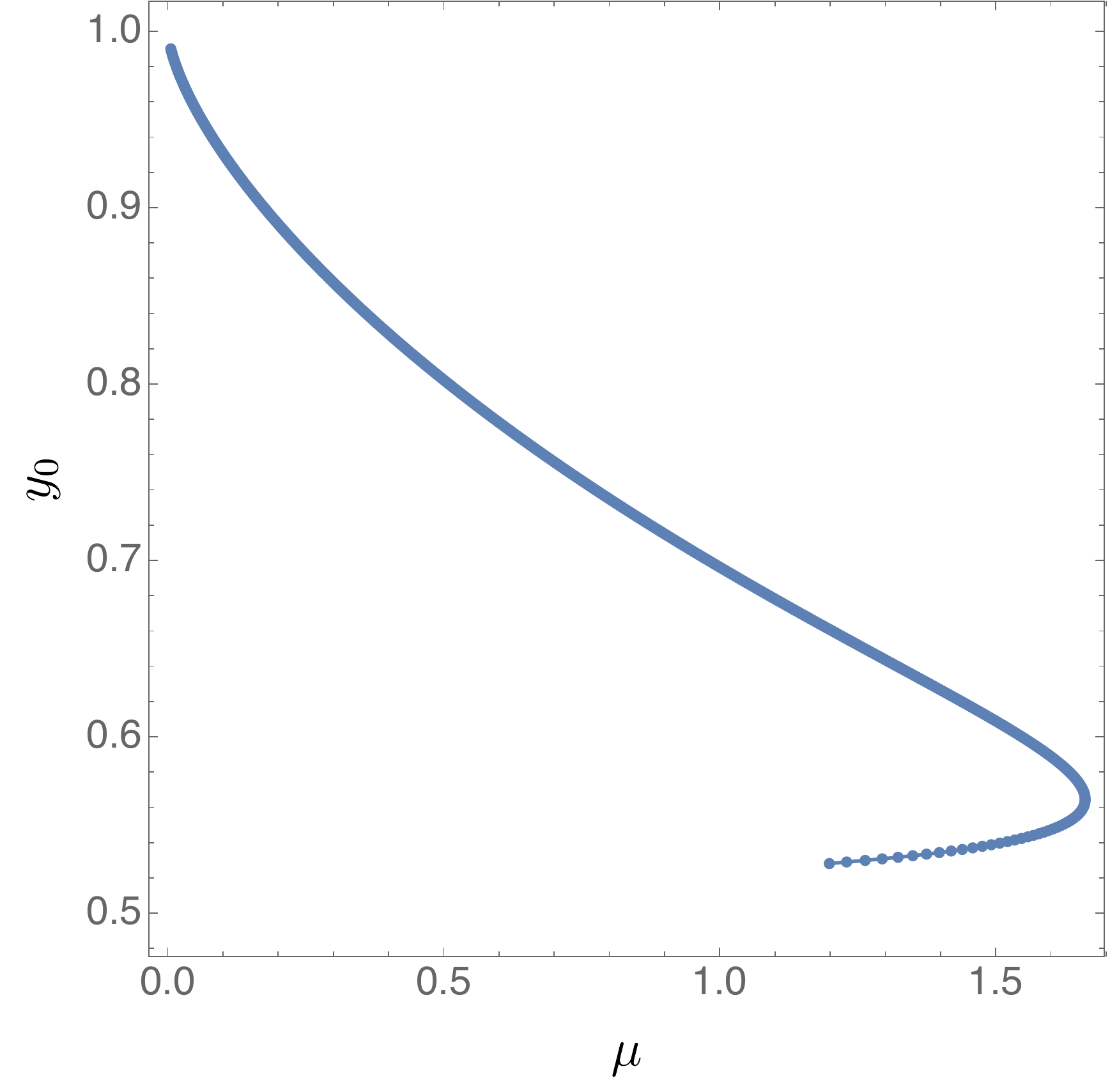}\\
\caption{No wormhole solutions seem to exist for $\mu>\mu_{\max}\approx1.66(2)<\sqrt{12}$, which in particular implies that no stable wormhole solution exist.}
\label{fig:massdeform}
\end{figure}

\section{Wormholes from type IIB theory}
\label{sec:IIB}

We can also find a truncation of a UV-complete scenario that is closely related to our Einstein-scalar model of section \ref{sec:scalars}, though this truncation will live in AdS$_3$ instead of AdS$_4$.  As described below, this model is a compactification of type IIB supergravity.  After discussing the \emph{Ans\"atze} for the fields in section \ref{sec:IIBtrunc}, we analyze the model as usual in the remaining subsections.

\subsection{A simple consistent truncation}
\label{sec:IIBtrunc}
We consider type IIB supergravity with only the ten dimensional metric ${}^{(10)}g$, Ramond-Ramond three-form $F_{(3)}\equiv \mathrm{d}A_{(2)}$ and dilaton $\phi$. The corresponding equations of motion are
\begin{subequations}
\begin{align}
& {}^{(10)}R_{AB} = \frac{1}{2}{}^{(10)}\nabla_A\phi{}^{(10)}\nabla_B\phi+\frac{e^\phi}{4}\left[F_{(3)\,ACD}F_{(3)\,B}^{\phantom{(3)B}CD}-\frac{{}^{(10)}g_{AB}}{12}F_{(3)\,CDE}F_{(3)}^{\phantom{(3)}\,CDE}\right]\,,
\\
& \mathrm{d}\left(e^{\phi}\star_{10} F_{(3)}\right)=0\,,
\\
&\Box \phi-\frac{e^{\phi}}{12} F_{(3)\,ABC}F_{(3)}^{\phantom{(3)}\,ABC}=0\,,
\end{align}
\label{eq:IIBSUGRA}%
\end{subequations}%
where $\star_{10}$ is the Hodge operation associated with the ten-dimensional metric ${}^{(10)}g$, ${}^{(10)}\nabla$ its associated metric-compatible connection and upper case Latin indices are ten-dimensional.

Consider the following ten-dimensional field configuration
\begin{subequations}
\begin{multline}
\mathrm{d}s^2 = (g_{ab}\mathrm{d}x^a\mathrm{d}x^b+L^2 \mathrm{d}\Omega_3^2)e^{\frac{\phi_4}{2}}+e^{-\frac{\phi_4}{2}}\Big[(e^{\sqrt{2}\phi_1}\mathrm{d}z_1^2+e^{-\sqrt{2}\phi_1}\mathrm{d}z_2^2)e^{\phi_3}
\\
+(e^{\sqrt{2}\phi_2}\mathrm{d}z_3^2+e^{-\sqrt{2}\phi_2}\mathrm{d}z_4^2)e^{-\phi_3}\Big]
\end{multline}
\begin{equation}
F_{(3)} = \frac{2\,i}{L}\mathrm{Vol}_{3}+2\,L^2\,\mathrm{d}^3\Omega_3
\end{equation}
and
\begin{equation}
\phi = \phi_4\, ,
\end{equation}
\label{eqs:full10d}%
\end{subequations}%
where $\mathrm{d}^3\Omega_3$ is the volume form on a round $3-$sphere, $\mathrm{Vol}_{3}$ is the volume form of the three-dimensional metric $g$, $\phi_i$, with $i=1,2,3,4$, depend only on the three-dimensional coordinates $x^a$ and the coordinates $\{z_1,z_2,z_3,z_4\}$ parametrise a 4-torus. Here, lower case Latin indices are three-dimensional. Inserting the \emph{Ans\"atze} (\ref{eqs:full10d}) into the $10$-dimensional equations equations of motion (\ref{eq:IIBSUGRA}) yields a set of three-dimensional equations for $g$, and $\phi_i$ which can be derived from the following three-dimensional action
\begin{equation}
S_{3d} = -\int_{\mathcal{M}}\mathrm{d}^3 x\sqrt{g}\left[R+\frac{1}{L^2}-\sum_{i=1}^4\nabla_a \phi_i \nabla^a \phi_i\right]\,.
\label{eq:add}
\end{equation}
This action will have to be supplemented by appropriate boundary terms. As expected, the three dimensional metric $g$ is asymptotically AdS$_3$. The scalars $\phi_i$ appear as massless AdS$_3$ scalars that are minimally coupled to gravity.

We wish to turn on a source for all the $\phi_i$, and we should thus consider adding boundary terms to (\ref{eq:add}) appropriate to such a choice and which render the on-shell action finite on solutions to corresponding equations of motion.  We now introduce Fefferman-Graham coordinates for $g$, where the conformal boundary is located at $z=0$. Since $\phi_i$ is massless, it will have in general a $\log z$ divergence close to the conformal boundary. In fact, the generic behaviour of $\phi_i$ close to the conformal boundary takes the form
\begin{equation}
\phi_i = V_i(x^\mu)+z^2\,Z_i(x^\mu)+z^2\,\log z\,\tilde{Z}_i(x^\mu)+\ldots\,,
\end{equation}
where $\mu$ runs over the boundary directions, and $\tilde{Z}_i$ is a function of $V_i(x^\mu)$ only. For this reason, the counterterms to be added to (\ref{eq:add}) will explicitly depend on a UV cut off $z=\varepsilon$. The boundary terms that lead to a well defined variational problem for the metric and scalar field and render the on-shell action finite read
\begin{multline}
\label{eq:3dtruncactionbdy}
S_c(\varepsilon)= -2\int_{\partial \mathcal{M}_\varepsilon} \mathrm{d}^2 x\sqrt{h}\;K+\frac{2}{L}\int_{\partial \mathcal{M}_\varepsilon} \mathrm{d}^2 x\sqrt{h}
\\
-L \log\frac{\varepsilon}{L}\sum_{i=1}^4\int_{\partial \mathcal{M}_\varepsilon} \mathrm{d}^2 x\sqrt{h} \tilde{\nabla}_{\mu}\phi_i\tilde{\nabla}^{\mu}\phi_i-L\log\frac{\varepsilon}{L}\int_{\partial \mathcal{M}_\varepsilon} \mathrm{d}^2 x\sqrt{h}\tilde{R}\,,
\end{multline}
where $h$ the induced metric on a surface of constant $z=\varepsilon$.  We denote this surface by  $\partial \mathcal{M}_\varepsilon$, and $\tilde{\nabla}$ is the natural metric-compatible connection on $(\partial \mathcal{M}_\varepsilon,h)$.  Furthermore,  $\tilde{R}$ the Ricci scalar on $\partial \mathcal{M}_\varepsilon$ and $K$ the trace of the extrinsic curvature associated to an outward-pointing normal to $\partial \mathcal{M}_\varepsilon$. The first term in \eqref{eq:3dtruncactionbdy} is the usual Gibbons-Hawking-York term, and the remaining are boundary counterterms. The total on-shell action is then
\begin{equation}
S_{\mathrm{on-shell}}=\lim_{\varepsilon\to0^+} \tilde{S}(\varepsilon)+S_c(\varepsilon)\,.
\end{equation}
where $\tilde{S}(\varepsilon)$ is obtained from ${}^{(3)}S$ in (\ref{eq:add}) by replacing $\mathcal{M}$ with $\mathcal{M}_\varepsilon$, which is obtained from $\mathcal{M}$ by chopping off from spacetime the region $z>\varepsilon$.

In the remainder of this section we will set $\phi_4=0$, which in particular restricts us to trivial dilaton fields. We stress however, that appendix \ref{app:torusboundaries} describes wormhole solutions with a toroidal boundary metric $\mathbb{T}^2$ for  which $\phi_4\neq0$.
\subsection{Symmetry}
The solutions we seek to construct enjoy spherically symmetric metrics, but the scalars $\phi_i$ will break said symmetry in a special way. In particular, we introduce standard polar and azimuthal coordinates $(\theta,\varphi)$ on the $S^2$, so that the standard metric on the $S^2$ reads
\begin{equation}
\mathrm{d}\Omega_2^2 = \mathrm{d}\theta^2+\sin^2\theta\,\mathrm{d}\varphi^2.
\end{equation}
We then take the scalars to satisfy
\begin{equation}
\phi_1 = \psi(r)\,\sin\theta\,\cos\varphi\,,\quad \phi_2 = \psi(r)\,\sin\theta\,\sin\varphi\quad \text{and}\quad \phi_3 = \psi(r)\,\cos \theta\,.
\end{equation}
These forms are chosen so that
\begin{equation}
\sum_{i=1}^3 \phi_i^2 = \psi(r)^2\,.
\end{equation}
For the metric we take
\begin{equation}
\mathrm{d}s^2 = \frac{\mathrm{d}r^2}{f(r)}+g(r)\mathrm{d}\Omega_2^2\,.
\label{eq:typeIIbgen}
\end{equation}

Wormhole solutions will have $f(r),g(r)>0$ throughout spacetime, whereas the disconnected solutions will have $f(r)= 1+\mathcal{O}(r^2)$, with $g(r)=\mathcal{O}(r^2)$ near $r=0$.
\subsection{D1 Branes}
We again wish to address potential brane nucleation instabilities. Since we are in type IIB and the only non-zero Ramond-Ramond field is the three-form $F_{(3)}$, the relevant branes are D$1$'s and D$5$'s. We shall present here results for the D1's, but the invariance of the background under Hodge-duality  of the Ramond-Ramond 3-form field strength and the trivial dilaton imply that equivalent results can be obtained for D$5$'s wrapped on the 4-torus (or for related D1D5 bound states).

For a generic spacetime $(\mathcal{M},g,A_{(2)})$, the Euclidean action of a probe D1 with world-volume coordinates $\sigma^{\dot{\mu}}$  (for $\dot{\mu}=1,2$) takes the simple form
\begin{equation}
S^{\mathrm{D1}}_{\pm\,E} = \int_{\mathcal{M}_2}\mathrm{d}^2\sigma\left[\sqrt{\mathrm{det}\left(g_{AB} \frac{\partial x^A}{\partial \sigma^{\dot{\mu}}} \frac{\partial x^B}{\partial \sigma^{\dot{\nu}}}\right)}\pm \frac{i}{2} \varepsilon^{\dot{\mu}\dot{\nu}} \frac{\partial x^A}{\partial \sigma^{\dot{\mu}}} \frac{\partial x^B}{\partial \sigma^{\dot{\nu}}} A_{(2)\,AB}\right]\,,
\end{equation}
where the lower and upper signs stand for D1 and $\overline{\mathrm{D}1}$, respectively. We will choose our brane world-volume coordinates to wrap the $S^2$, so that $\sigma^{\dot{1}} = \theta$, $\sigma^{\dot{2}}=\varphi$. We also introduce coordinates on the $S^3$ for which the metric is
\begin{equation}
\mathrm{d}\Omega_3^2 = \mathrm{d}\tilde{\theta}^2+\sin^2\tilde{\theta} \left(\mathrm{d}\hat{\theta}^2 +\sin^2\hat{\theta}\mathrm{d}\tilde{\varphi}^2\right),
\end{equation}
with $\tilde{\theta},\hat{\theta}\in(0,\pi)$ and $\tilde{\varphi}\in(0,2\pi)$.

Within our symmetry class, we can write $A_{(2)}$ locally as
\begin{equation}
A_{(2)} = \frac{2\,i}{L} \lambda(r) \sin \theta\,\mathrm{d}\theta \wedge \mathrm{d}\varphi+L^2\tilde{\lambda}(\tilde{\theta})\sin \hat{\theta}\,\mathrm{d}\hat{\theta}\wedge \mathrm{d}\tilde{\varphi}\,,
\end{equation}
with
\begin{equation}
\frac{\mathrm{d}\lambda}{\mathrm{d}r} = \frac{g(r)}{\sqrt{f(r)}}\,,\quad\text{and}\quad \frac{\mathrm{d}\tilde{\lambda}}{\mathrm{d}\tilde{\theta}}=\sin^2\tilde{\theta}\,.
\end{equation}
One can thus write $S^{\mathrm{D1}}_{E}$ as
\begin{equation}
S^{\mathrm{D1}}_{\pm\,E} =4 \pi  \left[g(r)\mp\frac{2}{L}\lambda (r)\right]\,.
\end{equation}
To simplify our analysis, we take as a boundary condition $\lambda(0)=0$ for both the wormholes and disconnected solutions. Using these boundary conditions, it is easy to check that $\lambda$ is an odd function of $r$. Thus, if we check that $S^{\mathrm{D1}}_{E}>0$ for the upper sign and for all values of $r$, we automatically guarantee positivity for the lower sign as well.

Even without actually solving the equations of motion for all values of $r$, we can obtain useful information by studying the asymptotic behavior of solutions. We first map everything to Fefferman-Graham coordinates, defined via
\begin{equation}
\frac{\mathrm{d}r}{\mathrm{d}z}=-\sqrt{f(r(z))}\frac{L}{z}\,
\end{equation}
with $r\,z = L^2$ as $z\to0$.
It is also useful to define
\begin{equation}
g = \frac{L^2}{z^2}G\,,
\end{equation}
in terms of which the equation for $\lambda$ becomes:
\begin{equation}
\frac{\mathrm{d}\lambda}{\mathrm{d}z}=-\frac{L^3}{z^3}G\,.
\label{eq:lambda}
\end{equation}

Solving the resulting equations of motion asymptotically yields
\begin{subequations}
\begin{align}
& G=1+\frac{1}{2} \left(V^2-1\right)z^2 +\mathcal{O}\left(z^4\right),
\\
&\psi = V+(\lambda^{(2)}+V \log z)z^2+\mathcal{O}\left(z^4\right),
\end{align}
\end{subequations}
where $\lambda^{(2)}$ is a constant. Integrating (\ref{eq:lambda}) then gives
\begin{equation}
\lambda = \frac{L^3}{2z^2}-\frac{L^3}{2}(V^2-1)\log z+\mathcal{O}(1)\,,
\end{equation}
which in turn yields
\begin{equation}
S^{\mathrm{D1}}_{+\,E} = L^2(V^2-1) \log z+\mathcal{O}(1)\,.
\end{equation}

This result suggests that any solution with $V>1$ will be unstable to spontaneously nucleating a D1 brane. The question is then whether we can find any wormhole solution with $V<1$. If so, we must then further check to see whether the probe brane action on the resulting background is positive definite for all $z$ (and not just near $z=0$).
\subsection{The disconnected phase}
We introduce radial coordinates, for which $g$ in (\ref{eq:typeIIbgen}) is given by
\begin{equation}
g(r) = r^2\,.
\end{equation}
To work with compact coordinates, we also introduce a new coordinate $y\in(0,1)$ so that
\begin{equation}
r = L \frac{y\sqrt{2-y^2}}{1-y^2}\,,
\end{equation}
with $y=1$ being the location of the conformal boundary and $y=0$ the centre where the $S^2$ shrinks to zero size. Regularity at $r=y=0$ demands that
\begin{equation}
\psi(r)=\mathcal{O}(r)=\mathcal{O}(y)\,.
\end{equation}
To sum up, we take the \emph{Ansatz}
\begin{equation}
\mathrm{d}s^2 = \frac{L^2}{(1-y^2)^2}\left[\frac{4 \mathrm{d}y^2}{\tilde{f}(y)\,(2-y^2)}+y^2(2-y^2) \mathrm{d}\Omega_2^2\right]\quad \text{and}\quad \psi = y\sqrt{2-y^2}\,q(y)\,,
\end{equation}
where have translated the $f(r)$ in (\ref{eq:typeIIbgen}) into a function $\tilde{f}(y) = f(r(y))$. The Einstein and Klein-Gordon equations yield
\begin{subequations}
\begin{align}
&\tilde{f}(y)=\frac{8-y^2 \left(2-y^2\right) \left(1-y^2\right)^2 \left[2 \left(1-y^2\right) q(y)+y \left(2-y^2\right) q^\prime(y)\right]^2}{8\left[1-y^2 \left(2-y^2\right) \left(1-y^2\right)^2 q(y)^2\right]}\,,\label{eq:f}
\\
&\frac{1-y^2}{y \sqrt{\tilde{f}(y)}}\left[\frac{y^2 \left(2-y^2\right)^{3/2}}{\sqrt{\tilde{f}(y)} \left(1-y^2\right)}\left(y \sqrt{2-y^2} q(y)\right)^\prime\right]^\prime-8\,q(y)=0\,.\label{eq:scalar}
\end{align}
\end{subequations}
Note that Eq.~(\ref{eq:scalar}) is a second order differential equation for $q$, since $\tilde{f}$ is given in terms of $q$ and $q^\prime$ in Eq.~(\ref{eq:f}). As boundary conditions, we take $q^\prime(0)=0$ (which follows from regularity at $y=0$) and $q(1)=V$\,.

\subsection{The wormhole phase}
For the wormhole phase, we take $g$ in (\ref{eq:typeIIbgen}) to have the form
\begin{equation}
g(r)=r^2+r_0^2
\end{equation}
with $r_0$ denoting the wormhole radius. As we shall see, it will correspond to the minimum size of the $S^2$. We change coordinates to
\begin{equation}
r = r_0 \frac{\sqrt{y}\sqrt{2-y}}{1-y}\,,
\end{equation}
where the $\mathbb{Z}_2$ symmetry plane of the wormhole solution is identified with $y=0$, and the conformal boundary is located at $y=1$. It is also convenient to define $r_0 \equiv y_0 L$\,. The relevant \emph{Ansatz} now reduces to
\begin{equation}
\mathrm{d}s^2 = \frac{L^2}{(1-y)^2}\left[\frac{\hat{f}(y)\,\mathrm{d}y^2}{(2-y)y}+y_0^2 \mathrm{d}\Omega_2^2\right]\,,\quad \text{and}\quad \psi = q(y)\,,
\end{equation}
where, once again, we rewrite $f$ in (\ref{eq:typeIIbgen}) in terms of $\hat{f}(y)$.

The Einstein equation and Klein gordon equations now yield
\begin{subequations}
\begin{align}
& \hat{f}(y) = \frac{y\,(2-y)\,y_0^2\,\left[2-(1-y)^2 q'(y)^2\right]}{2 \left[y_0^2-(1-y)^2 \left(q(y)^2-1\right)\right]}\,,
\\
& \frac{\sqrt{y}\sqrt{2-y}}{\sqrt{\hat{f}(y)}}\left[\frac{\sqrt{y}\sqrt{2-y}}{(1-y)\sqrt{\hat{f}(y)}}q^\prime(y)\right]^\prime-\frac{2 q(y)}{y_0^2 (1-y)}=0\,. \label{eq:qw}
\end{align}
\end{subequations}

Reflection symmetry around $y=0$, and smoothness of the corresponding solution imply
\begin{equation}
q(0) = \sqrt{1+y_0^2}\,,\quad \text{and}\quad q^\prime(0) = 2\frac{\sqrt{1+y_0^2}}{y_0^2}\,.
\end{equation}

At the conformal boundary, we wish to set $q(1)=V$. The strategy is now simple enough: we take Eq.~(\ref{eq:qw}) and integrate it all the way to $y=1$, where we read off $V$ for any value of $y_0$ we choose. We found it convenient to preform this integration using an implicit fourth order Runge-Kutta method.

\subsection{Results}
There are a couple of surprising results in this setup. First, we only find one branch of wormhole solutions, which we coin as large since it extends to arbitrarily large values of the source $V$. Nevertheless wormholes only seem to exist for $V\geq V^{\min}\approx 3.14$. These results can be seen on the left hand side panel of  Fig.~\ref{fig:typeIIbscalars} where we plot $y_0$ as a function of $V$. The fact that $V>1$ in order for the wormhole solution to exists immediately reveals that the wormholes we found suffer from nucleation instabilities associated with D$1$ and $\bar{\mathrm{D}}1$ branes. Finally, on the right hand side panel of Fig.~\ref{fig:typeIIbscalars} we plot the different in on-shell action
\begin{equation}
\Delta S_{3d}=2 S_{3d}^D-S_{3d}^W
\end{equation}
where $S_{3d}^D$ and $S_{3d}^W$ are the three-dimensional on-shell actions for the disconnected and wormhole solutions (respectively). We can see that the wormhole solution dominates over the disconnected solution for $V\geq V^{\mathrm{HP}}\approx  3.7373(8)$.
\begin{figure}[h]
\centering
\includegraphics[scale=0.4]{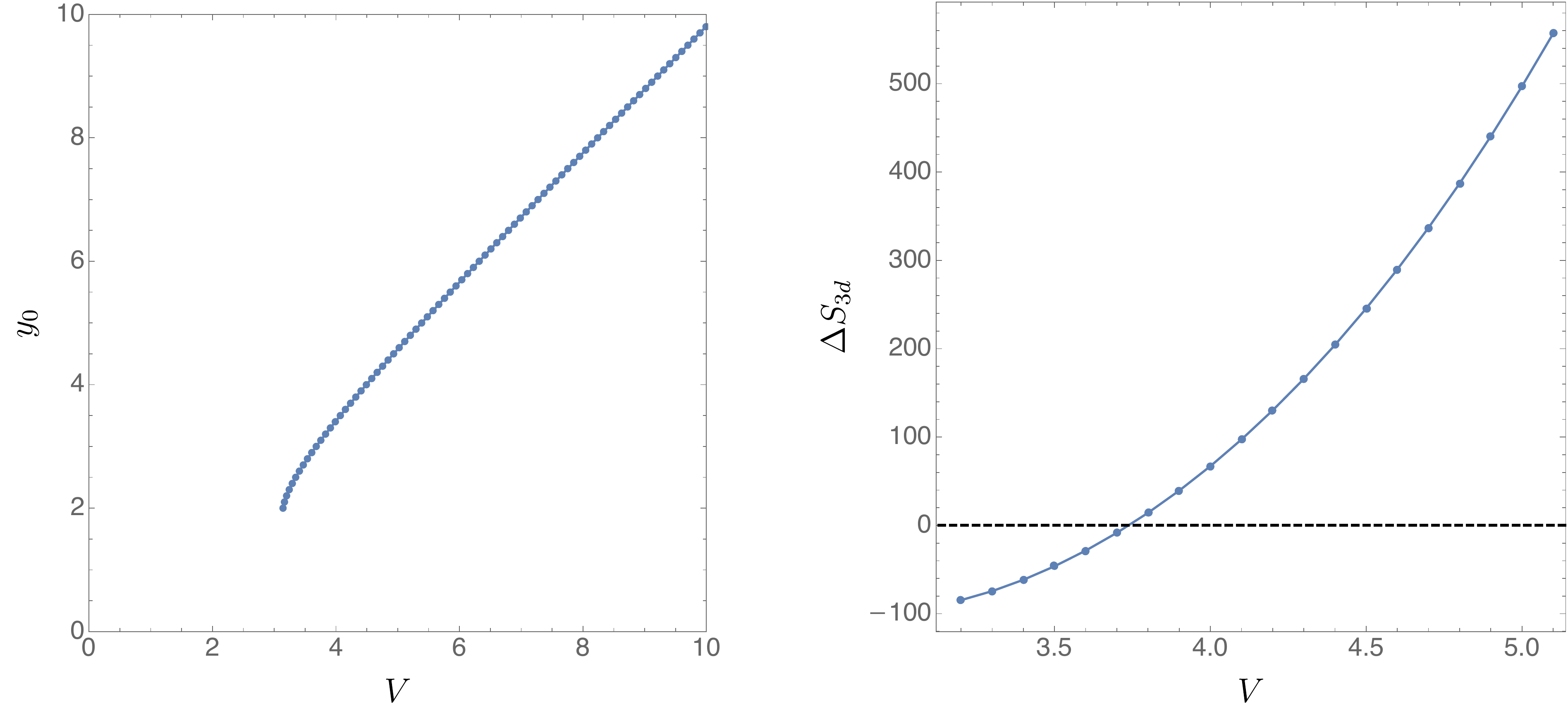}\\
\caption{{\bf Left panel:} radius of the wormhole solutions as a function of the source $V$. Wormholes only exists for $V>V^{\min}\approx 3.14$. {\bf Right panel:} difference in the Euclidean action $\Delta S_{3d}=2 S_{3d}^D-S_{3d}^W$ as a function of $V$. The wormhole solutions have a lower action for $V>V^{\mathrm{HP}}\approx 3.7373(8)$. Unlike the higher-dimensional examples, we find no evidence for a small wormhole branch.}
\label{fig:typeIIbscalars}
\end{figure}

We also studied the field theoretical stability of the wormhole solutions. The analysis is completely analogous to section \ref{sec:negascalarheading}, except it is easier because there are no tensor harmonics on $S^2$ and vectors harmonics can be obtained from scalar harmonics via the Hodge operation on gradients of the scalar harmonics. We found no negative modes, irrespectively of the value of $V$.

\section{Discussion}
\label{sec:disc}

Our work above explored the construction and stability of asymptotically anti-de Sitter Euclidean wormholes in a variety of models.  Indeed, we have studied many more models than were described above, but for brevity we limited our presentation to a few representative cases with spherical (or squashed sphere) boundaries.  A few low energy models with torus boundaries are discussed in appendix \ref{app:torusboundaries}, and a table showing a longer list of 22 string/M-theory compactifications and 14 ad hoc low energy models  and associated results obtained in a variety of dimensions is presented in appendix \ref{app:table}.  While not all issues were analyzed for all models in the table, we hope it will nevertheless be of use in guiding future studies.  Perhaps notably, our list does not include the model described in section 5 of \cite{Maldacena:2004rf} where the disconnected solution remains to be found in order to determine if the wormhole described there will dominate.

In simple ad hoc low-energy models, it was straightforward to find two-boundary Euclidean wormholes that dominate over disconnected solutions and which are stable (lacking negative modes) in the usual sense of Euclidean quantum gravity.  Similar results were found previously in the context of JT gravity coupled to matter \cite{Garcia-Garcia:2020ttf} and in studies of constrained wormholes in pure gravity \cite{Cotler:2020lxj}. In particular, resulting phase diagram was a direct analogue of the Hawking-Page phase transition for AdS-Schwarzschild black holes in which, for  boundary sources above some threshold, we find both a `large' and a `small' branch of wormhole solutions with the latter being stable and dominating over the disconnected solution for large enough sources.

We also studied two-boundary Euclidean wormholes in a variety of string and M-theory compactifications.  At first glance the solutions are generally similar to those in the ad hoc models, and we find several contexts where large wormholes dominate over the disconnected solutions we find and where the large wormholes are stable with respect to field-theoretic perturbations.  However,  wormholes in these UV-complete settings that are large enough to dominate over our disconnected solution always suffer from brane-nucleation instabilities (even when sources that one might hope would stabilize such instabilities are tuned to large values).  This implies the existence of additional solutions with lower action.  It is natural to expect that the lowest-action such solutions are again disconnected, but this remains to be studied in detail.  Including finite back-reaction from such branes would be a natural next step in understanding wormhole solutions in string theory.

The overall picture of UV-complete models is thus rather similar to that obtained by Maldacena and Maoz \cite{Maldacena:2004rf} with the following exceptions.  First, we have performed a thorough analysis of potential field-theoretic negative modes and shown our large wormholes to be free of such pathologies.  Second, we have identified subdominant wormholes that are free of both field-theoretic and brane-nucleation instabilities.  In particular, this was the case for  the $U(1)^2$ large wormholes with round boundaries and $4.352(8) \approx  \Phi_0^{M_2} > \Phi_0>\Phi_0^{\min} \approx 4.0162(5)$  described section \ref{sec:U12results}, though in other models  subdominant wormholes can suffer from brane nucleation instabilities as well (see e.g. section \ref{sec:IIB} and the discussion of general squashed boundaries in section \ref{sec:U12results}).  Third, section \ref{sec:massdeformed} investigated finite values of sources that provide mass-deformations of the form that were suggested in \cite{Maldacena:2004rf} to stabilize theories against brane-nucleation.  As predicted by \cite{Maldacena:2004rf}, this does in fact stabilize the theory in the sense that it removes the UV brane-nucleation instability near the asymptotically-AdS boundaries.  One thus expects that the disconnected solution is fully stabilized at such values of the deformation parameter.  However, we were able to find wormhole solutions only when the deformation is small enough that the UV remains unstable.   This is perhaps the strongest evidence yet that, at least without taking parameters to exponentially large values (see below), Euclidean wormholes will not dominate partition functions in UV-complete theories.

The interesting question is of course what such results imply for the AdS/CFT factorization problem described in the introduction.    To begin this discussion, we note that a brane-nucleation instability is really the statement that adding a brane to the given solution will lower the action.  Since branes are discrete, if these are the only ``instabilities'' this means that sufficiently small fluctuations around the solution must in fact {\it increase} the action.  So in a technical sense the wormhole saddles we found in the string-compactified models are in fact stable.  The point is simply that the wormholes constructed thus far will be sub-dominant saddles and will not control leading-order effects.

This result is natural even if one believes that bulk AdS gravity in UV-complete theories should be dual to an ensemble of quantum theories.  Had we found a case where a simple semi-classical wormhole dominates the computation of a partition function, this would have indicated that ensemble-fluctuations of that partition function are large, or at least that they are not particularly small when compared with its ensemble expectation-values\footnote{We thank Henry Maxfield for discussions on this point.}.  In other words, it would have implied that an ensemble dual to bulk string theory is not sharply peaked in the semi-classical limit.    On the other hand, evidence to date suggests that ensembles associated with quantum gravity are generally peaked very sharply indeed.  This is the case whether one looks at the ensembles associated with low-dimensional gravitational wormholes \cite{Saad:2018bqo,Saad:2019lba,Stanford:2019vob,Marolf:2020xie,Stanford:2020wkf,Marolf:2020rpm} or at fluctuations in ensembles \cite{Chamblin:1999hg} of states associated with black hole interiors\footnote{And indeed, these may be closely related \cite{Marolf:2020rpm}.}.
Indeed, the fluctuations associated with the double cone solution \cite{Saad:2018bqo} are visible only when one probes the fine structure of the associated ensemble by studying exponentially large times.  One might similarly expect that more standard partition functions become dominated by wormholes only at exponentially large sources, which is a regime that we have certainly not probed (and which may involve additional UV physics due large field values and strong gradients)! It is interesting that UV-complete models appear to generally achieve this expectation while ad hoc low-energy models often admit exceptions.

What then are the implications of our (sub-dominant) wormhole saddles in the string-compactified models?  We emphasize that they are in fact stable with respect to sufficiently small perturbations (within the truncations studied), so that they cannot be immediately dismissed.  This is in particular a technical advance beyond the analysis performed in \cite{Maldacena:2004rf}.

While in general the semi-classical approximation leads to a sum over all (stable) Euclidean saddles, in many contexts it would be dangerous to draw conclusions about physics based on sub-dominant saddles.  This is simply because the effects of sub-dominant saddles are small, and so in many cases can be easily dwarfed by even small corrections to the physics of dominant saddles.

However, the present context appears to be rather different.  In studying quantities like $\delta Z^2 : = \langle Z^2\rangle - \langle Z \rangle^2$ the contribution of any disconnected saddle will vanish identically, and without error.  So only contributions from connected saddles remain.  Unless such contributions fully cancel amongst themselves, $\delta Z^2$ will be non-zero.  Though we are certainly not able to analyze the possibility of a conspiracy that might enforce such cancellations, even in the string compactifications we studied the most naive interpretation of our sub-dominant wormhole saddles remains that they will make $\delta Z^2$ non-zero and require an ensemble of dual field theories.   This is of course also the picture implied by the dominance that of the double-cone wormholes of \cite{Saad:2018bqo} that appears to hold in late time computations of the spectral form factor in generic models\footnote{Though it would be interesting to further investigate higher-dimensional double cones in detail}.  The factorization problem of AdS/CFT thus remains far from resolved and will surely be the object of much future investigation.

\section*{Acknowledgments}
We thank Xi Dong, Jerome Gauntlett, Juan Maldacena, Henry Maxfield, Douglas Stanford and Edward Witten for useful conversations.
We also  thank the Kavli Institute for Theoretical Physics for its hospitality during a portion of this work. As a result, this research was supported in part by the National Science Foundation under Grant No. NSF PHY-1748958. D.~M. was primarily supported by NSF grant PHY-1801805 and funds from the University of California.  J.~E.~S. is supported in part by grant ST/T000694/1. J.~E.~S. was also supported by a J.~Robert Oppenheimer~Visiting~Professorship at the Institute for Advanced Studies in Princeton.
\appendix
\section{Symmetric matrices for the Einstein-$U(1)^3$ wormholes}
\subsection{\label{sec:crazy}The symmetric Matrix $\mathbb{V}$ for the scalars}
The symmetric matrix $\mathbb{V}$ is given by
\begin{equation}
 \mathbb{V}_{IJ}=(\mathbb{K}^{-1})_{IK}(\mathbb{K}^{-1})_{JM}\mathbb{V}^{KM}\,.
\end{equation}
We then have
\begin{equation}
\mathbb{V}_{22}=\frac{(\ell _S+1)}{64 \pi ^2 g^2 (\lambda _S-3) \lambda _S} \left[g^2 \left(\lambda _S-2\right) \lambda _S+24 \Phi ^2 \left(g L^2 \lambda _S+4 r_0^2L^2+4r_0^4\right)\right]\,.
\end{equation}
The remaining two components take a more complicated form.

For $\mathbb{V}_{11}$ we find
\begin{subequations}
\begin{equation}
\mathbb{V}_{11}=\frac{g \left(\ell _S+1\right)}{32 \pi ^2 L^2 r^3 f \left(\lambda _S-3\right) \left(g \lambda _S+24 L^2 \Phi ^2\right)^2}\left[\sum_{i=0}^3 \Phi^{2i}p_{11}^{(i)}(r)+\Phi\; \Phi^\prime\sum_{i=0}^2 \Phi^{2i}m_{11}^{(i)}(r)\right]\,,
\end{equation}
where
\begin{equation}
p^{(3)}_{11}(r)=-\frac{192 L^8 r \left(\lambda _S-4\right) \left(\lambda _S-3\right)}{g}\,,
\end{equation}
\begin{multline}
p^{(2)}_{11}(r)=-\frac{8 L^4 r}{g} \Big[6 g^2 (\lambda _S-4) (\lambda _S-3)+g L^2 (\lambda _S^3-14 \lambda _S^2+72)
\\
-6 g r^2 (\lambda _S-4) \lambda _S+3 r^2 (r^2-L^2)(\lambda _S-4) \lambda _S\Big]\,,
\end{multline}
\begin{multline}
p^{(1)}_{11}(r)=r \lambda _S \Big\{L^2 \left[2 g r^2 \lambda _S \left(2 \lambda _S-13\right)-2 g^2 \left(\lambda _S^2-14 \lambda _S+42\right)+r^4 \left(-2 \lambda _S^2+\lambda _S+12\right)\right]
\\+L^4 \left[g \left(\lambda _S^2-2 \lambda _S-60\right)+r^2 \left(2 \lambda _S^2-\lambda _S-12\right)\right]+24 g r_0^4 \left(\lambda _S-1\right)\Big\}\,,
\end{multline}
\begin{multline}
p^{(0)}_{11}(r)=-\frac{r \lambda _S^2}{8 L^2} \Big\{L^4 \left(\lambda _S^2+3 \lambda _S-18\right)-g^2 \left[12 L^2 r_0^2 \left(\lambda _S+1\right)+12 r_0^4 \left(\lambda _S+1\right)\right]
\\
-g^3 L^2 \left(\lambda _S^2+3 \lambda_S-18\right)+g L^2 r_0^2 \left(L^2+r_0^2\right) \left(\lambda _S^2-3 \lambda _S-18\right)+4 r_0^4 \left(L^2+r_0^2\right)^2 \left(\lambda _S+1\right)\Big\}\,,
\end{multline}
\begin{equation}
m_{11}^{(2)}(r)=-384 L^6 \left(2 g+L^2\right) \left(\lambda _S-3\right)\,,
\end{equation}
\begin{multline}
m_{11}^{(1)}(r)=-8 L^2 \lambda _S \Big\{3 g^2 L^2 \left(\lambda _S-3\right)+g \left[12 L^2 r_0^2-\left(\lambda _S-6\right)L^4+12 r_0^4\right]
\\
+5 L^2 r_0^2 \left(L^2+r_0^2\right) \left(\lambda _S-3\right)\Big\}\,,
\end{multline}
and
\begin{multline}
m_{11}^{(0)}(r)=\lambda _S^2 \Big[3 g^3 L^2 \left(\lambda _S+2\right)+3 g^2 L^4 \lambda _S+3 g^4-g L^2 r_0^2 \left(L^2+r_0^2\right) \left(3 \lambda _S-4\right)
\\
+r_0^4 \left(L^2+r_0^2\right)^2\Big]\,.
\end{multline}
\end{subequations}

While for $\mathbb{V}_{12}$ we have
\begin{subequations}
\begin{equation}
\mathbb{V}_{12}=\frac{(\ell _S+1)}{32 \pi ^2 L^2 r^2 f \left(\lambda _S-3\right) \left(g \lambda _S+24 L^2 \Phi ^2\right)^2}\left[\sum_{i=0}^3 \Phi^{2i}p_{12}^{(i)}(r)+\Phi\; \Phi^\prime\sum_{i=0}^2 \Phi^{2i}m_{12}^{(i)}(r)\right]\,,
\end{equation}
where
\begin{equation}
p_{12}^{(3)}(r)=2304 L^6 \left(2 g+L^2\right)\,,
\end{equation}
\begin{equation}
p_{12}^{(2)}(r)=-96 L^2 \Big\{g^2 L^2 (\lambda _S+6)+g \left[12 L^2 r_0^2+L^4 (\lambda _S+6)+12 r_0^4\right]
-L^2r_0^2(L^2+r_0^2) \lambda _S\Big\}\,,
\end{equation}
\begin{multline}
p_{12}^{(1)}(r)=-4 \lambda _S \Big\{15 g^3 L^2-g^2 \left[6 r_0^2 (L^2+r_0^2)+L^4 (\lambda _S-15)\right]
\\+g L^2 r_0^2 (L^2+r_0^2) (2 \lambda _S-21)+6 r_0^4 (L^2+r_0^2)^2\Big\}\,,
\end{multline}
\begin{multline}
p_{12}^{(0)}(r)=\frac{g \lambda _S^2}{2 L^2} \Big\{g^2 \left[6 L^2 r_0^2+L^4 (\lambda _S-3)+6 r_0^4\right]+g^3 L^2 (\lambda _S-3)
\\
-g L^2 r_0^2 (L^2+r_0^2) (\lambda _S-7)-2 r_0^4(L^2+r_0^2)^2\Big\}\,,
\end{multline}
\begin{equation}
m_{12}^{(2)}(r)=-1152 f L^8 r\,,
\end{equation}
\begin{equation}
m_{12}^{(1)}(r)=144 f^2 L^6 r^3 \lambda_S\,,
\end{equation}
and
\begin{equation}
m_{12}^{(0)}(r)=2 g r \lambda _S^2 \left[g^2 (7 L^2+3 r_0^2)+3 g^3+g (4 L^2-3 r_0^2) (L^2+r_0^2)-3 r_0^2 (L^2+r_0^2)^2\right]\,.
\end{equation}
\end{subequations}

\subsection{\label{sec:b}The symmetric Matrix $\mathbb{V}$ for the tensors}
The symmetric matrix $\mathbb{V}$ is given by
\begin{equation}
 \mathbb{V}_{IJ}=(\mathbb{K}^{-1})_{IK}(\mathbb{K}^{-1})_{JM}\mathbb{V}^{KM}\,.
\end{equation}
We then have
\begin{subequations}
\begin{multline}
\mathbb{V}_{11}=\frac{64 \pi ^2}{25 g^2 L^2} \Big\{25 g L^2 (m-1) m+[6 m (3 m+4)-17] r_0^2 \left(L^2+r_0^2\right)
\\
-24 L^4 (m-1) (3 m+7) \Phi ^2\Big\}\,,
\end{multline}
\begin{multline}
\mathbb{V}_{12}=-\frac{32 \sqrt{2} \pi ^2 (m-1)}{125 g^2   L^2 (6 m-1) }\Big\{25 g L^2 \left[8 (1-6 m) \Phi +17 m (m-1)^2\right]+
\\
17 (m-1) \left[8 L^4 \left(9 m^2-3 m+19\right) \Phi ^2+(6m-18 m^2-13) r_0^2 \left(L^2+r_0^2\right)\right]\Big\}
\end{multline}
\begin{multline}
\mathbb{V}_{22}=\frac{32 \pi ^2 (m-1)^2}{625 L^4 (g-6 g m)^2} \Big\{850 f g L^4 (1-6 m)^2 r \Phi '+1250 g^2 L^2 (1-6 m)^2
\\
+25 g \Big[36 L^2 (1-6 m)^2 r_0^2-144 (1-6 m)^2 \Phi ^2 L^4-340 (12 m^2 -8m+1) \Phi L^4+289 m (m-1)^3L^4
\\
+36 (1-6 m)^2 r_0^4\Big]+289 L^2 (m-1)^2 \left[(18 m^2-36 m-7) r_0^2 \left(L^2+r_0^2\right)-8 L^4 (3 m-8) (3 m+2) \Phi ^2\right]\Big\}
\end{multline}
\end{subequations}

\section{\label{eq:labc}Symmetric matrices for the scalars}
The symmetric matrix $\mathbb{V}$ is given by
\begin{equation}
 \mathbb{V}_{IJ}=(\mathbb{K}^{-1})_{IK}(\mathbb{K}^{-1})_{JM}\mathbb{V}^{KM}\,.
\end{equation}
We then have
\begin{equation}
 \mathbb{V}_{IJ}=\frac{1}{f \Pi ^2 y_0^4 n_y m_y^{2 \Delta } \, \left(m_y \Pi'^2+\lambda _S-3\right) \,d_{IJ}}\sum_{i=0}^{9}{\Pi^\prime}^i \mathbb{V}_{IJ}^{(i)}\,.
\end{equation}
with $\mathbb{V}_{11}^{(9)}=\mathbb{V}_{22}^{(9)}=0$, $n_y=y(2-y)$, $m_y=1-y$ and
\begin{align}
& d_{11}=\left(m_y \Pi'^2+\lambda _S-3\right)^2\,,
\\
& d_{12}=\left(m_y \Pi '^2+\lambda _S-3\right) \left[\Pi ^2 \left(m_y \Pi '^2-3\right)-m_y \Pi '^2-\lambda _S+3\right]\,,
\\
& d_{22}=\left(\lambda _S-3\right) \left[\Pi ^2 \left(m_y \Pi'^2-3\right)-m_y \Pi'^2-\lambda _S+3\right]\,.
\end{align}
Furthermore,
\begin{equation}
\mathbb{V}^{(8)}_{11}=y_0^4 \left[\Delta  \Pi ^4-(\Delta -1) \Pi ^2+1\right)]m_y n_y^2\,,
\end{equation}
\begin{equation}
\mathbb{V}^{(7)}_{11}=2 \Pi ^3 y_0^4 m_y n_y^2 \left(\Delta +\lambda _S\right)\,,
\end{equation}
\begin{multline}
\mathbb{V}^{(6)}_{11} =\Pi ^6 \left(2 f y_0^2 m_y^2 n_y \lambda _S-\Delta  f y_0^2 m_y^2 n_y-4 f y_0^2 m_y^2 n_y\right)
\\
+ \Pi ^4 \Big(2 \Delta  f y_0^2 m_y^2 n_y-3 f y_0^2 m_y^2 n_y \lambda _S+8 f y_0^2 m_y^2 n_y+4 \Delta  y_0^4 m_y n_y^2 \lambda _S-y_0^4 m_y n_y^2 \lambda _S^2
\\
+3 y_0^4 m_y n_y^2 \lambda _S+\Delta ^2 y_0^4 m_y n_y^2-12 \Delta  y_0^4 m_y n_y^2\Big)+\Pi ^2 \Big(f y_0^2 m_y^2 n_y \lambda _S-\Delta  f y_0^2 m_y^2 n_y
\\
-3 f y_0^4 m_y n_y-7 f y_0^2 m_y^2 n_y-3 \Delta  y_0^4 m_y n_y^2 \lambda _S+4 y_0^4 m_y n_y^2 \lambda_S-\Delta ^2 y_0^4 m_y n_y^2+12 \Delta  y_0^4 m_y n_y^2
\\
-6 y_0^4 m_y n_y^2\Big)+3 f y_0^4 m_y n_y+3 f y_0^2 m_y^2n_y+3 y_0^4 m_y n_y^2 \lambda _S-12 y_0^4 m_y n_y^2\,,
\end{multline}
\begin{multline}
\mathbb{V}^{(5)}_{11}=\Pi ^3 \left(4 f y_0^2 m_y^2 n_y \lambda _S+6 \Delta  y_0^4 m_y n_y^2 \lambda _S-6 y_0^4 m_y n_y^2 \lambda _S-18 \Delta  y_0^4 m_y n_y^2\right)
\\
-2 f \Pi^5 y_0^2 m_y^2 n_y \lambda _S\,,
\end{multline}
\begin{multline}
\mathbb{V}^{(4)}_{11} = \Pi ^4 (7 \Delta  f y_0^2 m_y^2 n_y \lambda _S-2 f^2 m_y^2 \lambda _S-6 f y_0^2 m_y^2 n_y \lambda _S^2+47 f y_0^2 m_y^2 n_y \lambda _S-18 \Delta  f y_0^2 m_y^2 n_y
\\
-72 f y_0^2 m_y^2 n_y+2 \Delta ^2 y_0^4 m_y n_y^2 \lambda _S+3 \Delta  y_0^4 m_y n_y^2 \lambda _S^2-30 \Delta  y_0^4 m_y n_y^2 \lambda _S+6 y_0^4 m_y n_y^2 \lambda _S^2-18 y_0^4 m_y n_y^2 \lambda _S
\\
-9 \Delta ^2 y_0^4 m_y n_y^2+54 \Delta  y_0^4 m_y  n_y^2)+\Pi ^6 (4 f^2 m_y^2 \lambda _S-4 \Delta  f y_0^2 m_y^2 n_y \lambda _S+4 f y_0^2 m_y^2 n_y \lambda _S^2-26 f y_0^2 m_y^2 n_y \lambda _S
\\
+9 \Delta  f y_0^2 m_y^2 n_y+36 f y_0^2 m_y^2 n_y)-2f^2 \Pi ^8 m_y^2 \lambda _S+\Pi ^2 (3 f y_0^2 m_y^2 n_y \lambda _S^2-3 \Delta  f y_0^2 m_y^2 n_y \lambda _S
\\
-21 f y_0^2 m_y^2 n_y \lambda _S+9 \Delta  f y_0^2 m_y^2 n_y+27 f y_0^4 m_y n_y+63 f y_0^2 m_y^2 n_y-3 \Delta ^2 y_0^4 m_y n_y^2 \lambda _S-3 \Delta  y_0^4 m_y n_y^2 \lambda _S^2
\\
+27 \Delta  y_0^4 m_y n_y^2 \lambda _S+2 y_0^4 m_y n_y^2 \lambda _S^2-15 y_0^4 m_y n_y^2 \lambda _S+9 \Delta ^2 y_0^4 m_y n_y^2-54\Delta  y_0^4 m_y n_y^2)+9 f y_0^4 m_y n_y \lambda _S
\\
+9 f y_0^2 m_y^2 n_y \lambda _S-27 f y_0^4 m_y n_y-27 f y_0^2 m_y^2 n_y+3 y_0^4 m_y n_y^2 \lambda _S^2-27 y_0^4 m_y n_y^2 \lambda _S+54 y_0^4 m_y n_y^2\,,
\end{multline}
\begin{multline}
\mathbb{V}^{(3)}_{11} = \Pi ^3 (6 \Delta  f y^2 y_0^2 m_y n_y \lambda _S-6 f^2 y_0^2 m_y^2 \lambda _S-6 f^2 m_y^3 \lambda _S+6 f y^2 y_0^2 m_y n_y \lambda _S
\\
+6 \Delta  f y_0^4 m_y n_y \lambda _S+6 \Delta  f y_0^2 m_y n_y \lambda_S-12 \Delta  f y y_0^2 m_y n_y \lambda _S-6 f y_0^4 m_y n_y \lambda _S^2
\\
+18 f y_0^4 m_y n_y \lambda _S+6 f y_0^2 m_y n_y \lambda _S-12 f y y_0^2 m_y n_y \lambda _S+4 \Delta  y_0^4 m_y n_y^2 \lambda _S^2
\\
-36\Delta  y_0^4 m_y n_y^2 \lambda _S+6 y_0^4 m_y n_y^2 \lambda _S^2-18 y_0^4 m_y n_y^2 \lambda _S+54 \Delta  y_0^4 m_y n_y^2)
\\
+\Pi ^5 (6 f^2 m_y^3 \lambda _S+6 f^2 y_0^2 m_y^2 \lambda _S)\,,
\end{multline}
\begin{multline}
\mathbb{V}^{(2)}_{11}=(9 f^2 m_y^2 \lambda _S-3 f^2 m_y^2 \lambda _S^2) \Pi ^8+(6 f^2 \lambda _S^2 m_y^2-108 f n_y y_0^2 m_y^2-27 f \Delta  n_y y_0^2 m_y^2
\\
-16 f n_y y_0^2 \lambda _S^2 m_y^2-3 f \Delta  n_y y_0^2 \lambda_S^2 m_y^2-18 f^2 \lambda _S m_y^2+84 f n_y y_0^2 \lambda _S m_y^2
\\
+18 f \Delta  n_y y_0^2 \lambda _S m_y^2) \Pi ^6+(27 \Delta ^2 m_y n_y^2 y_0^4-108 \Delta  m_y n_y^2 y_0^4+\Delta ^2 m_y n_y^2 \lambda_S^2 y_0^4
\\
-12 \Delta  m_y n_y^2 \lambda _S^2 y_0^4-9 m_y n_y^2 \lambda _S^2 y_0^4-12 \Delta ^2 m_y n_y^2 \lambda _S y_0^4+72 \Delta  m_y n_y^2 \lambda _S y_0^4
\\
+27 m_y n_y^2 \lambda _S y_0^4-f m_y^2 n_y \lambda _S^3 y_0^2+40 f m_y^2 n_y \lambda _S^2 y_0^2+6 f \Delta  m_y^2 n_y \lambda _S^2 y_0^2
\\
+216 f m_y^2 n_y y_0^2+54 f \Delta  m_y^2 n_y y_0^2-183 f m_y^2 n_y \lambda _S y_0^2-36 f \Delta  m_y^2 n_y \lambda _S y_0^2-3f^2 m_y^2 \lambda _S^2
\\
+9 f^2 m_y^2 \lambda _S) \Pi ^4+(27 \Delta ^2 m_y n_y^2 y_0^4-\Delta  m_y n_y^2 \lambda _S^3 y_0^4+108 \Delta  m_y n_y^2 y_0^4+54 m_y n_y^2 y_0^4
\\
-3 \Delta ^2 m_y n_y^2 \lambda_S^2 y_0^4+18 \Delta  m_y n_y^2 \lambda _S^2 y_0^4-3 f m_y n_y \lambda _S^2 y_0^4-81 f m_y n_y y_0^4
\\
+18 \Delta ^2 m_y n_y^2 \lambda _S y_0^4-81 \Delta  m_y n_y^2 \lambda _S y_0^4-18 m_y n_y^2 \lambda _S y_0^4+36 f m_y n_y \lambda _S y_0^4
\\
+3 f m_y^2 n_y \lambda _S^3 y_0^2-33 f m_y^2 n_y \lambda _S^2 y_0^2-3 f \Delta  m_y^2 n_y \lambda _S^2 y_0^2-189 f m_y^2 n_y y_0^2
\\
-27 f \Delta  m_y^2 n_y y_0^2+135 f m_y^2 n_y \lambda_S y_0^2+18 f \Delta  m_y^2 n_y \lambda _S y_0^2) \Pi ^2-108 m_y n_y^2 y_0^4
\\
+81 f m_y n_y y_0^4+m_y n_y^2 y_0^4 \lambda _S^3+81 f m_y^2 n_y y_0^2-18 m_y n_y^2 y_0^4 \lambda _S^2+9 f m_y n_y y_0^4 \lambda_S^2
\\
+9 f m_y^2 n_y y_0^2 \lambda _S^2+81 m_y n_y^2 y_0^4 \lambda _S-54 f m_y n_y y_0^4 \lambda _S-54 f m_y^2 n_y y_0^2 \lambda _S\,,
\end{multline}
\begin{multline}
\mathbb{V}^{(1)}_{11}=\Pi ^3 (12 f^2 y_0^2 m_y^2 \lambda _S^2-36 f^2 y_0^2 m_y^2 \lambda _S+12 f^2 m_y^3 \lambda _S^2-36 f^2 m_y^3 \lambda _S
\\
-6 \Delta  f y_0^4 m_y n_y \lambda _S^2+18 \Delta  f y_0^4 m_y n_y \lambda _S-6 \Delta  fy_0^2 m_y^2 n_y \lambda _S^2
\\
+18 \Delta  f y_0^2 m_y^2 n_y \lambda _S-18 f y_0^4 m_y n_y \lambda _S^2+54 f y_0^4 m_y n_y \lambda _S-4 f y_0^2 m_y^2 n_y \lambda _S^3
\\
-6 f y_0^2 m_y^2 n_y \lambda _S^2+54 f y_0^2m_y^2 n_y \lambda _S+12 \Delta  y_0^4 m_y n_y^2 \lambda _S^2-54 \Delta  y_0^4 m_y n_y^2 \lambda _S
\\
+18 y_0^4 m_y n_y^2 \lambda _S^2-54 y_0^4 m_y n_y^2 \lambda _S+54 \Delta  y_0^4 m_y n_y^2)+\Pi ^5 (36 f^2 m_y^3 \lambda _S-12f^2 m_y^3 \lambda _S^2-
\\
12 f^2 y_0^2 m_y^2 \lambda _S^2+36 f^2 y_0^2 m_y^2 \lambda _S+6 f y_0^2 m_y^2 n_y \lambda _S^2-18 f y_0^2 m_y^2 n_y \lambda _S)\,,
\end{multline}
\begin{multline}
\mathbb{V}^{(0)}_{11} = (12 f m_y n_y \lambda _S^2 y_0^2+3 f \Delta  m_y n_y \lambda _S^2 y_0^2+108 f m_y n_y y_0^2+27 f \Delta  m_y n_y y_0^2
\\
-72 f m_y n_y \lambda _S y_0^2-18 f \Delta  m_y n_y \lambda _S y_0^2) \Pi ^6+(-27 \Delta ^2 n_y^2 y_0^4+81 \Delta  n_y^2 y_0^4
\\
-3 \Delta ^2 n_y^2 \lambda _S^2 y_0^4+9 \Delta  n_y^2 \lambda _S^2 y_0^4+18 \Delta ^2 n_y^2 \lambda _S y_0^4-54 \Delta  n_y^2 \lambda _S y_0^4
\\
+7 f m_y n_y\lambda _S^3 y_0^2+f \Delta  m_y n_y \lambda _S^3 y_0^2-66 f m_y n_y \lambda _S^2 y_0^2-12 f \Delta  m_y n_y \lambda _S^2 y_0^2
\\
-216 f m_y n_y y_0^2-54 f \Delta  m_y n_y y_0^2+207 f m_y n_y \lambda _S y_0^2+45 f\Delta  m_y n_y \lambda _S y_0^2) \Pi ^4
\\
+(3 \Delta  n_y^2 \lambda _S^3 y_0^4-\Delta ^2 n_y^2 \lambda _S^3 y_0^4+27 \Delta ^2 n_y^2 y_0^4-81 \Delta  n_y^2 y_0^4
\\
-81 n_y^2 y_0^4-9 f^2 \lambda _S^2y_0^4+9 \Delta ^2 n_y^2 \lambda _S^2 y_0^4-27 \Delta  n_y^2 \lambda _S^2 y_0^4-18 n_y^2 \lambda _S^2 y_0^4
\\
+27 f n_y \lambda _S^2 y_0^4+81 f n_y y_0^4+27 f^2 \lambda _S y_0^4-27 \Delta ^2 n_y^2 \lambda _Sy_0^4+81 \Delta  n_y^2 \lambda _S y_0^4
\\
+81 n_y^2 \lambda _S y_0^4-108 f n_y \lambda _S y_0^4+f m_y n_y \lambda _S^4 y_0^2-13 f m_y n_y \lambda _S^3 y_0^2
\\
-f \Delta  m_y n_y \lambda _S^3 y_0^2-18 f^2 m_y \lambda_S^2 y_0^2+90 f m_y n_y \lambda _S^2 y_0^2+9 f \Delta  m_y n_y \lambda _S^2 y_0^2
\\
+189 f m_y n_y y_0^2+27 f \Delta  m_y n_y y_0^2+54 f^2 m_y \lambda _S y_0^2-243 f m_y n_y \lambda _S y_0^2-27 f \Delta  m_y n_y\lambda _S y_0^2
\\
-9 f^2 m_y^2 \lambda _S^2+27 f^2 m_y^2 \lambda _S) \Pi ^2+81 n_y^2 y_0^4-81 f n_y y_0^4-3 n_y^2 y_0^4 \lambda _S^3+3 f n_y y_0^4 \lambda _S^3
\\
+3 f m_y n_y y_0^2 \lambda _S^3-81 f m_y n_yy_0^2+27 n_y^2 y_0^4 \lambda _S^2-27 f n_y y_0^4 \lambda _S^2-27 f m_y n_y y_0^2 \lambda _S^2
\\
-81 n_y^2 y_0^4 \lambda _S+81 f n_y y_0^4 \lambda _S+81 f m_y n_y y_0^2 \lambda _S\,,
\end{multline}
\begin{equation}
\mathbb{V}^{(9)}_{12} = 2 \Pi ^5 y_0^4 m_y n_y^2 \lambda _S-2 \Pi ^3 y_0^4 m_y n_y^2 \lambda _S
\end{equation}
\begin{equation}
\mathbb{V}^{(8)}_{12}=3 \Pi ^4 y_0^4 m_y n_y^2 \lambda _S-\Pi ^2 y_0^4 m_y n_y^2 \lambda _S\,,
\end{equation}
\begin{multline}
\mathbb{V}^{(7)}_{12}=\Pi ^5 \left(12 f y_0^2 m_y^2 n_y \lambda _S-18 y_0^4 m_y n_y^2 \lambda _S\right)-6 f \Pi ^7 y_0^2 m_y^2 n_y \lambda _S\,,
\\
+\Pi ^3 \left(20 y_0^4 m_y n_y^2 \lambda _S-6 f y_0^2 m_y^2 n_y \lambda _S\right)\,,
\end{multline}
\begin{multline}
\mathbb{V}^{(6)}_{12}=-3 f \Pi ^6 y_0^2 m_y^2 n_y \lambda _S+\Pi ^4 (12 f y_0^4 m_y n_y \lambda _S+18 f y_0^2 m_y^2 n_y \lambda _S-y_0^4 m_y n_y^2 \lambda _S^2-27 y_0^4 m_y n_y^2 \lambda _S)
\\
+\Pi ^2 (6 y_0^4 m_y n_y^2 \lambda _S^2-12 fy_0^4 m_y n_y \lambda _S-15 f y_0^2 m_y^2 n_y \lambda _S+9 y_0^4 m_y n_y^2 \lambda _S)\,,
\end{multline}
\begin{multline}
\mathbb{V}^{(5)}_{12}=\Pi ^7 (40 f y_0^2 m_y^2 n_y \lambda _S-12 f^2 m_y^2 \lambda _S-2 f y_0^2 m_y^2 n_y \lambda _S^2)+\Pi ^5 (12 f^2 m_y^2 \lambda _S+6 f y_0^2 m_y^2 n_y \lambda _S^2
\\
-80 f y_0^2 m_y^2 n_y \lambda _S+54 y_0^4 m_y n_y^2 \lambda _S)+\Pi ^3 (18 f y_0^4 m_y n_y \lambda _S-4 f^2 m_y^2 \lambda _S+4 f y_0^2 m_y n_y^2 \lambda _S^2
\\
-4 f y_0^2 m_y n_y \lambda _S^2-58 f y_0^2 m_y n_y^2 \lambda _S+58 f y_0^2 m_y n_y \lambda _S+2 y_0^4 m_y n_y^2 \lambda _S^3-72 y_0^4 m_y n_y^2 \lambda _S)
\\
+4 f^2 \Pi ^9 m_y^2 \lambda _S+\Pi  (4 y_0^4 m_y n_y^2\lambda _S^2-12 f y_0^4 m_y n_y \lambda _S-12 f y_0^2 m_y^2 n_y \lambda _S)\,,
\end{multline}
\begin{multline}
\mathbb{V}^{(4)}_{12}=\Pi ^6 (f y_0^2 m_y^2 n_y \lambda _S^2-24 f^2 m_y^3 \lambda _S-24 f^2 y_0^2 m_y^2 \lambda _S+18 f y_0^2 m_y^2 n_y \lambda _S)
\\
+\Pi ^4 (48 f^2 y_0^2 m_y^2 \lambda _S+48 f^2 m_y^3 \lambda _S-72 f y_0^4 m_y n_y \lambda _S-7 f y_0^2 m_y^2 n_y \lambda _S^2-102 f y_0^2 m_y^2 n_y \lambda _S
\\
+6 y_0^4 m_y n_y^2 \lambda _S^2+81 y_0^4 m_y n_y^2 \lambda _S)+\Pi ^2 (72 f y_0^4 m_y n_y \lambda _S-24 f^2 y_0^2 m_y^2 \lambda _S-24 f^2 m_y^3 \lambda _S
\\
-4 f y_0^2 m_y n_y^2 \lambda _S^2+4 f y_0^2 m_y n_y \lambda _S^2-84 f y_0^2 m_y n_y^2 \lambda _S+84 f y_0^2 m_y n_y \lambda _S
\\
+7 y_0^4 m_y n_y^2 \lambda _S^3-36 y_0^4 m_y n_y^2 \lambda _S^2-27 y_0^4 m_y n_y^2 \lambda _S)\,,
\end{multline}
\begin{multline}
\mathbb{V}^{(3)}_{12}=\Pi ^7 (36 f^2 m_y^2 \lambda _S-8 f^2 m_y^2 \lambda _S^2+10 f y_0^2 m_y^2 n_y \lambda _S^2-78 f y_0^2 m_y^2 n_y \lambda _S)
\\
+\Pi ^5 (10 f^2 m_y^2 \lambda _S^2-36 f^2 m_y^2 \lambda _S-2 f y_0^2 m_y^2 n_y \lambda _S^3-30 f y_0^2 m_y^2 n_y \lambda _S^2+156 f y_0^2 m_y^2 n_y \lambda _S
\\
-54 y_0^4 m_y n_y^2 \lambda _S)+\Pi ^3 (36 f^2 y_0^4 m_y \lambda _S+72 f^2 y_0^2 m_y^2 \lambda _S-4 f^2 m_y^2 \lambda_S^2+36 f^2 m_y^3 \lambda _S+12 f^2 m_y^2 \lambda _S
\\
+6 f y_0^4 m_y n_y \lambda _S^2-108 f y_0^4 m_y n_y \lambda _S+2 f y_0^2 m_y^2 n_y \lambda _S^3+26 f y_0^2 m_y^2 n_y \lambda _S^2-186 f y_0^2 m_y^2 n_y \lambda_S
\\
-6 y_0^4 m_y n_y^2 \lambda _S^3+108 y_0^4 m_y n_y^2 \lambda _S)+\Pi  (-36 f^2 y_0^4 m_y \lambda _S-72 f^2 y_0^2 m_y^2 \lambda _S-36 f^2 m_y^3 \lambda _S
\\
+72 f y_0^4 m_y n_y \lambda _S+72 f y_0^2 m_y^2 n_y \lambda _S+4 y_0^4 m_y n_y^2 \lambda _S^3-24 y_0^4 m_y n_y^2 \lambda _S^2)+\Pi ^9 (2 f^2 m_y^2 \lambda _S^2-12 f^2 m_y^2 \lambda _S)\,,
\end{multline}
\begin{multline}
\mathbb{V}^{(2)}_{12}=\Pi ^6 (72 f^2 m_y^3 \lambda _S-12 f^2 m_y^3 \lambda _S^2-12 f^2 y_0^2 m_y^2 \lambda _S^2+72 f^2 y_0^2 m_y^2 \lambda _S-3 f y_0^2 m_y^2 n_y \lambda _S^2
\\
-27 f y_0^2 m_y^2 n_y \lambda _S)+\Pi ^4 (36 f^2 y_0^2 m_y^2 \lambda _S^2-144 f^2 y_0^2 m_y^2 \lambda _S+36 f^2 m_y^3 \lambda _S^2-144 f^2 m_y^3 \lambda _S
\\
+108 f y_0^4 m_y n_y \lambda _S-9 f y_0^2 m_y^2 n_y \lambda _S^3+33 f y_0^2 m_y^2 n_y \lambda_S^2+126 f y_0^2 m_y^2 n_y \lambda _S-9 y_0^4 m_y n_y^2 \lambda _S^2
\\
-81 y_0^4 m_y n_y^2 \lambda _S)+\Pi ^2 (72 f^2 y_0^2 m_y^2 \lambda _S-24 f^2 y_0^2 m_y^2 \lambda _S^2-24 f^2 m_y^3 \lambda _S^2+72 f^2 m_y^3 \lambda _S
\\
+12 f y_0^4 m_y n_y \lambda _S^3-108 f y_0^4 m_y n_y \lambda _S-21 f y_0^2 m_y n_y^2 \lambda _S^3+21 f y_0^2 m_y n_y \lambda _S^3+30 f y_0^2 m_y n_y^2 \lambda _S^2
\\
-30 f y_0^2 m_y n_y \lambda_S^2+99 f y_0^2 m_y n_y^2 \lambda _S -99 f y_0^2 m_y n_y \lambda _S-21 y_0^4 m_y n_y^2 \lambda _S^3+54 y_0^4 m_y n_y^2 \lambda _S^2+27 y_0^4 m_y n_y^2 \lambda _S)\,,
\end{multline}
\begin{multline}
\mathbb{V}^{(1)}_{12}=\Pi ^3 (108 f^2 y_0^4 m_y \lambda _S-18 f^2 y_0^4 m_y \lambda _S^2-36 f^2 y_0^2 m_y^2 \lambda _S^2+216 f^2 y_0^2 m_y^2 \lambda _S
\\
-18 f^2 m_y^3 \lambda _S^2+108 f^2 m_y^3 \lambda _S+18 f y_0^4 m_y n_y \lambda_S^2-162 f y_0^4 m_y n_y \lambda _S-4 f y_0^2 m_y^2 n_y \lambda _S^3
\\
+42 f y_0^2 m_y^2 n_y \lambda _S^2-198 f y_0^2 m_y^2 n_y \lambda _S+54 y_0^4 m_y n_y^2 \lambda _S)+\Pi  (36 f^2 y_0^4 m_y \lambda_S^2-108 f^2 y_0^4 m_y \lambda _S
\\
+72 f^2 y_0^2 m_y^2 \lambda _S^2-216 f^2 y_0^2 m_y^2 \lambda _S+36 f^2 m_y^3 \lambda _S^2-108 f^2 m_y^3 \lambda _S-12 f y_0^4 m_y n_y \lambda _S^3
\\
+108 f y_0^4 m_y n_y \lambda_S-12 f y_0^2 m_y^2 n_y \lambda _S^3+108 f y_0^2 m_y^2 n_y \lambda _S+12 y_0^4 m_y n_y^2 \lambda _S^3-36 y_0^4 m_y n_y^2 \lambda _S^2)
\\
+\Pi ^7 (12 f y_0^2 m_y^2 n_y \lambda _S^2-36 f y_0^2 m_y^2 n_y\lambda _S)+\Pi ^5 (4 f y_0^2 m_y^2 n_y \lambda _S^3-36 f y_0^2 m_y^2 n_y \lambda _S^2+72 f y_0^2 m_y^2 n_y \lambda _S)\,,
\end{multline}
\begin{equation}
\mathbb{V}^{(0)}_{12}=2 f \Pi ^2 y_0^2 m_y n_y \left(\lambda _S-3\right){}^2 \lambda _S \left(3 \Pi ^2+\lambda _S-3\right)\,,
\end{equation}
\begin{equation}
\mathbb{V}^{(8)}_{22}=\left(1-\Pi ^2\right) y_0^4 m_y n_y^2 \lambda _S \left[\Pi ^2 \left(3 \Delta -2 \lambda _S\right)+1\right]\,,
\end{equation}
\begin{equation}
\mathbb{V}^{(7)}_{22}=2 \Pi  (1-\Pi ^2) y_0^2 m_y n_y \lambda _S \left[f m_y+y_0^2 n_y \left(\Delta -2 \lambda _S\right)\right]\,,
\end{equation}
\begin{multline}
\mathbb{V}^{(6)}_{22}=\Pi ^4 (5 f y_0^2 m_y^2 n_y \lambda _S^2-6 \Delta  f y_0^2 m_y^2 n_y \lambda _S-\Delta ^2 y_0^4 m_y n_y^2 \lambda _S-4 \Delta  y_0^4 m_y n_y^2 \lambda _S^2
\\
+30 \Delta  y_0^4 m_y n_y^2 \lambda _S-y_0^4 m_yn_y^2 \lambda _S^3-9 y_0^4 m_y n_y^2 \lambda _S^2)+\Pi ^2 (3 \Delta  f y_0^2 m_y^2 n_y \lambda _S-9 f y_0^4 m_y n_y \lambda _S
\\
-3 f y_0^2 m_y^2 n_y \lambda _S^2-9 f y_0^2 m_y^2 n_y \lambda _S+\Delta ^2 y_0^4 m_y n_y^2 \lambda _S+7 \Delta  y_0^4 m_y n_y^2 \lambda _S^2-30 \Delta  y_0^4 m_y n_y^2 \lambda _S
\\
-4 y_0^4 m_y n_y^2 \lambda _S^3+18 y_0^4 m_y n_y^2 \lambda _S^2+12 y_0^4 m_y n_y^2\lambda _S)+\Pi ^6 (3 \Delta  f y_0^2 m_y^2 n_y \lambda _S-2 f y_0^2 m_y^2 n_y \lambda _S^2)
\\
+9 f y_0^4 m_y n_y \lambda _S+9 f y_0^2 m_y^2 n_y \lambda _S+y_0^4 m_y n_y^2 \lambda _S^2-12 y_0^4 m_yn_y^2 \lambda _S\,,
\end{multline}
\begin{multline}
\mathbb{V}^{(5)}_{22}=\Pi ^3 (12 f y_0^4 m_y n_y \lambda _S^2-12 f^2 y_0^2 m_y^2 \lambda _S-12 f^2 m_y^3 \lambda _S-6 \Delta  f y_0^4 m_y n_y \lambda _S
\\
-6 \Delta  f y_0^2 m_y^2 n_y \lambda _S+14 f y_0^2 m_y^2 n_y \lambda _S^2+18 fy_0^2 m_y^2 n_y \lambda _S-2 \Delta  y_0^4 m_y n_y^2 \lambda _S^2
\\
+18 \Delta  y_0^4 m_y n_y^2 \lambda _S-2 y_0^4 m_y n_y^2 \lambda _S^3-18 y_0^4 m_y n_y^2 \lambda _S^2)+\Pi  (6 f^2 y_0^2 m_y^2 \lambda_S+6 f^2 m_y^3 \lambda _S
\\
+6 \Delta  f y_0^4 m_y n_y \lambda _S+6 \Delta  f y_0^2 m_y^2 n_y \lambda _S-12 f y_0^4 m_y n_y \lambda _S^2-8 f y_0^2 m_y^2 n_y \lambda _S^2
\\
-18 f y_0^2 m_y^2 n_y \lambda _S+4 \Delta y_0^4 m_y n_y^2 \lambda _S^2-18 \Delta  y_0^4 m_y n_y^2 \lambda _S-8 y_0^4 m_y n_y^2 \lambda _S^3
\\
+36 y_0^4 m_y n_y^2 \lambda _S^2)+\Pi ^5 (6 f^2 m_y^3 \lambda _S+6 f^2 y_0^2 m_y^2 \lambda _S-4 f y_0^2m_y^2 n_y \lambda _S^2)\,,
\end{multline}
\begin{multline}
\mathbb{V}^{(4)}_{22}=(4 f m_y^2 n_y y_0^2 \lambda _S^3-6 f m_y^2 n_y y_0^2 \lambda _S^2+4 f \Delta  m_y^2 n_y y_0^2 \lambda _S^2-21 f \Delta  m_y^2 n_y y_0^2 \lambda _S) \Pi ^6
\\
+(6m_y n_y^2 \lambda _S^3 y_0^4-\Delta  m_y n_y^2 \lambda _S^3 y_0^4-2 \Delta ^2 m_y n_y^2 \lambda _S^2 y_0^4+30 \Delta  m_y n_y^2 \lambda _S^2 y_0^4
\\
+9 \Delta ^2 m_y n_y^2 \lambda _S y_0^4-108 \Delta  m_y n_y^2 \lambda _S y_0^4+2 f m_y^2 n_y \lambda_S^3 y_0^2-15 f m_y^2 n_y \lambda _S^2 y_0^2
\\
-11 f \Delta  m_y^2 n_y \lambda _S^2 y_0^2+42 f \Delta  m_y^2 n_y \lambda _S y_0^2) \Pi ^4+(5 \Delta  m_y n_y^2 \lambda _S^3y_0^4-2 m_y n_y^2 \lambda _S^4 y_0^4
\\
+22 m_y n_y^2 \lambda _S^3 y_0^4+3 \Delta ^2 m_y n_y^2 \lambda _S^2 y_0^4-51 \Delta  m_y n_y^2 \lambda _S^2 y_0^4-45 m_y n_y^2 \lambda _S^2 y_0^4
\\
-6 f m_y n_y \lambda _S^2 y_0^4-9 \Delta ^2 m_y n_y^2 \lambda_S y_0^4+108 \Delta  m_y n_y^2 \lambda _S y_0^4-54 m_y n_y^2 \lambda _S y_0^4-18 f^2 m_y \lambda _S y_0^4
\\
+81 f m_y n_y \lambda _S y_0^4-7 f m_y^2 n_y \lambda _S^3 y_0^2+15 f m_y^2 n_y \lambda _S^2 y_0^2+7 f\Delta  m_y^2 n_y \lambda _S^2 y_0^2-36 f^2 m_y^2 \lambda _S y_0^2
\\
+81 f m_y^2 n_y \lambda _S y_0^2-21 f \Delta  m_y^2 n_y \lambda _S y_0^2-18 f^2 m_y^3 \lambda _S) \Pi ^2-m_y n_y^2 y_0^4 \lambda _S^3-9m_yn_y^2 y_0^4 \lambda _S^2
\\
+15 f m_y n_y y_0^4 \lambda _S^2+15 f m_y^2 n_y y_0^2 \lambda _S^2+54 m_y n_y^2 y_0^4 \lambda _S+18 f^2 m_y y_0^4 \lambda _S-81 f m_y n_y y_0^4 \lambda _S
\\
+18 f^2 m_y^3 \lambda _S+36 f^2m_y^2 y_0^2 \lambda _S-81 f m_y^2 n_y y_0^2 \lambda _S\,,
\end{multline}
\begin{multline}
\mathbb{V}^{(3)}_{22}=\Pi ^3 (72 f^2 y_0^2 m_y^2 \lambda _S-18 f^2 y_0^2 m_y^2 \lambda _S^2-18 f^2 m_y^3 \lambda _S^2+72 f^2 m_y^3 \lambda _S
\\
-6 \Delta  f y_0^4 m_y n_y \lambda _S^2+36 \Delta  f y_0^4 m_y n_y \lambda _S-6 \Delta  f y_0^2 m_y^2 n_y \lambda _S^2+36 \Delta  f y_0^2 m_y^2 n_y \lambda _S
\\
-6 f y_0^4 m_y n_y \lambda _S^3-18 f y_0^4 m_y n_y \lambda _S^2-4 f y_0^2 m_y^2 n_y \lambda _S^3-12 f y_0^2 m_y^2 n_y \lambda _S^2-54 f y_0^2 m_y^2 n_y \lambda _S
\\
+12 \Delta  y_0^4 m_y n_y^2 \lambda _S^2-54 \Delta  y_0^4 m_y n_y^2 \lambda _S+12 y_0^4 m_y n_y^2 \lambda _S^3)+\Pi  (12 f^2 y_0^2 m_y^2 \lambda _S^2
\\
-36 f^2 y_0^2 m_y^2 \lambda_S+12 f^2 m_y^3 \lambda _S^2-36 f^2 m_y^3 \lambda _S+12 \Delta  f y_0^4 m_y n_y \lambda _S^2-36 \Delta  f y_0^4 m_y n_y \lambda _S
\\
+12 \Delta  f y_0^2 m_y^2 n_y \lambda _S^2-36 \Delta  f y_0^2 m_y^2 n_y \lambda_S-24 f y_0^4 m_y n_y \lambda _S^3+72 f y_0^4 m_y n_y \lambda _S^2
\\
-22 f y_0^2 m_y^2 n_y \lambda _S^3+48 f y_0^2 m_y^2 n_y \lambda _S^2+54 f y_0^2 m_y^2 n_y \lambda _S+2 \Delta  y_0^4 m_y n_y^2 \lambda _S^3
\\
-24\Delta  y_0^4 m_y n_y^2 \lambda _S^2+54 \Delta  y_0^4 m_y n_y^2 \lambda _S-4 y_0^4 m_y n_y^2 \lambda _S^4+48 y_0^4 m_y n_y^2 \lambda _S^3
\\
-108 y_0^4 m_y n_y^2 \lambda _S^2)+\Pi ^5 (6 f^2 m_y^3 \lambda_S^2-36 f^2 m_y^3 \lambda _S+6 f^2 y_0^2 m_y^2 \lambda _S^2-36 f^2 y_0^2 m_y^2 \lambda _S
\\
+2 f y_0^2 m_y^2 n_y \lambda _S^3+6 f y_0^2 m_y^2 n_y \lambda _S^2)\,,
\end{multline}
\begin{multline}
\mathbb{V}^{(2)}_{22}=(9 f^2 m_y^2 \lambda_S^2-3 f^2 m_y^2 \lambda_S^3) \Pi ^8+(6 f^2 m_y^2 \lambda_S^3-12 f m_y^2 n_y y_0^2 \lambda_S^3
\\
+f \Delta  m_y^2 n_y y_0^2 \lambda_S^3-18 f^2 m_y^2 \lambda_S^2+36 f m_y^2 n_y y_0^2 \lambda_S^2-18 f \Delta  m_y^2 n_y y_0^2 \lambda_S^2
\\
+45 f \Delta  m_y^2 n_y y_0^2 \lambda_S) \Pi ^6+(6 \Delta  m_y n_y^2 \lambda_S^3 y_0^4-9 m_y n_y^2\lambda_S^3 y_0^4-\Delta ^2 m_y n_y^2 \lambda_S^3 y_0^4
\\
+12 \Delta ^2 m_y n_y^2 \lambda_S^2 y_0^4-72 \Delta  m_y n_y^2 \lambda_S^2 y_0^4+27 m_y n_y^2 \lambda_S^2 y_0^4-27 \Delta ^2 m_y n_y^2 \lambda_S y_0^4+162 \Delta  m_y n_y^2 \lambda_Sy_0^4
\\
+3 f m_y^2 n_y \lambda_S^4 y_0^2-12 f m_y^2 n_y \lambda_S^3 y_0^2-6 f \Delta  m_y^2 n_y \lambda_S^3 y_0^2+9 f m_y^2 n_y \lambda_S^2 y_0^2+48 f \Delta  m_y^2 n_y \lambda_S^2 y_0^2
\\
-90 f \Delta  m_y^2 n_y\lambda_S y_0^2-3 f^2 m_y^2 \lambda_S^3+9 f^2 m_y^2 \lambda_S^2) \Pi ^4+(\Delta  m_y n_y^2 \lambda_S^4 y_0^4+6 m_y n_y^2 \lambda_S^4 y_0^4
\\
+3 \Delta ^2 m_y n_y^2 \lambda_S^3 y_0^4-24 \Delta  m_yn_y^2 \lambda_S^3 y_0^4-24 m_y n_y^2 \lambda_S^3 y_0^4-15 f m_y n_y \lambda_S^3 y_0^4-18 \Delta ^2 m_y n_y^2 \lambda_S^2 y_0^4
\\
+117 \Delta  m_y n_y^2 \lambda_S^2 y_0^4-18 f^2 m_y \lambda_S^2 y_0^4+90 f m_yn_y \lambda_S^2 y_0^4+27 \Delta ^2 m_y n_y^2 \lambda_S y_0^4-162 \Delta  m_y n_y^2 \lambda_S y_0^4
\\
+108 m_y n_y^2 \lambda_S y_0^4+108 f^2 m_y \lambda_S y_0^4-243 f m_y n_y \lambda_S y_0^4-5 f m_y^2 n_y\lambda_S^4 y_0^2+15 f m_y^2 n_y \lambda_S^3 y_0^2
\\
+5 f \Delta  m_y^2 n_y \lambda_S^3 y_0^2-36 f^2 m_y^2 \lambda_S^2 y_0^2+45 f m_y^2 n_y \lambda_S^2 y_0^2-30 f \Delta  m_y^2 n_y \lambda_S^2 y_0^2+216 f^2m_y^2 \lambda_S y_0^2
\\
-243 f m_y^2 n_y \lambda_S y_0^2+45 f \Delta  m_y^2 n_y \lambda_S y_0^2-18 f^2 m_y^3 \lambda_S^2+108 f^2 m_y^3 \lambda_S) \Pi ^2-m_y n_y^2 y_0^4 \lambda_S^4
\\
+6 m_y n_y^2 y_0^4\lambda_S^3+3 f m_y n_y y_0^4 \lambda_S^3+3 f m_y^2 n_y y_0^2 \lambda_S^3+27 m_y n_y^2 y_0^4 \lambda_S^2+36 f^2 m_y y_0^4 \lambda_S^2
\\
-90 f m_y n_y y_0^4 \lambda_S^2+36 f^2 m_y^3 \lambda_S^2+72 f^2 m_y^2y_0^2 \lambda_S^2-90 f m_y^2 n_y y_0^2 \lambda_S^2-108 m_y n_y^2 y_0^4 \lambda_S
\\
-108 f^2 m_y y_0^4 \lambda_S+243 f m_y n_y y_0^4 \lambda_S-108 f^2 m_y^3 \lambda_S-216 f^2 m_y^2 y_0^2 \lambda_S+243 f m_y^2n_y y_0^2 \lambda_S\,,
\end{multline}
\begin{multline}
\mathbb{V}^{(1)}_{22}=(72 f^2 \lambda_S^2 m_y^3-18 f^2 \lambda_S^3 m_y^3-54 f^2 \lambda_S m_y^3-18 f^2 y_0^2 \lambda_S^3 m_y^2+6 f n_y y_0^2 \lambda_S^3 m_y^2+72 f^2 y_0^2 \lambda_S^2 m_y^2
\\
-18 f n_y y_0^2 \lambda_S^2m_y^2-54 f^2 y_0^2 \lambda_S m_y^2) \Pi ^5+(18 m_y n_y^2 \lambda_S^3 y_0^4-18 f m_y n_y \lambda_S^3 y_0^4+18 \Delta  m_y n_y^2 \lambda_S^2 y_0^4
\\
-54 m_y n_y^2 \lambda_S^2 y_0^4+54 f m_y n_y\lambda_S^2 y_0^4-18 f \Delta  m_y n_y \lambda_S^2 y_0^4-54 \Delta  m_y n_y^2 \lambda_S y_0^4+54 f \Delta  m_y n_y \lambda_S y_0^4
\\
-4 f m_y^2 n_y \lambda_S^4 y_0^2+24 f^2 m_y^2 \lambda_S^3 y_0^2-12 f m_y^2n_y \lambda_S^3 y_0^2-108 f^2 m_y^2 \lambda_S^2 y_0^2+90 f m_y^2 n_y \lambda_S^2 y_0^2
\\
-18 f \Delta  m_y^2 n_y \lambda_S^2 y_0^2+108 f^2 m_y^2 \lambda_S y_0^2-54 f m_y^2 n_y \lambda_S y_0^2+54 f \Delta m_y^2 n_y \lambda_S y_0^2+24 f^2 m_y^3 \lambda_S^3
\\
-108 f^2 m_y^3 \lambda_S^2+108 f^2 m_y^3 \lambda_S) \Pi ^3+(12 f m_y n_y \lambda_S^4 y_0^4-12 m_y n_y^2 \lambda_S^4 y_0^4+6 \Delta  m_y n_y^2\lambda_S^3 y_0^4+72 m_y n_y^2 \lambda_S^3 y_0^4
\\
-72 f m_y n_y \lambda_S^3 y_0^4-6 f \Delta  m_y n_y \lambda_S^3 y_0^4-36 \Delta  m_y n_y^2 \lambda_S^2 y_0^4-108 m_y n_y^2 \lambda_S^2 y_0^4+108 f m_y n_y\lambda_S^2 y_0^4
\\
+36 f \Delta  m_y n_y \lambda_S^2 y_0^4+54 \Delta  m_y n_y^2 \lambda_S y_0^4-54 f \Delta  m_y n_y \lambda_S y_0^4+12 f y^2 m_y n_y \lambda_S^4 y_0^2+12 f m_y n_y \lambda_S^4 y_0^2
\\
-24 f y m_y n_y \lambda_S^4 y_0^2-6 f^2 m_y^2 \lambda_S^3 y_0^2-66 f y^2 m_y n_y \lambda_S^3 y_0^2-66 f m_y n_y \lambda_S^3 y_0^2+132 f y m_y n_y \lambda_S^3 y_0^2
\\
-6 f y^2 \Delta  m_y n_y \lambda_S^3 y_0^2-6 f\Delta  m_y n_y \lambda_S^3 y_0^2+12 f y \Delta  m_y n_y \lambda_S^3 y_0^2+36 f^2 m_y^2 \lambda_S^2 y_0^2
\\
+72 f y^2 m_y n_y \lambda_S^2 y_0^2+72 f m_y n_y \lambda_S^2 y_0^2-144 f y m_y n_y \lambda_S^2y_0^2+36 f y^2 \Delta  m_y n_y \lambda_S^2 y_0^2+36 f \Delta  m_y n_y \lambda_S^2 y_0^2
\\
-72 f y \Delta  m_y n_y \lambda_S^2 y_0^2-54 f^2 m_y^2 \lambda_S y_0^2+54 f y^2 m_y n_y \lambda_S y_0^2+54 f m_y n_y\lambda_S y_0^2-108 f y m_y n_y \lambda_S y_0^2
\\
-54 f y^2 \Delta  m_y n_y \lambda_S y_0^2-54 f \Delta  m_y n_y \lambda_S y_0^2+108 f y \Delta  m_y n_y \lambda_S y_0^2-6 f^2 m_y^3 \lambda_S^3+36 f^2 m_y^3\lambda_S^2
\\
-54 f^2 m_y^3 \lambda_S) \Pi\,,
\end{multline}
\begin{multline}
\mathbb{V}^{(0)}_{22}=(18 f \Delta  m_y n_y y_0^2 \lambda_S^2-3 f \Delta  m_y n_y y_0^2 \lambda_S^3-27 f \Delta  m_y n_y y_0^2 \lambda_S) \Pi ^6+(3 \Delta ^2 n_y^2 \lambda_S^3 y_0^4
\\
-9 \Delta  n_y^2 \lambda_S^3y_0^4-18 \Delta ^2 n_y^2 \lambda_S^2 y_0^4+54 \Delta  n_y^2 \lambda_S^2 y_0^4+27 \Delta ^2 n_y^2 \lambda_S y_0^4-81 \Delta  n_y^2 \lambda_S y_0^4
\\
-3 f m_y n_y \lambda_S^4 y_0^2-f \Delta  m_y n_y \lambda_S^4y_0^2+18 f m_y n_y \lambda_S^3 y_0^2+12 f \Delta  m_y n_y \lambda_S^3 y_0^2-27 f m_y n_y \lambda_S^2 y_0^2
\\
-45 f \Delta  m_y n_y \lambda_S^2 y_0^2+54 f \Delta  m_y n_y \lambda_S y_0^2) \Pi ^4+(\Delta ^2 n_y^2 y_0^4 \lambda_S^4-fm_y n_y y_0^2 \lambda_S^5-3 \Delta  n_y^2 y_0^4 \lambda_S^4
\\
+9 f m_y n_y y_0^2 \lambda_S^4+f \Delta  m_y n_y y_0^2 \lambda_S^4-27 f^2 y_0^4 \lambda_S^3-9 \Delta ^2 n_y^2y_0^4 \lambda_S^3+27 \Delta  n_y^2 y_0^4 \lambda_S^3-18 n_y^2 y_0^4 \lambda_S^3
\\
+45 f n_y y_0^4 \lambda_S^3-27 f^2 m_y^2 \lambda_S^3-54 f^2 m_y y_0^2 \lambda_S^3+18 f m_y n_y y_0^2 \lambda_S^3-9 f \Delta m_y n_y y_0^2 \lambda_S^3+135 f^2 y_0^4 \lambda_S^2
\\
+27 \Delta ^2 n_y^2 y_0^4 \lambda_S^2-81 \Delta  n_y^2 y_0^4 \lambda_S^2+81 n_y^2 y_0^4 \lambda_S^2-216 f n_y y_0^4 \lambda_S^2+135 f^2 m_y^2 \lambda_S^2+270 f^2 m_y y_0^2 \lambda_S^2
\\
-189 f m_y n_y y_0^2 \lambda_S^2+27 f \Delta  m_y n_y y_0^2 \lambda_S^2-162 f^2 y_0^4 \lambda_S-27 \Delta ^2 n_y^2 y_0^4 \lambda_S+81 \Delta  n_y^2 y_0^4 \lambda_S
\\
-81n_y^2 y_0^4 \lambda_S+243 f n_y y_0^4 \lambda_S-162 f^2 m_y^2 \lambda_S-324 f^2 m_y y_0^2 \lambda_S+243 f m_y n_y y_0^2 \lambda_S
\\
-27 f \Delta  m_y n_y y_0^2 \lambda_S) \Pi ^2+3 n_y^2 y_0^4 \lambda_S^4-3 f n_y y_0^4 \lambda_S^4-3 f m_y n_y y_0^2 \lambda_S^4+18 f^2 y_0^4 \lambda_S^3-9 n_y^2 y_0^4 \lambda_S^3
\\
-9 f n_y y_0^4 \lambda_S^3+18 f^2 m_y^2 \lambda_S^3+36 f^2 m_y y_0^2 \lambda_S^3-9 f m_y n_yy_0^2 \lambda_S^3-108 f^2 y_0^4 \lambda_S^2-27 n_y^2 y_0^4 \lambda_S^2
\\
+135 f n_y y_0^4 \lambda_S^2-108 f^2 m_y^2 \lambda_S^2-216 f^2 m_y y_0^2 \lambda_S^2+135 f m_y n_y y_0^2 \lambda_S^2+162 f^2 y_0^4\lambda_S
\\
+81 n_y^2 y_0^4 \lambda_S-243 f n_y y_0^4 \lambda_S+162 f^2 m_y^2 \lambda_S+324 f^2 m_y y_0^2 \lambda_S-243 f m_y n_y y_0^2 \lambda_S\,.
\end{multline}

\section{Wormholes with toroidal boundaries in simple low-energy theories}
\label{app:torusboundaries}

This appendix collects some results regarding wormholes with toroidal boundaries.  As in the spherical case, we consider both the $U(1)^3$ theory of section \ref{sec:u13} (see section \ref{sec:u13torus}) and a scalar theory (see section \ref{sec:scalartorus}), though now the latter will contain three complex scalars.  These results are less complete than for the spherical-boundary cases of sections \ref{sec:u13} and \ref{sec:scalars}, in part because we construct only wormhole solutions and do not construct a disconnected solution.  Indeed, in the torus case a smooth disconnected solution must feature a preferred cycle of the torus that shrinks to zero size while the other cycles remain finite (much as in the familar AdS soliton \cite{Witten:1998zw,Horowitz:1998ha}).  This requires a metric \emph{Ansatz} that breaks the discrete symmetries we impose below.

Despite this lack of completeness, the results below indicate that toroidal solutions are broadly similar to those with spherical boundaries.  In particular, in the scalar case we again find a Hawking-Page-like structure for the wormhole phases, and in particular the large wormhole branch is stable.  In contrast, in the $U(1)^3$ case we identify only a single branch of wormhole solutions which we find exists for arbitrarily small values of appropriate boundary sources.

\subsection{$U(1)^3$ with a toroidal boundary}
\label{sec:u13torus}
As a short aside we mention that there is a very simple four-dimensional example of a wormhole with three gauge fields when the boundary metric is a torus. We take the same theory as in (\ref{eq:simpleu13}) but with three gauge fields of the form
\begin{equation}
A^I = L\,B_0\,\frac{\varepsilon^{IJK}}{2}x_J\,\mathrm{d}x_K
\end{equation}
with $B_0$ constant, and a metric of the form
\begin{equation}
\mathrm{d}s^2 = \frac{\mathrm{d}r^2}{f}+(r^2+r_0^2)\left(\mathrm{d}x_1^2+\mathrm{d}x_2^2+\mathrm{d}x_3^2\right)\,.
\end{equation}
A solution exists provided
\begin{equation}
f(r)=\frac{r^2+2r_0^2}{L^2}\quad\text{and}\quad B_0 = \frac{r_0^2}{L^2}\,.
\end{equation}
In this section we are assuming that $x_1$, $x_2$ and $x_3$ are periodic coordinates with periods $\Delta x_1$, $\Delta x_2$, $\Delta x_3$. In this context we believe that the solution we have found is the unique connected geometry, though we have not found a way to construct a disconnected solution. It is a simple exercise to compute the regulated on-shell action for this solution which yields
\begin{equation}
S=\frac{8 \sqrt{2} K\left(\frac{1}{2}\right) r_0^3}{L}\Delta x_1\Delta x_2\Delta x_3\,.
\end{equation}
We have not attempted to study the negative modes of this solution, though we believe that this wormhole will again be stable. The reason for this is that the infinite-radius limit of the $U(1)^3$ wormholes with a spherical boundary we constructed are connected coincides with the $\Delta x_i\rightarrow \infty$ limit of the wormholes discussed in this section. Since for the spherical wormholes we found no negative modes, we expect the same to hold here.
\subsection{Wormholes sourced by scalar fields with a toroidal boundary}
\label{sec:scalartorus}
We now consider wormholes sourced by scalar fields which have a toroidal boundary. If we want to keep isotropy and homogeneity, we seem to need at least three complex scalar fields, which we label by $\psi_I$ and collectively assemble in a vector $\vec{\psi}$. We will proceed much as in section \ref{sec:scalars}, using the simple low-energy action
\begin{equation}
S=-\int_{\mathcal{M}} \mathrm{d}^4 x\sqrt{g}\left[R+\frac{6}{L^2}-2 (\nabla_a \vec{\psi})\cdot  (\nabla^a \vec{\psi})^*-2 \mu^2 \vec{\psi}\cdot \vec{\psi}^*\right]-2\int_{\partial \mathcal{M}} \mathrm{d}^3 x\sqrt{h}\;K+S^{\mu^2}_{\mathcal{B}}\,.
\label{eq:actionscalargeneraltorus}
\end{equation}
Here $L$ is the four-dimensional AdS length scale and ${}^*$ denotes complex conjugation, the second term is the usual Gibbons-Hawking term and $S^{\mu^2}_{\mathcal{B}}$ the boundary counter-term to make the action finite and the variational problem well defined from the perspective of $\vec{\psi}$. We again consider both the massless example and the effective mass that would descirbe conformal coupling, so the boundary terms $S^{\mu^2}_{\mathcal{B}}$ are chosen as in Section \ref{sec:scalars} (but now with three complex scalar fields). The Einstein equation and scalar field equation (ignoring boundary terms) derived from this action read
\begin{subequations}
\label{eqs:eqspsi}
\begin{align}
&R_{ab}-\frac{R}{2}g_{ab}-\frac{3}{L^2}g_{ab}=2 \,\nabla_{(a} \vec{\psi}\cdot \nabla_{b)} \vec{\psi}^*-g_{ab} \nabla_c \vec{\psi}\cdot \nabla^c\vec{\psi}^*-\mu^2\,g_{ab}\,\vec{\psi}\cdot \vec{\psi}^*\,,
\\
& \Box \vec{\psi}=\mu^2 \vec{\psi}\,.
\end{align}
\end{subequations}

Before describing our (numerical) solutions with torus boundary,  we would like to comment on a simple analytic wormhole that arises for the massless case when the torus boundary is replaced by $\mathbb{R}^3$. For this solution one considers the configuration
\begin{subequations}
\begin{equation}
\vec{\psi}=\left[
\begin{array}{c}
\Phi_0\,x_1
\\
\Phi_0\,x_2
\\
\Phi_0\,x_3
\end{array}\right],
\end{equation}
with metric
\begin{equation}
\mathrm{d}s^2 = \frac{\mathrm{d}r^2}{f(r)}+(r^2+r_0^2)(\mathrm{d}x_1^1+\mathrm{d}x_1^2+\mathrm{d}x_1^3)\,,
\label{eq:lineansatztoroidal}
\end{equation}
taking $r_0=\Phi_0\,L$ and
\begin{equation}
f=\frac{r^2+r_0^2}{L^2}\,.
\end{equation}
\end{subequations}%
This shows rather explicitly that wormhole solutions can exist for any value of $A_0$, though the solution does not satisfy standard boundary conditions at large $x_i$.

To fix this, we introduce harmonic dependence on $x_i$ in the scalar field \emph{Ansatz} and also consider $\mu\neq0$ to write
\begin{equation}
\vec{\psi}=\vec{X}_k\,\psi(r)\,,
\end{equation}
with
\begin{equation}
\vec{X}_k=\left[
\begin{array}{c}
e^{i\,k\,x_1}
\\
e^{i\,k\,x_2}
\\
e^{i\,k\,x_3}
\end{array}\right].
\end{equation}
We also consider the metric
\begin{equation}
\mathrm{d}s^2 = \frac{\mathrm{d}r^2}{f(r)}+(r^2+r_0^2)(\mathrm{d}x_1^2+\mathrm{d}x_2^2+\mathrm{d}x_3^2)\, ,
\end{equation}
where $f$, $\psi$ and $r_0$ are to be determined numerically for a given source $\Phi_0$ associated with the boundary value of $\psi$. In performing such numerics it is wise to use a compact coordinate, so we introduce
\begin{equation}
y=1-\frac{r_0}{\sqrt{r^2+r_0^2}}\,.
\end{equation}
The conformal boundary is now located at $y=1$, and the $\mathbb{Z}_2$ symmetry plane at $y=0$.

We are interested in the case where $x_1$, $x_2$ and $x_3$ span a cubic 3-torus $\mathbb{T}^3$ with period $\ell = 2\pi/k$. By construction, the torus has minimal volume at $r=y=0$. It is a simple exercise to determine $f$ from the Einstein equation to find
\begin{equation}
f= \frac{L^2 \left[k^2 (1-y)^2+r_0^2 \mu ^2\right] \psi ^2-r_0^2}{L^2 (2-y) (1-y)^2 y \left[(1-y)^2 \psi'^2-1\right]}\,.
\end{equation}
Since we are interested in solutions for which $\psi$ is smooth at $y=0$ and $f$ is finite there, we need to have
\begin{equation}
L^2(k^2+r_0^2\mu^2) \psi(0)^2=r_0^2\,.
\end{equation}
At this stage we introduce $\psi(0)=A_0$ and write $r_0$ in terms of $A_0$ in the equation for $\psi$.  This yields
\begin{subequations}
\begin{equation}
m_y^2 \psi''+\frac{p_y^{(0,1,1)}}{\psi}+m_y p_y^{(2,1,2)}\psi'-m_y^2\,p^{(0,1,1)}\frac{\psi'^2}{\psi}-m_y^3p_y^{(3,2,3)}\psi'^3=0\,,
\label{eq:onescalartoiralode}
\end{equation}
where
\begin{equation}
p_y^{(a,b,c)}=\frac{a\,A_0^2-(b\,m_y^2+c\,A_0^2\,L^2\,\mu^2)\psi^2}{A_0^2-(m_y^2+A_0^2\,L^2\,\mu^2)\psi^2}\,,
\end{equation}
\end{subequations}
and we again recall that $m_y=1-y$. The equation for $\psi$ depends only on $A_0$, and the dependence in $r_0$ and $k_1$ has dropped out. This is to be expected. If the boundary metric is flat, there is a residual gauge freedom when one simultaneously scales all the $x_i$ and uses conformal invariance. A priori one might have thought that the wormholes we seek to construct formed a two-dimensional family of solutions parametrised by $\Phi_0$ and $k$ with $r_0$ being fixed by the former. However, due to conformal invariance this is not the case, and instead only the ratio $V/k^{\Delta_-}$ is physically meaningful in the bulk. One might erroneously think that this should have reduced the moduli space of solutions in the spherical case to 0 dimensions. However, the sphere there introduces a new scale in the problem which cannot be removed.


Since we do not construct a disconnected solution for comparison,
computing the on-shell action is not of much interest.  Instead, we will focus on trying to understand whether wormholes in this class of theories exist to arbitrary small values of $\Phi_0$ and whether they are free of negative modes. The answer to the first question appears to be negative. Even for wormholes with toroidal boundary conditions we find a minimal critical amplitude $V/k^{\Delta_-}$ above which they can exist. This is perhaps surprising, as we find no smooth solutions at all within our \emph{ansatz} for  $V/k$ less than this critical value. For the massless case we find that wormholes only exist for $V\geq V_{\min}\approx 1.7107(3)$ (see left hand side of Fig.~\ref{fig:toroidalscalar}), whereas for the conformal case we need $V/k >(V/k)_{\min}\approx 11.2529(7)$ (see right hand side of Fig.~\ref{fig:toroidalscalar}). Just as for the spherical case, for each value of $V\geq V_{\min}$ two wormhole solutions exist. We again call the phase with smallest $y_0/k$ (with $y_0\equiv r_0/L$) the small wormhole phase and we call the phase with largest $y_0/k$ the large wormhole phase. Precisely at $V/k^{\Delta_-}_{\min}$ we have $y_0/k=(y_0/k)_{\min}$ with $(y_0/k)_{\min}\approx 1.10146(5)$ and $(y_0/k)_{\min}\approx 0.49202(4)$ for the massless and conformal cases, respectively. The small wormhole phase is shown in Fig.~\ref{fig:toroidalscalar} as orange squares, while the large wormhole phase is represented by the blue disks.
\begin{figure}[h]
\centering
\includegraphics[width =\textwidth]{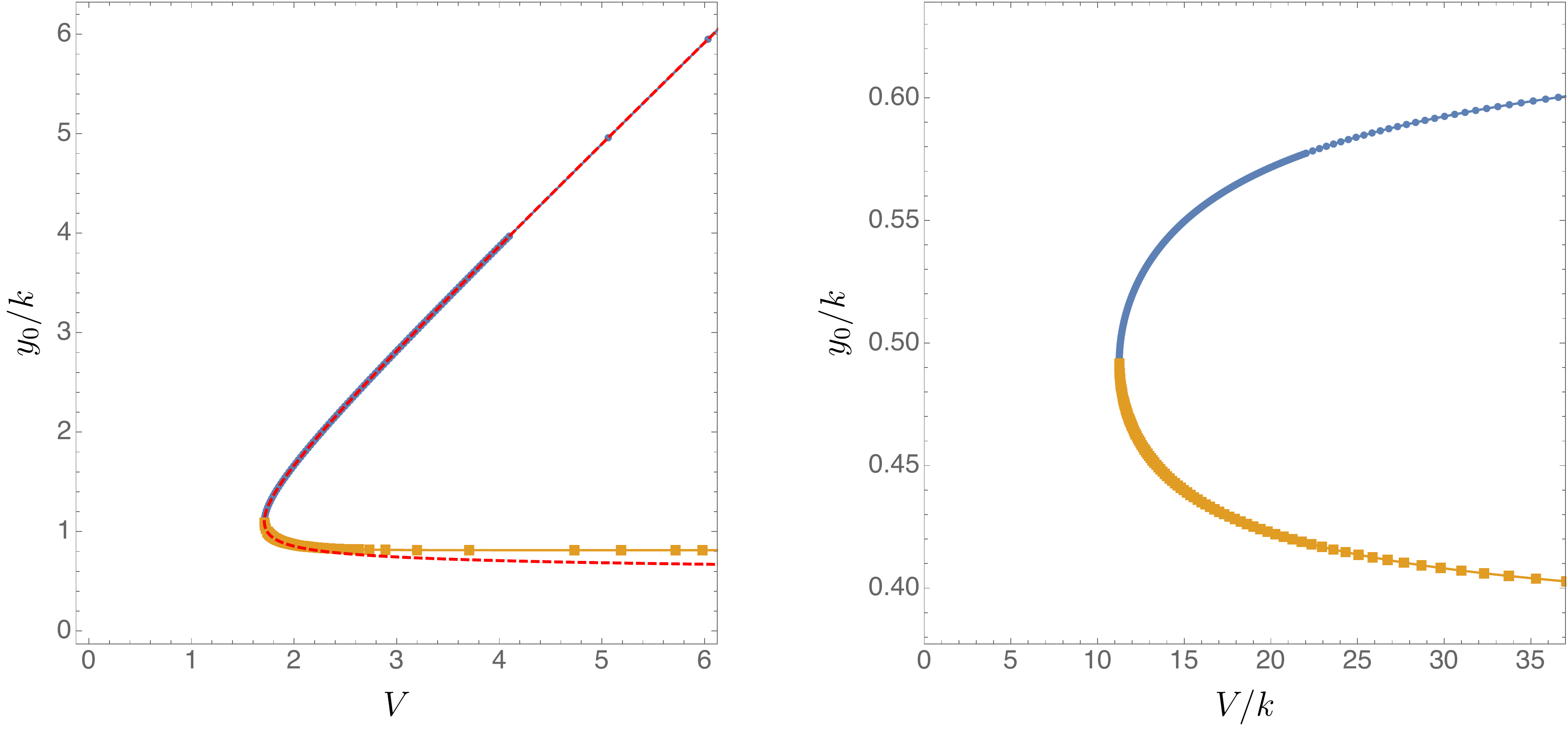}
\caption{Wormholes with toroidal boundary conditions. {\bf Left panel}: wormholes sourced by massless scalars, with the dashed red line being analytically generated by Eq.~(\ref{eq:expansionA0}) with terms up to $i=8$. {\bf Right panel}: wormholes generated by conformally coupled scalars.  In both panels, the small wormhole phase is represented by orange squares, while large wormholes are given by blue disks.}
\label{fig:toroidalscalar}
\end{figure}

For the massless case we can actually do better and find a uniform expansion for $\psi$ in powers of $1/A_0$. This tuns out to be a very useful expansion because this allows us to check our numerics. The expansion takes the rather simple form
\begin{equation}
\psi = A_0+\sum_{i=0}^{+\infty}\frac{1}{A_0^{2i+1}}\psi^{(i)}\left(\frac{r}{r_0}\right)\,.
\label{eq:expansionA0}
\end{equation}
We have carried out this expansion to $i=8$, but for the sake of brevity we only present here results for $i=0,1$
\begin{align}
& \psi^{(0)}(z)=\frac{1}{2}\frac{z^2}{1+z^2}\,,
\\
& \psi^{(1)}(z)=\frac{1}{48} \left[\frac{z^2 \left(9+27 z^2+14 z^4\right)}{\left(1+z^2\right)^3}-\frac{6 z}{1+z^2} \arctan z-3 \arctan^2z\right]\,.
\end{align}
On the left panel of Fig.~\ref{fig:toroidalscalar} we compare our analytic expansion up to $i=8$, with the numerical data and find excellent agreement for a large range of $A_0$.
\subsubsection{Negative modes with toroidal boundaries}
Studying perturbations of wormholes with planar boundaries turns out to be schematicall similar to studying those with spherical boundaries, though it is easier in practice. Again, we expand all our perturbations in terms of scalar, vector and tensor harmonics on $\mathbb{T}^3$ obeying to
\begin{subequations}
\begin{equation}
\Box_{\mathbb{T}^3} \mathbb{S}^{k_S}+k_S^2 \mathbb{S}^{k_S}=0\,,
\end{equation}
\begin{equation}
\Box_{\mathbb{T}^3} \mathbb{S}^{k_V}_{i}+k_V^2 \mathbb{S}^{k_V}_{i}=0\quad \text{with}\quad \Grad^i \mathbb{S}_i^{k_V}=0\,,
\end{equation}
and
\begin{equation}
\Box_{\mathbb{T}^3} \mathbb{S}^{k_T}_{ij}+k_T^2 \mathbb{S}^{k_T}_i=0\,,\quad\text{with}\quad  \Grad^i \mathbb{S}_{ij}^{k_T}=0\quad \text{and}\quad \mathbbm{g}^{ij}\mathbb{S}^{k_T}_{ij}=0\,,
\end{equation}%
respectively. Modes with $k_S=k_V=0$ have to be studied separately, but $k_T=0$ can be obtained from the result with $k_T=0$.
\end{subequations}
\paragraph{Scalar-derived perturbations with $k_S=0$}
This section is similar in many respects to the section where we studied the negative mode of a spherically symmetric wormhole with respect to scalar-derived perturbations with $\ell=0$. We shall see that the large wormhole phase has no negative modes, whereas the small wormhole phase does seem to possess such a mode.

Our perturbations read
\begin{subequations}
\begin{equation}
\delta \mathrm{d}s^2 = \frac{L^2}{(1-y)^2}\left[\delta f(y) \frac{\mathrm{d}y^2}{(2-y)y}+\delta p(y) (\mathrm{d}x_1^2+\mathrm{d}x_2^2+\mathrm{d}x_3^2)\right]\,,
\end{equation}
and for the scalars
\begin{equation}
\delta \vec{\psi} = \vec{X}_k\,\delta \psi\,.
\end{equation}
\end{subequations}
Under an infinitesimal diffeomorphism $\xi = \xi_y\,\mathrm{d}y$ these perturbations transform as
\begin{subequations}
\begin{align}
& \delta f = \frac{2 (1-y)}{L^2}\left[1-4 y+2 y^2-\frac{y (2-y) (1-y) f'}{2 f}\right] \xi _y+\frac{2 y (2-y) (1-y)^2}{L^2} \xi _y'\,,
\\
& \delta p= \frac{2 (2-y) (1-y) y y_0^2}{L^2 f} \xi _y\,,
\\
& \delta \psi =\frac{(2-y) (1-y)^2 y \psi '}{L^2 f} \,\xi _y\,.
\end{align}
\label{eqs:transcalartoroidalk0}
\end{subequations}%

By now the procedure is familiar. We first expand the action (\ref{eq:actionscalargeneraltorus}) to second order in the perturbations. The resulting action, $S^{(2)}$ is a function of $\delta \psi$, $\delta f$ and their first derivatives with respect to $y$. Additionally, $S^{(2)}$ is also a function of $\delta p$ and its first and second derivatives. We first integrate by parts the term proportional to $\delta p^{\prime\prime}$ with the resulting boundary term cancelling the Gibbons-Hawking-York perturbed boundary action. $S^{(2)}$ is now a function of $\delta \psi$, $\delta f$, $\delta p$ and their first derivatives. One can also integrate one more my parts terms proportional to $\delta f^\prime$ (whose boundary terms cancel with the perturbed boundary counter terms appearing in (\ref{eq:actionscalargeneraltorus}) ). The second order action $S^{(2)}$ is then a function $\delta \psi$, $\delta p$ and their first derivatives and of $\delta f$. Crucially, $\delta f$ enters the action algebraically.  This means we can perform the Gaussian integral over $\delta f$ (again using the Wick rotation described in section \ref{sec:neggen}) and find an effective action $\check{S}^{(2)}$ that is a function of $\delta \psi$, $\delta p$ and their first derivatives only. At this stage we introduce the gauge invariant quantity
\begin{equation}
Q=\sqrt{6}\left[\delta \psi-\frac{(1-y)\psi^\prime}{2 y_0^2}\delta p\right]\,,
\end{equation}
where the factor of $\sqrt{6}$ was chosen for later convenience of presentation. Clearly $Q$ is invariant under the infinitesimal gauge transformations (\ref{eqs:transcalartoroidalk0}). Solving the above relation with respect to $\delta \psi-$ gives an action for $Q$, where the dependence in $\delta p$ completely drops out because of gauge invariance. The final action for $Q$ reads
\begin{equation}
\check{S}^{(2)}= 2\,L^2 \ell^3 \int_0^{+\infty}\mathrm{d}y\,\frac{ y_0^3 \sqrt{f}}{\sqrt{2-y} (1-y)^4 \sqrt{y}} \left[\frac{(2-y) (1-y)^2 y}{1-(1-y)^2 \psi'^2}\frac{Q'^2}{f}+V Q^2\right]\,,
\label{eq:quadraticzeromodetoroidal}
\end{equation}
where
\begin{equation}
V= \frac{1}{\psi ^2}\left\{\frac{(2-y) y}{f}+\frac{2 f \left[y_0^2-k^2 (1-y)^3 \psi ^3 \psi '\right]}{(2-y) y y_0^2 \left[1-(1-y)^2 \psi '^2\right]^2}-\frac{3}{1-(1-y)^2 \psi '^2}\right\}\,.
\end{equation}
As expected, $V$ is not positive definite for all wormholes. The combination $1-(1-y)^2 \psi '^2$ which appears multiplying the kinetic term for $Q$ is positive definite so long as $y_0/k=1$\footnote{This can be analytically proved by manipulating the scalar equation (\ref{eq:onescalartoiralode}) and numerically checked to be the case for the first $100$ digits.} and $y_0/k\gtrsim 0.4529(0)$, for the massless and conformally coupled cases, respectively. In particular, for the large wormhole branch $1-(1-y)^2 \psi '^2$ is positive definite.

To proceed, we use numerical methods. We first note that in the original $r$ coordinates of (\ref{eq:lineansatztoroidal}), $V$ would have been even around $r=0$. This means perturbations that are even and odd with respect to $r=0$ will be orthogonal, so we can study them separately. In terms of the $y$ coordinates, these correspond to very distinct behaviours near $y=0$. Namely, in the odd sector we have $Q\sim \sqrt{y}$ near the origin, while in the even sector $Q$ admits a regular Taylor expansion around $y=0$. We have not found any negative mode on the odd sector of perturbations.

To search for negative modes $\lambda$, we integrate (\ref{eq:lineansatztoroidal}) by parts and set
\begin{equation}
- \frac{\sqrt{2-y} (1-y)^4 \sqrt{y}}{\sqrt{f}} \left[\frac{1}{(1-y)^2} \frac{\sqrt{2-y}\sqrt{y}}{1-(1-y)^2 \psi'^2}\frac{Q'}{\sqrt{f}}\right]^\prime+V Q =\lambda Q\,.
\end{equation}
Near the boundary, we find that admits two possible boundary behaviours
\begin{equation}
Q = C_+ (1-y)^{\frac{3}{2}+\sqrt{\left(\Delta-\frac{3}{2}\right)^2-\lambda }}\left[1+\ldots\right]+C_- (1-y)^{\frac{3}{2}-\sqrt{\left(\Delta-\frac{3}{2} \right)^2-\lambda }}\left[1+\ldots\right]\,,
\end{equation}
with normalisability demanding we set $C_-=0$. We thus have a well defined Sturm-Liouville problem, which we can readily solve numerically.

The results of this analysis can be seen in Fig.~\ref{fig:negativetoroidal} where we plot $\lambda$ as a function of $y_0/y_0^{\min}$ for the massless (left panel) and conformal (right panel) cases. In both cases, a negative mode exists for the small wormhole phase, but becomes positive on the large wormhole phase. This establishes that large wormholes are stable with respect to scalar-derived perturbations with $k_S=0$.
\begin{figure}[h]
\centering
\includegraphics[width =\textwidth]{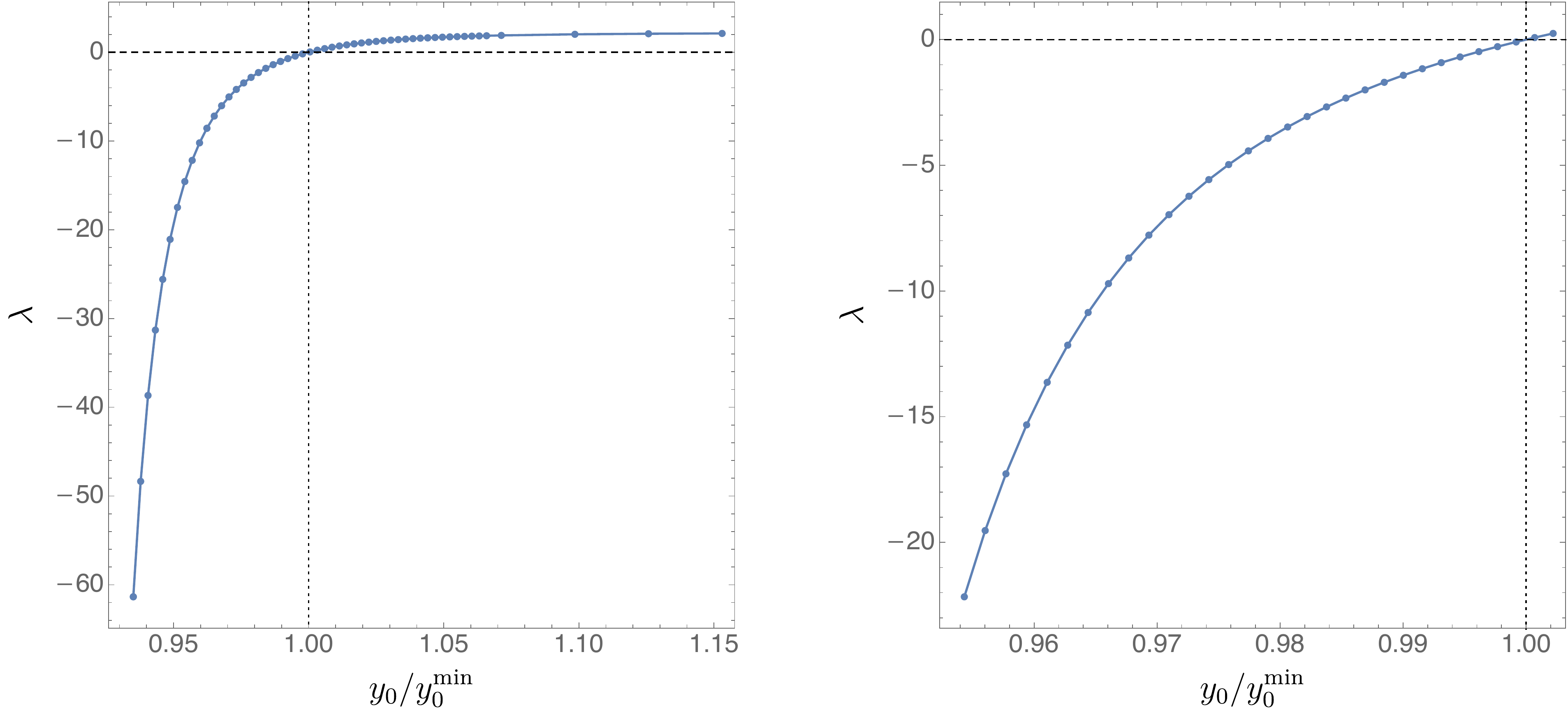}
\caption{Negative mode for the $k=0$ perturbations as a function of $y_0/y_0^{\min}$. {\bf Left panel}: negative modes of wormholes sourced by massless scalars. {\bf Right panel}: negative modes of wormholes generated by conformally coupled scalars.}
\label{fig:negativetoroidal}
\end{figure}
\paragraph{Scalar-derived perturbations with $k_S\neq0$}
We have chosen our boundary metric to be a torus so our fundamental scalar harmonic takes a very simple form
\begin{equation}
\mathbb{S}^{k_S} = \cos (\mathbf{k}_S\cdot \mathbf{x}+\gamma_S)\,,
\end{equation}
with $k_S = |\mathbf{k}_S|$ and $\mathbf{x}=\{x_1,x_2,x_3\}$. Note also that $\mathbf{k}_S=k\{n_1,n_2,n_3\}$, where $n_i$ are integers because our 3-torus is cubic. Finally, $\gamma_S$ is an arbitrary phase which will play no role. While we were able to keep all $n_i$ distinct, it is clear the most dangerous sector occurs when we take all $n_i=n$. This is the sector we present here. We also define $k_S=\sqrt{3}\kappa$, with $\kappa = n\,k$.

The metric perturbations take the already familiar form
\begin{subequations}
\begin{equation}
\delta \mathrm{d}s^2_{k_S} = h^{k_S}_{yy}(y)\,\mathbb{S}^{k_S}\,\mathrm{d}y^2+2 h^{k_S}_y(y)\,\Grad^i\mathbb{S}^{k_S}\,\mathrm{d}y\, \mathrm{d}x^{i}+H_T^{k_S}(y)\,\mathbb{S}^{k_S}_{ij}\,\mathrm{d}x^i \mathrm{d}x^j+H_L^{k_S}(y)\,\mathbb{S}^{k_S}\,\mathbbm{g}_{ij}\mathrm{d}x^i\mathrm{d}x^j
\label{eq:grandtorus}
\end{equation}
where
\begin{equation}
\mathbb{S}^{k_S}_{ij}=\Grad_i \Grad_j \mathbb{S}^{k_S}-\frac{\mathbbm{g}_{ij}}{3} \Grad^k \Grad_k \mathbb{S}^{k_S}\,,
\end{equation}
while for the scalar perturbation we choose
\begin{equation}
\delta\vec{\psi}_{k_S} = \vec{X}_k\,B_{k_S}(y)\,\mathbb{S}^{k_S}+(\Grad^i\mathbb{S}^{k_S}\Grad_i \vec{X}_k)\,A_{k_S}(y)\,.
\end{equation}
\end{subequations}%
In the above $\mathbbm{g}$ is the metric on $\mathbb{T}^3$ and $\Grad$ its associated metric preserving connection.

Under an infinitesimal diffeomorphism of the form
\begin{equation}
\xi^{k_S} = \xi^{k_S}_y(y)\,\mathrm{d}y+L^{k_S}_y(y)\,\Grad_i \mathbb{S}^{k_S}\,\mathrm{d}x^i
\end{equation}
the metric and scalar perturbations transform as
\begin{subequations}
\begin{align}
& \delta A_{k_S} = \frac{(1-y)^2\,\psi}{L^2\,y_0^2}L^{k_S}_y\,,
\\
& \delta B_{k_S} = \frac{(2-y)y(1-y)^2\,\psi^\prime}{L^2\,f}\xi^{k_S}_y\,,
\\
& \delta h_{yy}^{k_S} = \frac{2 \left(1-4 y+2 y^2\right)}{y (2-y) (1-y)} \xi _y^{k_S}-\frac{f'}{f}\xi _y^{k_S} +2 \xi _y^{k_S}{}'\,,
\\
& \delta h_{y}^{k_S} = \xi _y^{k_S}+L_y^{k_S}{}'-\frac{2}{1-y}L_y^{k_S}\,,
\\
& \delta H_{T}^{k_S} = 2L_y^{k_S}\,,
\\
& \delta H_{L}^{k_S} = \frac{2 (2-y) y y_0^2}{(1-y) f} \xi _y^{k_S}-2 \kappa ^2 L_y^{k_S}\,.
\end{align}
\label{eq:traroruskscalar2}
\end{subequations}%
Had we taken all $n_i$ distinct, we would have to consider three different perturbations similar to $A_{k_S}$ and $B_{k_S}$, parametrising each of the complex scalars in $\vec{\psi}$.

The remaining procedure is very similar to what we have seen when studying negative modes of the womholes sourced by the scalars with a spherical boundary. First, we write the second order action $S^{(2)}$  in first order form by integrating by parts and find it it can be written in term of $A_{k_S}$, $B_{k_S}$, $H_{L}^{k_S}$, $H_{T}^{k_S}$, $h_{y}^{k_S}$ and their first derivatives. However, $h_{yy}^{k_S}$ appears algebraically and we again apply the Wick-rotation procedure of section \ref{sec:neggen}.  We can thus perform the Gaussian integral and find a new second order action $\tilde{S}^{(2)}$ that is a function of $A_{k_S}$, $B_{k_S}$, $H_{L}^{k_S}$, $H_{T}^{k_S}$ and their first derivatives, but now $h_y^{k_S}$ enters algebraically and again we can perform the corresponding Gaussian integral finding an action $\check{S}^{(2)}$ that is a function of $A_{k_S}$, $B_{k_S}$, $H_{L}^{k_S}$, $H_{T}^{k_S}$ and their first derivatives.

At this point we introduce gauge invariant variables $Q^{k_S}_1$ and $Q^{k_S}_2$ defined by
\begin{subequations}
\begin{align}
&Q^{k_S}_1 = A_{k_S}-\frac{(1-y)^2\,\psi}{2L^2 y_0^2}H_T^{k_S}\,,
\\
&Q^{k_S}_2 = B_{k_S}-\frac{(1-y)^3\,\kappa^2\,\psi'}{2L^2 y_0^2}H_T^{k_S}-\frac{(1-y)^3\,\psi'}{2L^2 y_0^2}H_L^{k_S}\,.
\end{align}
\end{subequations}
which are invariant under the infinitesimal transformations (\ref{eq:traroruskscalar2}). Solving the above relations with respect to $A_{k_S}$ and $B_{k_S}$ and inputting those in $\check{S}^{(2)}$ gives an action for $Q^{k_S}_1$, $Q^{k_S}_2$ and their first derivatives only. The dependence in $H_T^{k_S}$ and $H_L^{k_S}$, after using the equations of motion for $\psi$, completely drops out by virtue of gauge invariance. To ease presentation we define further
\begin{align}
& Q^{k_S}_1 = \frac{1}{2 \sqrt{6}\,\pi ^{3/2} \sqrt{k}}m_y^\Delta\,\psi\,\left[q_1^{k_S}-\frac{(1-y)^2}{n^2}\psi^\prime q_2^{k_S}\right]\,,
\\
& Q^{k_S}_2 = \frac{1}{2 \sqrt{6}\,\pi ^{3/2} \sqrt{k}}m_y^\Delta\,q_2^{k_S}\,.
\end{align}
The second order action $\check{S}^{(2)}$ can then be written as
\begin{equation}
\check{S}^{(2)}=2L^2 \int_0^{+\infty}\mathrm{d}y\frac{y_0^3m_y^{2\Delta-4}\sqrt{f}}{\sqrt{n_y}}\left[\frac{m_y^2 n_y}{f} q_i^{k_S}{}'\mathbb{K}^{ij}q_j^{k_S}{}'+q_i^{k_S} \mathbb{V}^{ij}q_j^{k_S} \right]\,,
\end{equation}
where
\begin{equation}
\mathbb{K}^{-1}=
\left[
\begin{array}{cc}
 \frac{1}{n^4} \left(1+\frac{n^2}{\psi ^2}+3 m_y^2 \psi'^2\right)& \frac{2 m_y \psi'}{n^2} \\
 \frac{2 m_y \psi'}{n^2} & 1 \\
\end{array}
\right]\,.
\end{equation}
It is a simple exercise to show that $\mathbb{K}$ is positive definite so long as $1-(1-y)^2 \psi'^2$ is positive. However, we have argued that this is the case for all large wormhole. It then all boils down to the positivity properties of $\mathbb{V}$, whose explicit form we present in section \ref{eq:labd}. This easy to study numerically, and we find that, for $|n|=1$\footnote{Other values of $|n|>1$ are even more stable.}, $\mathbb{V}$ is positive definite for $y_0/k\gtrsim0.4173(5)$ and $y_0/k\geq0.9207(9)$ for the conformal and massless cases, respectively. Both these values are well within the small wormhole branch, so these results establish stability in the large wormhole branch.

\subparagraph{\label{eq:labd}Symmetric matrices for the scalars with toroidal boundary conditions}

The symmetric matrix $\mathbb{V}$ is given by
\begin{equation}
 \mathbb{V}_{IJ}=(\mathbb{K}^{-1})_{IK}(\mathbb{K}^{-1})_{JM}\mathbb{V}^{KM}\,.
\end{equation}
We then have
\begin{equation}
 \mathbb{V}_{IJ}=\frac{1}{A_0^2 f \psi ^2\,d_{IJ}}\sum_{i=0}^{6}{\psi^\prime}^i \mathbb{V}_{IJ}^{(i)}\,.
\end{equation}
with $\mathbb{V}_{12}^{(6)}=\mathbb{V}_{22}^{(6)}=\mathbb{V}_{12}^{(5)}=0$, $n_y=y(2-y)$, $m_y=1-y$ and
\begin{align}
& d_{11}=A_0^2\,n^4\,\psi\,n_y\,,
\\
& d_{12}=A_0^2\,n^2\,n_y\,,
\\
& d_{22}=1\,.
\end{align}
Furthermore
\begin{equation}
\mathbb{V}_{11}^{(6)}=72 A_0^4 \psi ^3 m_y^6 n_y^2\,,
\end{equation}
\begin{equation}
\mathbb{V}_{11}^{(5)}=48 A_0^4 \psi ^2 m_y^5 n_y^2\,,
\end{equation}
\begin{multline}
\mathbb{V}_{11}^{(4)}=A_0^4 \left(42 \Delta ^2 f \psi ^5 m_y^6 n_y-126 \Delta  f \psi ^5 m_y^6 n_y-27 \Delta  \psi ^3 m_y^4 n_y^2-72 \psi ^3 m_y^4 n_y^2+4 \psi  m_y^4 n_y^2\right)
\\
-42 A_0^2 f \psi ^5 m_y^6 n_y\,,
\end{multline}
\begin{multline}
\mathbb{V}_{11}^{(3)}=A_0^4 \left(16 \Delta ^2 f \psi ^4 m_y^5 n_y-48 \Delta  f \psi ^4 m_y^5 n_y+48 f \psi ^2 m_y^3 n_y-6 \Delta  \psi ^2 m_y^3 n_y^2-48 \psi ^2 m_y^3 n_y^2\right)
\\
-16 A_0^2 f \psi ^4 m_y^5 n_y\,,
\end{multline}
\begin{multline}
\mathbb{V}_{11}^{(2)}=A_0^2 \left(36 \Delta  f^2 \psi ^7 m_y^6-12 \Delta ^2 f^2 \psi ^7 m_y^6+9 f n^2 \psi ^3 m_y^4 n_y+9 \Delta  f \psi ^5 m_y^4 n_y+14 f \psi ^5 m_y^4 n_y\right)
\\
+A_0^4 (6 \Delta ^4 f^2 \psi ^7 m_y^6-36 \Delta ^3 f^2 \psi ^7 m_y^6+54 \Delta ^2 f^2 \psi ^7 m_y^6-9 \Delta ^2 f n^2 \psi ^3 m_y^4 n_y
\\
+27 \Delta  f n^2 \psi ^3 m_y^4 n_y-9 \Delta ^3 f \psi ^5 m_y^4 n_y+13 \Delta ^2 f \psi ^5 m_y^4 n_y+42 \Delta  f \psi ^5 m_y^4 n_y+4 f \psi  m_y^2 n_y
\\
-3 \Delta  n^2 \psi  m_y^2 n_y^2-3 \Delta ^2 \psi ^3 m_y^2 n_y^2+24 \Delta  \psi ^3 m_y^2 n_y^2-4 \psi  m_y^2 n_y^2)+6 f^2 \psi ^7 m_y^6\,,
\end{multline}
\begin{multline}
\mathbb{V}_{11}^{(1)}=A_0^4 (6 \Delta ^2 f^2 \psi ^4 m_y^3-18 \Delta  f^2 \psi ^4 m_y^3+4 \Delta ^2 f n^2 \psi ^2 m_y^3 n_y-12 \Delta  f n^2 \psi ^2 m_y^3 n_y
\\
-6 \Delta  f \psi ^2 m_y n_y+2 \Delta  n^2 m_y n_y^2+6 \Delta  \psi ^2m_y n_y^2)-A_0^2 \left(6 f^2 \psi ^4 m_y^3+4 f n^2 \psi ^2 m_y^3 n_y\right)\,,
\end{multline}
\begin{multline}
\mathbb{V}_{11}^{(0)}=A_0^4 (21 \Delta  f n^2 \psi ^3 m_y^2 n_y-\Delta ^3 f n^2 \psi ^3 m_y^2 n_y-4 \Delta ^2 f n^2 \psi ^3 m_y^2 n_y-3 \Delta ^2 f n^4 \psi  m_y^2 n_y
\\
+9 \Delta  f n^4 \psi  m_y^2 n_y+\Delta ^3 (-f) \psi ^5 m_y^2 n_y-\Delta ^2 f \psi ^5 m_y^2 n_y+12 \Delta  f \psi ^5 m_y^2 n_y-\Delta ^2 n^2 \psi  n_y^2
\\
+3 \Delta  n^2 \psi  n_y^2-\Delta ^2 \psi ^3 n_y^2+3 \Delta  \psi ^3 n_y^2)+A_0^2 (\Delta  f n^2 \psi ^3 m_y^2 n_y+7 f n^2 \psi ^3 m_y^2 n_y+3 f n^4 \psi  m_y^2 n_y
\\
+\Delta  f \psi ^5 m_y^2 n_y+4 f \psi ^5 m_y^2 n_y)\,,
\end{multline}
\begin{equation}
\mathbb{V}_{12}^{(5)}= 36 A_0^4 \psi ^2 m_y^5 n_y^2\,,
\end{equation}
\begin{equation}
\mathbb{V}_{12}^{(4)}=24 A_0^4 \psi  m_y^4 n_y^2\,,
\end{equation}
\begin{multline}
\mathbb{V}_{12}^{(3)}=A_0^4 \left(18 \Delta ^2 f \psi ^4 m_y^5 n_y-54 \Delta  f \psi ^4 m_y^5 n_y-12 \Delta  \psi ^2 m_y^3 n_y^2-36 \psi ^2 m_y^3 n_y^2+2 m_y^3 n_y^2\right)
\\
-18 A_0^2 f \psi ^4 m_y^5 n_y\,,
\end{multline}
\begin{multline}
\mathbb{V}_{12}^{(2)}=A_0^4 \left(8 \Delta ^2 f \psi ^3 m_y^4 n_y-24 \Delta  f \psi ^3 m_y^4 n_y+24 f \psi  m_y^2 n_y-2 \Delta  \psi  m_y^2 n_y^2-24 \psi  m_y^2 n_y^2\right)
\\
-8 A_0^2 f \psi ^3 m_y^4 n_y\,,
\end{multline}
\begin{multline}
\mathbb{V}_{12}^{(1)}=A_0^2 (12 \Delta  f^2 \psi ^6 m_y^5-4 \Delta ^2 f^2 \psi ^6 m_y^5+6 f n^2 \psi ^2 m_y^3 n_y+4 \Delta  f \psi ^4 m_y^3 n_y+6 f \psi ^4 m_y^3 n_y)
\\
+A_0^4 (2 \Delta ^4 f^2 \psi ^6 m_y^5-12 \Delta ^3 f^2 \psi ^6 m_y^5+18 \Delta ^2 f^2 \psi ^6 m_y^5-6 \Delta ^2 f n^2 \psi ^2 m_y^3 n_y
\\
+18 \Delta  f n^2 \psi ^2 m_y^3 n_y -4 \Delta ^3 f \psi ^4 m_y^3 n_y+6 \Delta ^2 f \psi ^4 m_y^3 n_y+18 \Delta  f \psi ^4 m_y^3 n_y+2 f m_y n_y
\\
-2 \Delta ^2 \psi ^2 m_y n_y^2 +12 \Delta  \psi ^2 m_y n_y^2-2 m_y n_y^2) +2 f^2 \psi ^6 m_y^5\,,
\end{multline}
\begin{multline}
\mathbb{V}_{12}^{(0)}=A_0^4 (2 \Delta ^2 f^2 \psi ^3 m_y^2-6 \Delta  f^2 \psi ^3 m_y^2+2 \Delta ^2 f n^2 \psi  m_y^2 n_y-6 \Delta  f n^2 \psi  m_y^2 n_y
\\
-2 \Delta  f \psi  n_y+2 \Delta  \psi  n_y^2)-A_0^2 \left(2 f^2 \psi ^3m_y^2+2 f n^2 \psi  m_y^2 n_y\right)
\end{multline}
\begin{equation}
\mathbb{V}_{22}^{(4)}=18 A_0^2 \psi ^2 m_y^4 n_y\,,
\end{equation}
\begin{equation}
\mathbb{V}_{22}^{(3)}=12 A_0^2 \psi  m_y^3 n_y\,,
\end{equation}
\begin{equation}
\mathbb{V}_{22}^{(2)}=A_0^2 (6 \Delta ^2 f \psi ^4 m_y^4-18 \Delta  f \psi ^4 m_y^4-3 \Delta  \psi ^2 m_y^2 n_y-18 \psi ^2 m_y^2 n_y+m_y^2 n_y) -6 f \psi ^4 m_y^4\,,
\end{equation}
\begin{equation}
\mathbb{V}_{22}^{(1)}=A_0^2 (4 \Delta ^2 f \psi ^3 m_y^3-12 \Delta  f \psi ^3 m_y^3+12 f \psi  m_y-12 \psi  m_y n_y)-4 f \psi ^3 m_y^3
\end{equation}
\begin{multline}
\mathbb{V}_{22}^{(0)}=A_0^2 (9 \Delta  f n^2 \psi ^2 m_y^2-3 \Delta ^2 f n^2 \psi ^2 m_y^2+\Delta ^3 (-f) \psi ^4 m_y^2+3 \Delta ^2 f \psi ^4 m_y^2+f-\Delta ^2 \psi ^2 n_y
\\
+3 \Delta  \psi ^2 n_y-n_y)+3 f n^2 \psi ^2 m_y^2+\Delta  f \psi ^4 m_y^2\,.
\end{multline}

\paragraph{Vector-derived perturbations with $k_V=0$}
This mode is special, and must be studied separately. It is the analogue of the mode with $\ell_V=1$ in the spherical case.  Again we find that the associated $S^{(2)}$ vanishes identically after integrating out $h_y$ (whose quadratic coefficient has the correct sign to make the integral over $h_y$ converge). Thus the other parts of this mode are pure-gauge.
\paragraph{Vector-derived perturbations with $k_V\neq0$}
Vector derived perturbations with $k_V\neq0$ follow a similar pattern to the spherical case. Recall that vector harmonics on $\mathbb{T}^3$ must be transverse, i.e. $\Grad_i \mathbb{S}^{i}=0$ and obey to
\begin{equation}
\Box_{\mathbb{T}^3}\mathbb{S}^{k_V}_{i}+k_V^2\,\mathbb{S}^{k_V}_{i}=0\,.
\end{equation}
One such example is for instance
\begin{equation}
\mathbb{S}^{k_V}_i \mathrm{d}x^i = \cos(k_V x_1+\gamma)\mathrm{d}x_3\,,
\end{equation}
where $\gamma$ is an unimportant phase and $k_V = n k$. From $\mathbb{S}_i^{k_V}$ we can construct the following symmetric tensor
\begin{equation}
\mathbb{S}^{k_V}_{ij}=\Grad_i\mathbb{S}^{k_V}_j+\Grad_j\mathbb{S}^{k_V}_i\,,
\end{equation}
which we use to build the most general vector-derived metric perturbation. Vector-derived perturbation with $k_V$ read
\begin{subequations}
\begin{equation}
\delta \mathrm{d}s^2_{k_V} = 2 h^{k_V}_y(y)\,\mathbb{S}_i^{k_V}\,\mathrm{d}y\, \mathrm{d}x^{i}+H_T^{k_V}(y)\,\mathbb{S}^{k_V}_{ij}\,\mathrm{d}x^i \mathrm{d}x^j
\end{equation}
while for the scalar perturbation we choose
\begin{equation}
\delta\vec{\Pi}_{k_V} = (\mathbb{S}_i^{k_V}\Grad^i \vec{X}_k)\,A_{k_V}(y)\,.
\end{equation}
\end{subequations}
The most general vector-derived infinitesimal diffeomorphism can be constructed via
\begin{equation}
\xi^{\ell_V} = L_y^{k_V} \mathbb{S}^{k_V}_i \mathrm{d}x^i\,.
\end{equation}
The above infinitesimal diffeomorphism induces the following gauge transformations
\begin{subequations}
\begin{align}
\delta h_y^{k_V} &=-\frac{2 L_y^{k_V}}{1-y}+{L_y^{k_V}}'\,,
\\
\delta H_T^{k_V} &=L_y^{\ell_V}\,,
\\
\delta A^{k_V}& =\frac{(1-y)^2\psi}{L^2 y_0^2}L_y^{k_V}\,.
\end{align}
\end{subequations}

By now the procedure should be very familiar. We expand the action to second order in perturbations, and write it in first order form. To do this, we have to integrate by parts, and the boundary terms cancel with the perturbed Gibbons-Hawking-York term. It turns out that $h_y^{k_S}$ enters the second order action $S^{(2)}$ algebraically, which means we can do the Gaussian integral and find a new action $\check{S}^{(2)}$ which is a function of $H_T^{k_S}$ and $A_{k_S}$ only. At this stage we introduce a gauge invariant variable $Q_{k_V}$ defined through the relation
\begin{equation}
A_{k_V}=\frac{\sqrt{k} m_y^{\Delta }}{2 \sqrt{2} \pi ^{3/2}} \sqrt{1+\frac{4 \psi ^2}{n^2}}Q_{k_V} +\frac{(1-y)^2\psi}{L^2 y_0^2}H_T^{k_V}
\end{equation}
which leads to the following second order action
\begin{equation}
\check{S}^{(2)}=2 L^2 \int_0^{\infty}\mathrm{d}y\,\frac{\sqrt{f} y_0^3 m_y^{2 (\Delta -2)}}{\sqrt{n_y}}\left[\frac{m_y^2 n_y}{f}{Q_{k_V}'}^2+V Q_{k_V}^2\right]
\end{equation}
with
\begin{multline}
V = m_y^2 \left(\frac{1}{A_0^2}-\mu ^2 L^2\right) \left(n^2+4 \psi ^2+\Delta  \psi^2\right)-\frac{n_y \mu ^2 L^2}{f}+\frac{n^2}{\left(n^2+4 \psi ^2\right) \psi ^2}
\\
-\frac{n_y}{f} \left\{3 \Delta  m_y^2 \psi '^2+\frac{1}{\left(n^2+4 \psi ^2\right) \psi^2}\left[n^2-\frac{m_y^2 n^2 \left(n^2-8 \psi ^2\right) \psi '^2}{n^2+4 \psi ^2}\right]\right\}\,,
\end{multline}
where we recall that $n_y\equiv y(2-y)$, $m_y \equiv 1-y$ and $k_V = n\,k$. Though it is not apparent from the above expression, it turns out that $V$ seems positive definite for all wormholes we have constructed. In particular, it appears positive for the large wormholes, thus establishing stability with respect to vector-derived perturbations in this sector as well.
\paragraph{Tensor-derived perturbations}
We now come to the easiest sector of perturbations. The building blocks for perturbations in this sector are given by tensor harmonics on $\mathbb{T}^3$, which obey to
\begin{equation}
\Box_{\mathbb{T}^3}\mathbb{S}^{k_T}_{ij}+k_T^2 \, \mathbb{S}^{k_T}_{ij}=0
\end{equation}
with $\Grad^i \mathbb{S}^{k_T}_{ij}=0$ and $\mathbbm{g}^{ij}\mathbb{S}^{k_T}_{ij}=0$. An example of such an harmonic is
\begin{equation}
\mathbb{S}^{k_T}_{ij}\mathrm{d}x^i \mathrm{d}x^ j = \cos(k_T\,x_1+\gamma)\mathrm{d}x_2\,\mathrm{d}x_3
\end{equation}
where $\gamma$ is again an arbitrary phase and $k_T = n k$.

The metric perturbation is simply given by
\begin{equation}
\delta \mathrm{d}s^2_{k_T} =\frac{\sqrt{2} k^{3/2} L^2 y_0^2}{\pi ^{3/2}\,m_y^2}\, H_{k_T}(y)\,\mathbb{S}^{k_T}_{ij}\mathrm{d}x^i \mathrm{d}x^ j\,,
\end{equation}
while for $\psi$ we demand $\delta \psi = 0$. In this case $H_{k_T}$ is automatically gauge invariant, and bringing the quadratic action to first order form yields
\begin{equation}
\check{S}^{(2)} = 2L^2 \int_0^{+\infty}\mathrm{d}y\,\frac{\sqrt{f} y_0^3}{m_y^4\sqrt{n_y}}\left[\frac{m_y^2 n_y}{f}{H_{k_T}'}^2+V H_{k_T}^2\right]
\end{equation}
with
\begin{equation}
V = \frac{k^2 m_y^2}{y_0^2} \left(n^2+4 \psi ^2\right)\,,
\end{equation}
which is manifestly positive. This thus establishes that no negative modes exist in the tensor sector of perturbations.

\section{Table with longer list of models studied}
\label{app:table}
This appendix provides a table (below) listing the 22 string/M-theory compactifications and the 14 ad hoc low energy models that we have studied most fully, along with the results obtained.  The methods applied to obtain these results are much like those explained in detail in the main text and in appendix \ref{app:torusboundaries} for the cases reported there.  Models 1-4, 20, 23, 27, and the $d=4$ version of 26  were analyzed in the main text, and models 24, 29, and the $d=4$ version of 28 were discussed in appendix \ref{app:torusboundaries}.

In most cases, models are described by giving the bibliography reference that describes them.
The consistent truncations leading to the models labelled by IIB were discussed in section \ref{sec:IIB}. Models labelled by $E$ are pure gravity models, labelled by $ES$ are Einstein-scalar models and labelled by $EM$ are Einstein-Maxwell models. The model labelled by $M^\dagger$ consists of a reduction of 11-dimensional supergravity on $\mathrm{AdS}_3\times S^2\times \mathbb{T}^6$ of the form
\begin{subequations}
\begin{multline}
\mathrm{d}s^2 = g_{ab}\mathrm{d}x^a\mathrm{d}x^b+L^2 \mathrm{d}\Omega_2^2+e^{\sqrt{2}\phi_1}\mathrm{d}x_1^2+e^{-\sqrt{2}\phi_1}\mathrm{d}x_2^2
\\
+e^{\sqrt{2}\phi_2}\mathrm{d}x_3^2+e^{-\sqrt{2}\phi_2}\mathrm{d}x_4^2+e^{\sqrt{2}\phi_3}\mathrm{d}x_5^2+e^{-\sqrt{2}\phi_3}\mathrm{d}x_6^2
\end{multline}
\begin{equation}
F_{(4)} = \frac{L}{2} \mathrm{d}^2 \Omega_2 \wedge \left(\mathrm{d}z_1\wedge \mathrm{d}z_2+\mathrm{d}z_3\wedge \mathrm{d}z_4+\mathrm{d}z_5\wedge \mathrm{d}z_6\right)\,.
\end{equation}
where $\mathrm{d}^2 \Omega_2$ is the two-dimensional volume form on the round $S^2$.  Perhaps notably, our list does not include the model described in section 5 of \cite{Maldacena:2004rf} where the disconnected solution remains to be found in order to determine if the wormhole described there will dominate.
\end{subequations}

\vfil \eject

\begin{center}
\begin{table}[h]
\scriptsize
\begin{tabular}{|@{\makebox[2em][c]{\rownumber\space}} |c|c|c|c|c|c|c|c|}
\hline\hline
\multicolumn{1}{|@{\makebox[2em][c]{\#}} | c|}{\bf Model} & \bf Fields & $\partial\mathcal{M}$& \bf E& \bf WD & \bf NM & \bf BN
\\
\hline\hline
 \cite{Cvetic:1999au} &  $\phi=\chi=0, A^{i}_{(1)}=\Phi\,\delta^{i}_1, \widetilde{A}^{i}_{(1)}=\Phi\,\delta^{i}_1, i = 1,2,3$ & $S^3$& Y &Y & ND & YD
 \\
 \hline
  \cite{Cvetic:1999au} &  $\phi=\chi=0, A^{i}_{(1)}=\Phi\,\delta^{i}_1, \widetilde{A}^{i}_{(1)}=\Phi\,\delta^{i}_1, i = 1,2,3$ & $\widetilde{S}^3$& Y &Y & ND & YD
    \\
   \hline
  \cite{Cvetic:1999au} &  $\phi\neq0,\chi\neq0, A^{i}_{(1)}= \widetilde{A}^{i}_{(1)}=0, i = 1,2,3$ & $\mathbb{H}^3$& Y &? & ? & Y
    \\
   \hline
  \cite{Cvetic:1999au} &  $\phi\neq0,\chi\neq0, A^{i}_{(1)}= \widetilde{A}^{i}_{(1)}=0, i = 1,2,3$ & $\widetilde{\mathbb{H}}^3$& Y &? & ? & Y
   \\
 \hline
 \cite{Azizi:2016noi} &  $A^{i}=\Phi\,\sigma_i$, $\varphi_1=\varphi_2=-\varphi_3, \sigma_i = A^{4}= 0,i=1,2,3$ & $S^3$& Y &Y & ND & YD
 \\
\hline
 \cite{Azizi:2016noi} &  $A^{i}=\Phi\,\sigma_i$, $\varphi_1=\varphi_2=-\varphi_3, \sigma_i = A^{4}= 0,i=1,2,3$ & $\widetilde{S}^3$& Y &Y & ND & YD
 \\
\hline
 \cite{Azizi:2016noi} &  $A^{i}=\Phi,\varphi_1=\varphi_2=-\varphi_3, \sigma_i = A^{4}= 0, i=1,2,3$ & $\mathbb{T}^3$& Y &? & ND & Y
  \\
\hline
 \cite{Azizi:2016noi} &  $A^{i}=A^{4}=0,\varphi_i=\varphi, \sigma_i = k\,x_i,i=1,2,3$ & $\mathbb{T}^3$& N &\cellcolor{Gray} & \cellcolor{Gray} & \cellcolor{Gray}
      \\
   \hline
  \cite{Girardello:1999bd,Bobev:2018eer} &  $\phi,\chi$ & $\mathbb{H}^4$& Y &? & ? & Y
      \\
   \hline
  \cite{Bobev:2016nua,Arav:2020obl} &  $z_4=-z_3=-z_2, \beta_1=\beta_2=0$ & $\mathbb{T}^4$& N &\cellcolor{Gray} & \cellcolor{Gray} & \cellcolor{Gray}
 \\
   \hline
  \cite{Bobev:2016nua,Arav:2020obl} &  $z_1=z_2=-z_3=-z_4, \beta_1\neq0,\beta_2=0$ & $\mathbb{T}^4$& N &\cellcolor{Gray} & \cellcolor{Gray} & \cellcolor{Gray}
   \\
   \hline
  \cite{Bobev:2016nua,Arav:2020obl} &  $z_1=z_3,z_2=z_4=\beta_2=0, \beta_1\neq0$ & $\mathbb{T}^4$& N &\cellcolor{Gray} & \cellcolor{Gray} & \cellcolor{Gray}
  \\
\hline
 \cite{Cvetic:1999xp} &  $A^{i}=B_0\,\mathbb{A}_1,\varphi_1=\varphi_2=0, i=1,2,3$ & $\mathbb{CP}_1\times \mathbb{H}^2$& Y &? & ? & Y
   \\
\hline
 \cite{Cvetic:1999xp} &  $A^{i}=B_0\,x_1 \mathrm{d}x_2,\varphi_1=\varphi_2=0, i=1,2,3$ & $\mathbb{T}^4$& N &\cellcolor{Gray} & \cellcolor{Gray} &\cellcolor{Gray}
   \\
\hline
 \cite{Cvetic:1999xp} &  $A^{i}=B_0\,\mathbb{A}_1,\varphi_1=\varphi_2=0, i=1,2,3$ & $\mathbb{CP}_1\times \mathbb{T}^2$& N &\cellcolor{Gray} & \cellcolor{Gray} &\cellcolor{Gray}
   \\
\hline
 \cite{Cvetic:1999xp} &  $A^{i}=\Phi\,\hat{\sigma}_i,\varphi_1=\varphi_2=0, i=1,2,3$ & $S^1\times S^3$& N &\cellcolor{Gray} & \cellcolor{Gray} &\cellcolor{Gray}
    \\
\hline
 \cite{Cvetic:1999xp} &  $A^{i}=\Phi\,\hat{\sigma}_i,\varphi_1=\varphi_2=0, i=1,2,3$ & $S^1\times \widetilde{S}^3$& N &\cellcolor{Gray} & \cellcolor{Gray} &\cellcolor{Gray}
      \\
\hline
 \cite{Pope:1985bu} &  $A^{i}=\Phi_i \hat{\sigma}_i, i=1,2,3$ & $S^3$& Y &N & Y &?
     \\
\hline
 \cite{Pope:1985bu} &  $A^{i}=\Phi_1 \hat{\sigma}_i,A^{i}=\Phi_2 \hat{\sigma}_3, i=1,2$ & $S^3$& Y &N & Y &?
 \\
 \hline
IIB &  $\phi_i= \Phi \mathbf{x}_i,\phi_4=0, |\mathbf{x}|=1, i = 1,2,3$ & $S^2$& Y &Y & N &Y
\\
\hline
IIB &  $\phi_1+i \phi_2= \Phi e^{i k x_1}, \phi_3+i \phi_4= \Phi e^{i k x_2}$ & $\mathbb{T}^2$& Y &? & N &Y
\\
\hline
$M^\dagger$&$\phi_i= \Phi \mathbf{x}_i, |\mathbf{x}|=1, i = 1,2,3$ & $S^2$& Y &Y & N &Y
\\
\hline\hline
$EM$&$d=3, A_i = \Phi_i \hat{\sigma}_i,i=1,2,3$ & $S^3$& Y &Y & ND &\cellcolor{Gray}
\\
\hline
$EM$&$d=3, 2\,A^I = L\,B_0\,\varepsilon^{IJK}\,x_J\,\mathrm{d}x_K,I=1,2,3$ & $\mathbb{T}^3$& Y &Y & N &\cellcolor{Gray}
\\
\hline
$ES$&$d=3,4, 5,\phi_i= \Phi \mathbf{x}_i, |\mathbf{x}|=1, i = 1,\ldots,d-1,\mu^2=0$ & $S^{d}$& Y &Y & ND &\cellcolor{Gray}
\\
\hline
$ES$&$d=3,\phi_i= \Phi \mathbf{x}_i, |\mathbf{x}|=1, i = 1,\ldots,3,\mu^2=-2$ & $S^3$& Y &Y & ND &\cellcolor{Gray}
\\
\hline
$ES$&$d=4,\phi_i= \Phi \mathbf{x}_i, |\mathbf{x}|=1, i = 1,\ldots,4,\mu^2=-3$ & $S^4$& Y &Y & ND &\cellcolor{Gray}
\\
\hline
$ES$ &  $d=3,4,5, \phi_{i}+i \phi_{i+1}= \Phi e^{i k x_i}, i =1,\dots,d-1,\mu^2=0$ & $\mathbb{T}^{d}$& Y &Y & ND &\cellcolor{Gray}
\\
\hline
$ES$ &  $d=3, \phi_{i}+i \phi_{i+1}= \Phi e^{i k x_i}, i =1,2,3,\mu^2=-2$ & $\mathbb{T}^{3}$& Y &Y & ND &\cellcolor{Gray}
\\
\hline
$ES$ &  $d=4, \phi_{i}+i \phi_{i+1}= \Phi e^{i k x_i}, i =1,2,3,4,\mu^2=-3$ & $\mathbb{T}^{4}$& Y &Y & ND &\cellcolor{Gray}
\\
\hline
$EM$&$d=4, F_1=\star F_2$ & $S^2\times S^2$& Y &? & N &\cellcolor{Gray}
\\
\hline
$E$&$d=3$ & $\widetilde{S}^3$ & N &\cellcolor{Gray} & \cellcolor{Gray} &\cellcolor{Gray}
\\
\hline
$EM$&$d=3,A = \Phi \hat{\sigma}_3$ & $\widetilde{S}^3$ & N &\cellcolor{Gray} & \cellcolor{Gray} &\cellcolor{Gray}
\\
\hline
$EM$&$d=4$ & $\mathbb{CP}_2$ & Y &? & N &\cellcolor{Gray}
\\
\hline
$EM$&$d=6$ & $\mathbb{CP}_3$ & Y &? & N &\cellcolor{Gray}
\\
\hline
$EM$&$d=3$ & $S^1\times S^2$& N &\cellcolor{Gray} & \cellcolor{Gray} &\cellcolor{Gray}
\\
\hline\hline
\end{tabular}
\caption{The columns are as follows:  {\bf Model} gives the citation where to find the given supergravity model or indicates the model as described above, ${\bf Fields}$ lists the supergravity fields taken to be non-trivial, $\partial\mathcal{M}$ gives the boundary manifold, {\bf E} states whether co-homogeneity one wormholes exist in the model, {\bf WD} states whether wormholes ever dominate over the disconnected solution, {\bf NM} states whether the model has field-theoretic negative modes and {\bf BN} states whether the model suffers from brane nucleation instabilities. In most cases Y/N indicates yess/no.  Question marks (?) indicate issues not investigated or not resolved.  In the columns marked by {\bf NM} and {\bf BN} the possible entries are: Y - all wormholes have brane nucleation instabilities/ field theoretical negative modes; YD - all wormholes that dominate over the disconnected solutions have brane nucleation instabilities/ field theoretical negative modes; N - all wormholes are free of brane nucleation instabilities/ field theoretical negative modes; ND - all wormholes that dominate over the disconnected solutions are free of brane nucleation instabilities/ field theoretical negative modes. Boundary manifolds marked with $\widetilde{\,}$ are squashed at the boundary: e.g. $\widetilde{S^3}$ denotes a boundary squashed $S^3$. Entries filled in grey represent situations for which it does not make sense to fill the respective entry. Model 9 is closely related to the mass deformation of $\mathcal{N}=4$ SYM studied in \cite{Buchel:2004rr}.}
\end{table}
\end{center}

\bibliographystyle{JHEP}
\bibliography{bibfile}

\end{document}